\documentclass[12pt]{article}
\usepackage{graphicx,amssymb,amsfonts,latexsym,amsmath,amsthm,times}
\usepackage{epsfig}
\usepackage{fancyhdr}
\usepackage{color}
\usepackage[english]{babel}
\usepackage{subfigure}
\numberwithin{equation}{section}

\newtheorem{theorem}{Theorem}

\newcommand{\D}{\mathrm{d}}

\begin{document}

\footnotesize {\flushleft \mbox{\bf \textit{Math. Model. Nat.
Phenom.}}}
 \\
\mbox{\textit{{\bf Vol. 6, No. 5, 2011, pp. 184-262}}}

\thispagestyle{plain}

\vspace*{2cm} \normalsize \centerline{\Large \bf Quasichemical
Models \\ of Multicomponent Nonlinear Diffusion}

\vspace*{1cm}

\centerline{\bf Alexander N. Gorban$^a$\footnote{Corresponding
author. E-mail: ag153@le.ac.uk}, Hrachya P. Sargsyan$^b$ and
Hafiz A. Wahab$^a$ }

\vspace*{0.5cm}


\centerline{$^a$ Department of Mathematics, University of
Leicester, Leicester, LE1 7RH, UK}

\centerline{$^b$ UNESCO Chair -- Life Sciences International
Postgraduate} \centerline{Educational Center (LSIPEC), Yerevan,
Republic of Armenia}


\vspace*{1cm}

\noindent {\bf Abstract.} Diffusion preserves the positivity of
concentrations, therefore, multicomponent diffusion should be
nonlinear if there exist non-diagonal terms. The vast variety
of nonlinear multicomponent diffusion equations should be
ordered and special tools are needed to provide the systematic
construction of the nonlinear diffusion equations for
multicomponent mixtures with significant interaction between
components. We develop an approach to nonlinear multicomponent
diffusion based on the idea of the reaction mechanism borrowed
from chemical kinetics.

Chemical kinetics gave rise to very seminal tools for the
modeling of processes. This is the stoichiometric algebra
supplemented by the simple kinetic law. The results of this
invention are now applied in many areas of science, from
particle physics to sociology. In our work we extend the area
of applications onto nonlinear multicomponent diffusion.

We demonstrate, how the mechanism based approach to
multicomponent diffusion can be included into the general
thermodynamic framework, and prove the corresponding
dissipation inequalities. To satisfy  thermodynamic
restrictions, the kinetic law of an elementary process cannot
have an arbitrary form. For the general kinetic law  (the
generalized Mass Action Law), additional conditions are proved.
The cell--jump formalism gives an intuitively clear
representation of the elementary transport processes and, at
the same time, produces kinetic finite elements, a tool for
numerical simulation.

\vspace*{0.5cm}




\tableofcontents
\newpage

\section{Introduction}

\subsection{Linear Diffusion: from Graham and Fick \\ to Einstein,
Onsager and Teorell}

\subsubsection{Fick's Law}

The first prominent equation of diffusion is Fick's Law.
According to this law, the diffusion flux $J$ is proportional
to the antigradient of the concentration $c$:
\begin{equation}\label{PreFick}
J=-D \mathrm{grad} c\, .
\end{equation}
The time derivative of the concentration is the negative of the
divergence of the flux:
\begin{equation}\label{Fick}
\frac{\partial c}{\partial t}=-\mathrm{div}J=D \Delta c\, ,
\end{equation}
where $\Delta$ is the Laplace operator.

This statement is closely related to the Gauss--Ostrogradskii
theorem
\begin{equation}\label{Gauss--Ostrogradskii}
\iiint\limits_V\left(\mathrm{div}J\right)\, \D
V=\iint\limits_{S}\!\!\!\!\!\!\!\!\!\!\!\subset\!\supset J\cdot
n\,\D S \, .
\end{equation}
The left side is an integral over the volume $V$, the right
side is the surface integral over the boundary $S$ of the
volume $V$, $S=\partial V$, and $n$ is the outward pointing
unit normal field of $S$. The right-hand side represents the
total flow across the boundary ``out of the volume $V$". The
theorem was first discovered by J. L. Lagrange in 1762, then
later independently rediscovered by C. F. Gauss in 1813, by G.
Green in 1825 and in 1831 by M. V. Ostrogradsky. According to
this theorem, the diffusion equation $\partial_t
c=-\mathrm{div}J$ is just a conservation law: all changes of
concentration are caused by the flux only.

In his work on diffusion law, A. Fick used the conservation of
matter and the analogy between diffusion and Fourier's law for heat
conduction (1822), or Ohm's law for electricity (1827). Development
of the fundamental law of diffusion was inspired by the Graham's
investigations  on the diffusion of salts in water
\cite{Graham1850}, in which he studied and compared the
diffusibility of different salts.

Before his study of diffusion in liquids, Graham studied diffusion
in gases (1833). In 1863, J. C. Maxwell calculated the diffusion
coefficients in gases from the Graham data. The results are amazing:
``His coefficient of diffusion of CO$_2$ in air is accurate at
$\pm$5\%. Isn't it extraordinary?" \cite{History}.

Maxwell's theory of diffusion was based on gas kinetics and
mean fee path estimates.

\subsubsection{Einstein's Mobility}

 In his theory of Brownian motion, A.
Einstein \cite{Einstein1905} developed the microscopic theory
of the diffusion coefficient for diluted particles in a liquid
and compared two processes: the motion of particles in a liquid
under a constant external force $K$, and diffusion. For a given
$K$, each particle has the average velocity $\mathfrak{m} K$
where the coefficient $\mathfrak{m}$ characterizes mobility of
particles. (We use $\mathfrak{m}$ for mobility and reserve
$\mu$ for chemical potential.) For spherical particles in
liquid,
\begin{equation}\label{mobility}
\mathfrak{m}=\frac{1}{6\pi \eta r}\, ,
\end{equation}
where $\eta$ is the viscosity of the liquid, $r$ is  the radius
of particles, and ${6\pi \eta r}$ is the Stokes friction force.

This approach results in a very useful relation for the
diffusion coefficient:
\begin{equation}\label{EinstSmolForm}
D=\mathfrak{m} \frac{RT}{N_A}=\mathfrak{m} k_B T\, ,
\end{equation}
where $R$ is the gas constant, $N_A$ is the Avogadro constant,
$k_B$ is the Boltzmann constant. The coefficient $\mathfrak{m}$
is called mobility or the Einstein mobility.

Graham's experimental research was extended to solids by W. C.
Roberts-Austen \cite{Roberts-Austen1896}. He used Fick's
equation to determine the diffusion coefficient
\cite{History2}. In 1922, S. Dushman and I. Langmuir
\cite{DushmanLangmuir1922} proposed to use the Arrhenius law
for diffusion coefficient:
\begin{equation}
D=D_0 \exp{(-Q/kT)}\, ,
\end{equation}
where $Q$ is a constant, which we now  recognize as the
activation Gibbs energy of diffusion $\Delta G$. More
precisely, $\Delta G$ includes two terms: $\Delta G= \Delta H -
T\Delta S$, where $\Delta H$ is the activation enthalpy and
$\Delta S$ is the activation entropy.

They checked this law by their own experiments with the
diffusion of thorium through tungsten and found satisfactory
agreement. Even better agreement was found with the published
results of W. C. Roberts-Austen's experiments.

\subsubsection{Teorell Formula}

\begin{figure}
\centering
\includegraphics[width=0.7\textwidth]{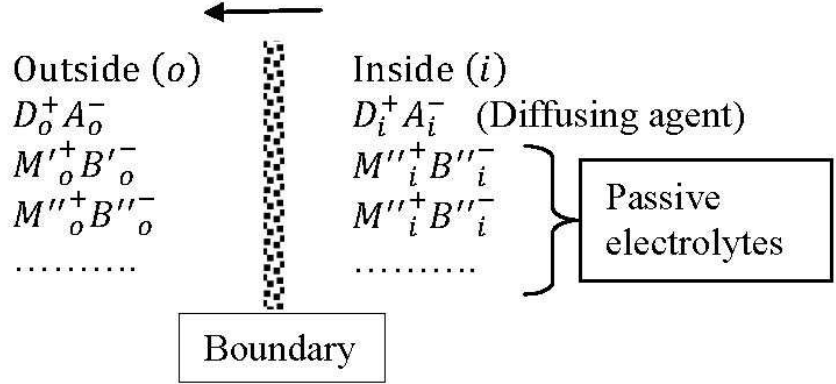}
\caption{Teorell's model \label{fig:energy} \cite{Teorell1935}. A system of ideally dilute solutions in water of binary
univalent strong electrolytes $DA$, $M'B'$, $M''B''$,...  is considered. The boundary membrane is permeable for all ions (cations $D$, $M$,... and anions $A$, $B$,...) but the movement of water
is prevented. The ionic concentrations outside are kept constant. Also $D_i^+$ is maintained constant. Accordingly, $DA$ is a
steadily diffusing electrolyte. No other electric field is present besides that due to diffusion potential. No chemical reaction takes place. The steady state of a system of this nature
with a steady diffusion is characterized by a constant ratio series: $\frac{M_i'}{M_o'}=\frac{M_i''}{M_o''}=\cdots = \frac{B_i'}{B_o'}=\frac{B_i''}{M_o''}=\cdots$.}
\end{figure}

The mobility--based approach was further applied by T. Teorell
\cite{Teorell1935}. In 1935, he studied the diffusion of ions
through a membrane (Fig.~\ref{fig:energy}). He considered a
system of an ideally dilute solution of binary univalent strong
electrolysis at the same temperature in water. The boundary is
considered to be a membrane with strong a electrolyte in the
presence of water. The solutions are assumed to be kept
homogeneous on both sides of the membrane up to the boundary by
some form of convection. The ionic mobilities within the
membrane are assumed constant and may be different, and the
membrane is not permeable for water. Heat effects, special
membrane effects and chemical reactions are ignored.

He formulated the essence of his approach in the formula:

\noindent {\bf the flux is equal to
mobility$\times$concentration$\times$force per gram ion}.

This is the so-called Teorell formula.

The force consists of two parts:
\begin{enumerate}
\item Diffusion force caused by concentration gradient: $-
    RT \frac{1}{c}\frac{\D c}{\D x}.$
\item Electrostatic force caused by electric potential
    gradient:  $q \frac{\D \varphi}{\D x}.$
\end{enumerate}
Here $R$ is the gas constant, $T$ is the absolute temperature, $c$
is the concentration, $q$ is the charge and $\varphi$ is the
electric potential.

In these notations, the Teorell formula for the flux $J$ is
\begin{equation}
J=\mathfrak{m} c\left(- \frac{RT}{c}\frac{\D c}{\D x} + q\frac{\D \varphi}{\D
x}\right) \label{Teorell}
\end{equation}
($\mathfrak{m}$ denotes mobility; here we slightly modernize
notations). It may be worthwhile to introduce the reference
equilibrium concentrations vector $c^*$ and write the diffusion
force in the form
\begin{equation}
- \frac{RT}{c}\frac{\D c}{\D x}=- RT \frac{\D \ln(c/c^*)}{\D x} \, .
\end{equation}
This expression allowed Teorell to find the concentration jump
and the electric potential across the membrane caused by the
joint action of diffusion and the electric field, when
mobilities of various components are different.

\subsubsection{Onsager's Linear Phenomenology}

In 1931, L. Onsager \cite{Onsager1,Onsager2} included diffusion in
the general context of linear non-equilibrium thermodynamics. For
multi-component diffusion,
\begin{equation}
J_i=\sum_j L_{ij} X_j \, ,
\end{equation}
where $J_i$ is the flux of the $i$th component and $X_j$ is the
$j$th thermodynamic force (for pure diffusion, this is the
space antigradient of the $j$th chemical potential divided by
$T$). After linearization near equilibrium, this approach gives
for perfect systems (for which the chemical potential is
$RT\ln(c/c^*)$):
\begin{equation}\label{OnsagerForm}
\begin{split}
&X_j=-\frac{1}{T}\mathrm{grad}\mu_j=-\frac{R}{c^*_j}\mathrm{grad}c_j ; \; \\ & J_i=-\sum_j L_{ij}
\frac{R}{c^*_j}\mathrm{grad} c_j;  \\
 &\frac{\partial c_i}{\partial t}=-\mathrm{div} J_i = {R} \sum_j L_{ij} \frac{\Delta c_j}{c^*_j}
 \, ,
\end{split}
\end{equation}
where $c^*_j$ are equilibrium constants ($c^*$ is the point of
linearization), deviations of $c_j$ from $c^*_j$ are assumed to
be small, $\Delta $ is the Laplace operator, and
$L_{ij}=L_{ji}$ is the matrix of the coefficients. Its symmetry
follows from microreversibility.

The system (\ref{OnsagerForm}) has one attractive property. Let us
consider this system in a bounded domain $V$ with smooth boundary
and with zero fluxes through its boundary: $(n,\mathrm{grad}c_j)=0$
at any point of $\partial V$ at any time ($n$ is the vector of the
outer normal). The positive quadratic functional
\begin{equation}
S_2=\frac{1}{2}\sum_i\int_V \frac{(c_i(x)-c^*_i)^2}{c_i^*} \, \D x
\end{equation}
is the second-order approximation to the relative entropy (or the
so-called Kulback-Leubler divergence, see the review paper
\cite{GGJ2010})
\begin{equation}
S_{KL}=\sum_i \int_V c(x) \ln \left(\frac{c_i(x)}{c_i^*}\right) \,
\D x \, .
\end{equation}
Let us calculate the time derivative of $S_2$ due to the system
(\ref{OnsagerForm}). Using the Gauss--Ostrogradskii formula
(\ref{Gauss--Ostrogradskii}) we get for the positive
semidefinite matrix $L$:
\begin{equation}\label{OnsENtrProdPerfect}
\frac{\D S_2}{\D t}= - R\int_V \sum_{ij} {L_{ij}} \left( \frac{\nabla
c_i}{c_i^*},\frac{\nabla c_j}{c_j^*}\right) \, \D x \leq 0 \, ,
\end{equation}
where $\left( \frac{\nabla c_i}{c_i^*},\frac{\nabla
c_j}{c_j^*}\right)$ is the inner product of the space vectors.

Therefore, $\frac{\D S_2}{\D t}\leq 0$ if the symmetric
coefficient matrix $L_{ij}$ is positive semidefinite (this
means that for any vector $\xi$ the following inequality holds:
$\sum_{ij} L_{ij} \xi_i \xi_j \geq 0$).

The Onsager form of the diffusion equations is correct near the
equilibrium but violates the obvious physical requirement: the
diffusion flux of the $i$th component is zero if its
concentration has zero value: {\it the flux vanishes with the
concentration}. The Teorell formula satisfies this requirement.
Fick's law also satisfies this requirement in the following
sense: if for nonnegative smooth $c(x)$ the concentration
vanishes at some points, then at these points the flux vanishes
too (because these points are minimizers of concentration and
the gradient vanishes there).

For isotropic non-perfect systems we have to use the generalized
thermodynamic forces in Onsager's form of the diffusion law:
\begin{equation}
X_j=-\left.\frac{\partial f}{\partial
c_j}\right|_{c=c^*}\mathrm{grad}c_j \, ,
\end{equation}
where $Tf(c,T)$ is the free energy density.

Let us denote $\Phi_{ij}=(\partial^2 f /
\partial c_i \partial c_j)_{c=c^*}$. In this notation,
\begin{equation}\label{OnsagerFormNonperf}
\begin{split}
&X_j=-\sum_k \Phi_{jk} \mathrm{grad}c_k ; \;  J_i=\sum_j L_{ij}
X_j=-\sum_k \left(\sum_j L_{ij}\Phi_{jk}\right)\mathrm{grad}c_k ;  \\
 &\frac{\partial c_i}{\partial t}=-\mathrm{div} J_i = \sum_k \left(\sum_j L_{ij}\Phi_{jk}\right)\Delta
 c_k
 \, .
\end{split}
\end{equation}
The quadratic form $$F_2=\frac{1}{2}\int \sum_{jk} \Phi_{jk}
(c_j-c_j^*)(c_k-c_k^*)\, \D x$$ is positive definite because
$F$ is strictly convex. For positive definite $L$, $F_2$
decreases in time due to diffusion. Indeed, in a bounded domain
$V$ with a smooth boundary and without fluxes through the
boundary we get analogously to (\ref{OnsENtrProdPerfect}):
\begin{equation}\label{OnsENtrProdNonPerfect}
\frac{\D F_2}{\D t}=
 -\int_V \sum_{ij} \left(\sum_{k}\Phi_{ik} \nabla c_k\right) {L_{ij}}
  \left(\sum_{l} \Phi_{jl} \nabla c_l\right) \, \D x  \leq 0 \, .
\end{equation}
For non-isotropic diffusion (for example, in crystals), the
coefficients $L$ have two pairs of indexes: $L_{i\alpha\,
j\beta}$, where $i,j$ correspond to components and $\alpha ,
\beta$ correspond to the space coordinates. The forces and
fluxes also have these two indexes and
$$J_{i \alpha}=\sum_{j\beta} L_{i\alpha\, j\beta} X_{j\beta}\, .$$

In all cases, the diffusion equations in Onsager's form do not
describe the non-diagonal terms (the influence of gradients
$c_i$ on fluxes of $c_j$ for $i\neq j$) properly near zeros of
concentrations. These equations are applicable near a reference
point $c^*>0$ only.

{\bf Non-diagonal diffusion must be non-linear.} This simple
remark is so important that we will explain it in detail. Let
diffusion be non-diagonal and linear: $$\partial_t c_i =\sum_j
D_{ij} \Delta c_j \, .$$ Assume that $D_{12}\neq 0$ and
consider the state with $c_2= \ldots = c_n=0$. At this state,
$$\partial_t c_2 = D_{12} \Delta c_1\, .$$ If $D_{12} \Delta
c_1(x) < 0$ at some points then $c_2(x)$ becomes negative at
these point in a short time. Therefore, linear non-diagonal
diffusion does not preserve positivity of concentrations.

\subsection{Mechanisms of Nonlinear Diffusion}

\subsubsection{Jumps on the Surface}

In 1980, A.N, Gorban, V.I. Bykov and G.S. Yablonskii
\cite{GorBykYab1980} proposed a model for diffusion in monolayers of
reagents on the surface of a catalyst, which is based on the jumps
of the reagents on the nearest free places. This model was used for
CO on Pt oxidation under low gas pressure.

The system includes several reagents $A_1,A_2,\ldots A_n$ on
the surface. Their surface concentrations are $c_1,c_2,\ldots
c_n$. The surface is a lattice of the adsorbtion places.  Each
reagent molecule fills a place on the surface. Some of the
places are free. We use $Z=A_0$ for a free place and the
concentration is $z=c_0$. The sum of all $c_i$ (including free
places) is constant: $$\sum_{i=0}^n c_i=b=\mathrm{const} \, .$$

The jump model gives for the diffusion flux of $A_i$ ($i=1,\ldots,
n$):
\begin{equation}\label{ExchangeFlux}
J_i=-D_i[z \nabla c_i - c_i \nabla z] \, .
\end{equation}
Therefore, the corresponding diffusion equation is:
\begin{equation}
\frac{\partial c_i}{\partial t}=- \mathrm{div}J_i=D_i[z \Delta c_i -
c_i \Delta z] \, .
\end{equation}

Due to the conservation law, $$z=b-\sum_{i=1}^n c_i \, ,$$ and we
have the system of $n$ diffusion equations:
\begin{equation}\label{Excangediff}
\begin{split}
&J_i=-D_i\left[\left(b-\sum_{i=1}^n c_i\right) \nabla c_i + c_i
\nabla \left(\sum_{i=1}^n c_i\right)\right] \\
&\frac{\partial c_i}{\partial t}=D_i\left[\left(b-\sum_{i=1}^n
c_i\right) \Delta c_i + c_i \Delta \left(\sum_{i=1}^n
c_i\right)\right] \, .
\end{split}
\end{equation}
It is straightforward to check that when $c_i\geq 0$ for all
$x$ then $\partial_t c_i \geq 0$ for $c_i=0$. This is a
necessary condition for preservation of positivity.

If we assume that all particles can exchange their positions
with their closest neighbors then a simple generalization of
(\ref{ExchangeFlux}), (\ref{Excangediff}) appears:
\begin{equation}\label{ExchangeGeneralDiff}
\begin{split}
&J_i=-\sum_j D_{ij}[c_j \nabla c_i - c_i \nabla c_j] \, ; \\
&\frac{\partial c_i}{\partial t}=\sum_j D_{ij}[c_j \Delta c_i - c_i
\Delta c_j] \, ,
\end{split}
\end{equation}
where $D_{ij} = D_{ji} \geq 0$ is a symmetric matrix of
coefficients which characterize the intensities of jumps.

The entropic Lyapunov functional for (\ref{Excangediff}),
(\ref{ExchangeGeneralDiff}) has a simple traditional form of perfect
relative entropy: for any reference vector of concentrations $c^*$
($c^*_i \geq 0$)
\begin{equation}
S_{KL}=\int \sum_i c_i \ln\left(\frac{c_i}{c_i^*}\right) \, \D x \,
.
\end{equation}
Remark: the free place entropy should be obligatorily included
into $S_{KL}$.

Simple algebra gives that in a bounded domain $V$ with a smooth
boundary and without fluxes through boundaries
\begin{equation}\label{ThermodynamicInequalityExc}
\frac{\D S_{KL}}{\D t}=-\sum_{ij} D_{ij}\int_V \left(\frac{c_i}{c_j}
\nabla c_j - \frac{c_j}{c_i} \nabla c_i\right)^2\, \D x \leq 0 \, .
\end{equation}
This inequality provides the Lyapunov stability of diffusion.

It is worth mentioning that the thermodynamic inequality
(\ref{ThermodynamicInequalityExc}) requires only the
non-negativity of coefficients $D_{ij}$ and does not imply any
requirements on the matrix $D$ as a whole (like positive
definiteness). Another form of the thermodynamic inequality
makes the formula for the entropy production more transparent:

\begin{equation}\label{ThermodynamicInequalityExc2}
\frac{\D S_{KL}}{\D t}=-\sum_{ij} D_{ij}\int_V \left(c_i \nabla \ln
\left(\frac{c_j}{c_j^*}\right) - c_j \nabla \ln
\left(\frac{c_i}{c_i^*}\right)\right)^2\, \D x \leq 0 \, .
\end{equation}

This system of models was further developed by A.N. Gorban and H.P.
Sargsyan \cite{GorbanSrk1986} and published in a book
\cite{Ocherki1986}.

\subsubsection{Diffusion in Solids as Reaction: from Frenkel to Eyring}

The physical idea of  the quasi-chemical representation of
diffusion in solids belongs to Yakov Frenkel
\cite{Frenkel1924,Frenkel1925}. He introduced both the vacancy
and the interstitial mechanisms of diffusion and found some
rate constants from experimental data.

Thirty years later, F. C. Frank and D. Turnbull developed the
Frenkel theory further \cite{FrankTurnbull1925}. They studied
the diffusion of copper in germanium. This diffusivity is very
rapid. They proposed that the copper could be dissolved in two
states, interstitial and substitutional. For the interstitial
state the solubility of copper is two orders of magnitude less
and the diffusivity many orders of magnitude greater than in
the substitutional state. The conversion of these states is
effected by lattice vacancies.

The quasi-chemical theory of diffusion and viscosity was developed
also by H. Eyring with co-authors \cite{EyringAtAlDiff1941}. Eyring
developed the theory of absolute reaction rates for chemical
reactions in gases \cite{Eyring1935} and in condensed phase
\cite{Wynne-Jones1935} and then applied these ideas to transport
phenomena.

In this theory, the transport process is represented by an
ensemble of elementary events. Each elementary event is
represented by the creation or disintegration of an activated
complex. The rate of the elementary process is given by the
concentration of activated complexes, multiplied by the rate at
which they decompose.

The main constructive hypothesis is that it is possible to
calculate the concentration of activated complexes by
equilibrium statistical thermodynamics: the complex
concentration is in quasi-equilibrium with the stable
components. Each complex has its ``internal translational"
degree of freedom. On the surface of potential energy this
corresponds to the ``reaction path". Complexes move along this
path. The velocity of this motion is assumed to be just a
thermal velocity and is proportional to $\sqrt{T}$.

The additional reaction path degree of freedom has its own
kinetic energy and, therefore, increases the complex heat
capacity. We have to take this into account in the calculation
of the equilibrium constant.

Collective models of diffusion were proposed too. One of the
earliest collective model is the Z. Jeffries ``ring mechanism"
with 4 or more atoms. More on the history of solid-state
diffusion is presented in the review \cite{Tuijn1997} and in a
modern textbook \cite{Mehrer2007}.

On the surface, there are various mechanisms for collective
diffusion \cite{OuraAtAl2003} as well. Elementary events for
these mechanisms involve many atoms simultaneously. A dynamic
description of nonlinear multicomponent diffusion requires a
unified framework that should satisfy basic physical
principles.

\subsubsection{Ginzburg--Landau Free energy and Cahn--Hilliard
equation}

The processes of phase separation has remained for a long time
an important source of problems and ideas for the theory of
nonlinear diffusion. The analogue for Fick's equation is the
Cahn--Hilliard equation \cite{CahnHilliard1958}.

The Cahn--Hilliard equation in its simplest form has the
standard Onsager form, the flux is proportional to the force,
the force is the gradient of the chemical potential:
\begin{equation}\label{simplestDiff}
J=-D \nabla \mu \, , \; \frac{\partial c}{\partial t}= -\mathrm{div}
J=D \Delta \mu \, .
\end{equation}
If we compare this equation to the Teorell formula then we
immediately find the missed factor $c$ (concentration). We will
return to the problem of the proper prefactor in the
Cahn-Hilliard equations later. The main specificity of the
Cahn-Hilliard equations is the form of the free energy and the
chemical potential \cite{CahnHilliard1958,Cahn1959} (the
Ginzburg--Landau form):
\begin{equation}\label{GinzLandFree}
f=f^c(c)+\gamma (\nabla c)^2\, , \; F=\int f \, \D x \, , \;
\mu=\frac{\partial f^c(c)}{\partial c}- \gamma \Delta c \, .
\end{equation}
The term $\gamma (\nabla c)^2$ in the  free energy penalizes
ovr sharp gradients and, in particular, models the interface
energy.

According to (\ref{simplestDiff}) and (\ref{GinzLandFree}), the
Cahn--Hilliard equation reads
\begin{equation}
J=-D\nabla \left(\frac{\partial f^c(c)}{\partial c} - \gamma \Delta
c\right)\, , \;
 \frac{\partial c}{\partial t}=D \Delta \left(\frac{\partial f^c(c)}{\partial c} - \gamma \Delta
 c\right) \, .
\end{equation}
If the chemical part of the free energy, $f^c(c)$ is not convex
then phase separation is possible. If at the point $c$ this
function is concave (spinodal) then without the term with
$\Delta^2$ the constant solution $c(x)=c$ for the diffusion
equation becomes unstable to any perturbation (negative
diffusion coefficient).  The term $-\gamma \Delta^2 c$ in the
right hand part of the Cahn--Hilliard equation regularizes
solutions and the existence theorem was proved for the
initial--boundary value problem given smooth initial data
\cite{Eliott1986}. The proof relies essentially on the
existence of a Lyapunov functional $F$ (\ref{GinzLandFree}).

The time derivative of $F$ in a domain $V$ with a smooth
boundary and without external fluxes ($(J,n)=0$ on the
boundary, where $n$ is the vector of outer normal to the
boundary) is
\begin{equation} \label{CahnEntropyProd}
\begin{split}
\frac{\D F}{\D t}&=\int_V \mu \partial_t c  \, \D x=
\int_V \mu \mathrm{div} J  \, \D x \\ &= -
\int_V  (\nabla \mu , J)  \, \D x =
- D \int_V  (\nabla \mu)^2  \, \D x \leq 0 \, .
\end{split}
\end{equation}
If we correct the Cahn--Hilliard equation by the ``Teorell"
factor $c$ then the dissipation inequality $\dot{F}\leq 0$
(\ref{CahnEntropyProd}) persists: Due to the Teorell formula
\begin{equation}\label{CahnHillTeorell}
\begin{split}
&J=-\mathfrak{m} c \nabla \mu = -\mathfrak{m}c  \nabla \left(\frac{\partial
f^c(c)}{\partial c} - \gamma \Delta c\right)\, , \\
 &\frac{\partial c}{\partial t}=-\mathrm{div}J=\mathfrak{m} \, \mathrm{div} \left( c\, \mathrm{grad} \left(\frac{\partial f^c(c)}{\partial c} - \gamma \Delta
 c\right)\right) \, .
\end{split}
\end{equation}
Here, $\mathfrak{m}$ is the Einstein mobility. This equation
should be called ``The Cahn--Hilliard--Teorell" equation. In
accordance with (\ref{CahnHillTeorell}),
\begin{equation} \label{CahnTeorEntropyProd}
\begin{split}
\frac{\D F}{\D t}&=\int_V \mu \partial_t c  \, \D x=
\int_V \mu \mathrm{div} J  \, \D x \\ &= -
\int_V  (\nabla \mu , J)  \, \D x =
- \mathfrak{m} \int_V c (\nabla \mu)^2  \, \D x \leq 0 \, .
\end{split}
\end{equation}
This dissipation inequality allows us to transfer all the
results about solutions of the Cahn--Hilliard equation to the
Cahn--Hilliard--Teorell equation.

\subsubsection{Teorell Formula for Non-perfect Systems}

It seems very natural that the flux is proportional to the
concentration of particles: the average velocity is
proportional to the force and the total flux is the product of
the average velocity and the amount of moving particles. This
could be proved in the framework of non-equilibrium
thermodynamics and the theory of absolute reaction rates when
the concentration of moving particles is small. In perfect
gases or in dilute solutions the chemical potential is $$\mu=RT
\ln c + \mu_0 \, ,$$ where $\mu_0$ does not depend on $c$ (it
is a function of $T$ and the state of the environment). In this
case, we neglect the interaction between moving particles and
use the Teorell formula (exactly as Einstein did in his theory
of Brownian motion 30 years before Teorell
\cite{Einstein1905}).

When the concentration of moving particles $c$ is not small
enough then the formula for perfect chemical potential is no
longer valid and in front of $c$ in the flux a special {\it
activity coefficient} $\alpha$ appears.

Such coefficients were introduced for diffusion by Eyring at al
in 1941 \cite{EyringAtAlDiff1941} and were used systematically
for the theory of nonlinear diffusion by Gorban at al in the
1980--1986 \cite{Ocherki1986}. Roughly speaking, the activity
$$a=\exp\left(\frac{\mu}{RT}\right) \, $$
should substitute the concentration $c$ in the Teorell formula
with the proper renormalization of the mobility coefficients:
$$J=\mathfrak{m}' a (-\nabla \mu + (\mbox{external force per gram particle})) \, .$$
The renormalized coefficient $\mathfrak{m}'$ is defined by the
condition: $(\mathfrak{m}' a) / (\mathfrak{m} c) \to 1$ for
$c\to 0$, $\nabla c \to 0$. This means that

$$\mathfrak{m}'=\mathfrak{m}
\exp\left(-\frac{\mu_0}{RT}\right)\, \mbox{ where }
\mu_0=\lim_{c \to 0}( \mu-RT\ln c) \, , $$
or we can write the {\em Teorell formula for non-perfect
systems} using the usual Einstein mobility $\mathfrak{m}$
defined for small concentrations and this standard value of
chemical potential, $\mu_0$:
\begin{equation}\label{nonperfectTeorell}
J=\mathfrak{m} \exp\left(\frac{\mu-\mu_0}{RT}\right)
(-\nabla \mu + (\mbox{external force per gram particle})) \, .
\end{equation}
This formula is the main analogue of Fick's law for
monomolecular diffusion in non-perfect media.

For the Cahn--Hilliard--Teorell equation
(\ref{CahnHillTeorell}), the Teorell formula for non-perfect
systems (\ref{nonperfectTeorell}) significantly changes  the
diffusion coefficient: the regularizing gradient term should
appear in the activity coefficient.

More details about activity coefficients in thermodynamics can
be found, for example, in Chapter 9 of the classical book
\cite{Denbigh1981}.

In the phase separation problem, the components are definitely
non-perfect and the further correction of the
Cahn--Hilliard--Teorell equation by the activity coefficients
is necessary.

The problem of the extension of the Cahn--Hilliard approach to
multicomponent diffusion was discussed by various authors
\cite{MaierMultiCahn2000,BlesgenMultiCahn2005}. Elastic forces
and plasticity are also taken into account
\cite{BlesgenMultiCahn2005,FratzlPenroseLebowitz1999}.
Nevertheless, the problem of the proper equations of
multicomponent nonlinear diffusion in highly non-homogeneous
condensed phases is still open. From our point of view, there
is no single ``proper model" and the variety of possible models
is very rich. In our work, we attempt to formulate the proper
language for the description of the universe of these models
similarly to chemical kinetics models.

\subsection{Main Ideas}

\subsubsection{Mechanisms as Collections of Elementary Processes}

A complex process can be disassembled into several elementary
processes. The dependence of the process rate (the flux) on the
state (concentrations, chemical potentials and their gradients)
is simple for elementary processes.  The model of the whole
process is assembled from these elementary ``details".

This idea was developed in chemical kinetics. In 1862--1867,
Guldberg and Waage proposed the mass action law for
equilibrium. In 1879 they developed the mass action law for
dynamics. This idea was developed further by many researchers
and after several dozen  years it was transformed into a
technology for the representation of complex processes: A
complex reaction is represented as an ensemble of elementary
reactions. The reaction rate has a simple monomial dependence
on concentration.

Van't Hoff \cite{VantHoff1884} called the reactions that
satisfy the mass action law ``normal transformations" and found
that ``normal transformations take place very rarely". Now we
can say that most reactions are complex and the interaction of
several elementary reactions causes non-trivial complex
(``abnormal") behavior.

Van't Hoff did not study complex reactions by disassembling
them into several elementary reactions. Therefore, he was
disappointed with the mass action law and finally wrote: ``As a
theoretical foundation I did not accept the concept of mass
action, I had to abandon this concept in the course of my
experiments...". (``J'ai adopt\'e pour la th\'eorie, non la
notion des masses actives, notion que j'ai d\^u abandonner dans
le cours de mes exp\'eriences, ..." \cite{VantHoff1884}, p. 7.)

The set of elementary reactions which constitute a complex
reactions is called the {\em reaction mechanism}. A mechanic
analogue is obvious: the elementary reactions are the details
of the mechanism that represents the complex reaction. This
notion was, finally, introduced into chemistry in the 20th
century due to the efforts of M. Bodenstein.

M. Boudart described the  ``century of Bodenstein" in his paper
\cite{Boudart2000}: ``First came the data, then the rate
equation, and finally the fitting of the data into the rate
equation by means of a hypothesized mechanism with rate
constants chosen for the best fit."

The great success of this approach was the theory of chain
reactions. N.N. Semenov was awarded the Nobel prize for this
theory \cite{Semenov1956}. The modern theory of complex
chemical reactions is based on the idea of the detailed
reaction mechanism and the simple kinetic law of elementary
reactions \cite{Yab1991}.

Many authors proposed various particular mechanisms of
nonlinear diffusion. One of our goals in this work is to repeat
the way of chemical kinetics in application to multicomponent
diffusion and to create a comprehensive theory of the
mechanisms of diffusion.

\subsubsection{Discrete Kinetic Models and Lattice Automata}

In the 1940s, S. Ulam and J. von Neumann proposed networks of
interconnected finite-state automata for the modeling of
complex systems. In the first period of study, research was
focused on the abilities of these networks and, in particular,
on the ability of self-reproduction \cite{VonNeumannBurks1966}.

The behavior of these cellular automata is so variable and
surprising and its complexity is so high that Ulam proposed the
idea of the {\em computational experiment}: we should regard
our invention as a new sort of reality and study it by the
experimental approach, as physics or chemistry does.

Cellular automata were invented as an intellectual journey but
soon were recognized as efficient tools for modeling
\cite{Toffoli1987}.

Feynman's attention to automata with local interactions as a
tool for simulating physics, attracted much attention to this
area: ``Therefore my question is, can physics be simulated by a
universal computer? I would like to have the elements of this
computer locally interconnected, and therefore sort of think
about cellular automata as an example (but I don't want to
force it). But I do want something involved with the locality
of interaction. I would not like to think of a very enormous
computer with arbitrary interconnections throughout the entire
thing" \cite{Feynman1982}.

The idea of modeling the natural world in terms of the behavior
of sets of rules that can be embodied in simple automata with
local interactions is now an important part of science.
Sometimes it is called ``the new science" to distinguish this
approach from classical modeling by equations
\cite{Wolfram2002}.

For the modeling of transport processes the lattice gas
automata \cite{Wolf-Gladrow2000,Chopard1998} were invented and
the lattice Boltzmann methods \cite{Succi2001} became very
popular: they are flexible and efficient. At the same time, the
lattice Boltzmann methods are very simple for programming and
parallelization.

The essence of the lattice Boltzmann methods was formulated by
S.~Succi in the following maxim: ``Nonlinearity is local,
non-locality is linear'' \cite{SucciLecture2006}. We should
even strengthen this statement. Non-locality (a) is linear; (b)
is exactly and explicitly solvable for all time steps; (c)
space discretization is an exact operation.

The lattice Boltzmann method is a discrete velocity method. The
finite set of velocity vectors $\{v_i\}$ ($i=1,...m$) is
selected, and a fluid is described by associating, with each
velocity $v_i$, a single-particle distribution function
$f_i=f_i(x,t)$ which is evolved by advection and interaction
(collision) on a fixed computational lattice. The values $f_i$
are named \emph{populations}. If we look at all lattice
Boltzmann models, one finds that there are two steps: free
flight for time $\delta t$ and a local collision operation.

The free flight transformation for continuous space is
\begin{equation*}
f_i(x,t+\delta t) = f_i(x-v_i \delta t ,t).
\end{equation*}
After the free flight step the collision step follows:
\begin{equation}\label{coll}
f_i(x) \mapsto F_i (\{f_j(x)\}),
\end{equation}
or in the vector form
\begin{equation*}
f(x) \mapsto F (f(x)).
\end{equation*}
Here, the  \emph{collision operator} $F$ is the set of
functions $F_i (\{f_j\})$ ($i=1,...m$). Each function $F_i$
depends on all $f_j$ ($j=1,...m$): new values of the
populations $f_i$ at a point $x$ are known functions of all
previous population values at the same point.

The lattice Boltzmann chain ``free flight $\to$ collision $\to$
free flight $\to$ collision $\dotsb$'' can be exactly
restricted onto any space lattice which is invariant with
respect to space shifts of the vectors $v_i \delta t$
($i=1,\dotsc,m$). Indeed, free flight transforms the population
values at sites of the lattice into the population values at
sites of the same lattice. The collision operator~\eqref{coll}
acts pointwise at each lattice site separately. Much effort has
been applied to answer the questions: ``how does the lattice
Boltzmann chain approximate the transport equation for the
moments $M$?'', and ``how does one construct the lattice
Boltzmann model for a given macroscopic transport phenomenon?''
(a review is presented in the book~\cite{Succi2001}).

The lattice Boltzmann models should describe the macroscopic
dynamic, i.e., the dynamic of macroscopic variables. The
macroscopic variables $M_\ell(x)$ are some linear functions of
the population values at the same point: $M_\ell (x)=\sum_i
m_{\ell i} f_i(x)$, or in the vector form, $M(x)=m(f(x))$. The
macroscopic variables are invariants of collisions:
\begin{equation*}
    \sum_i m_{\ell i} f_i=\sum_i m_{\ell i} F_i (\{f_j\}) \qquad \text{(or $m(f)=m(F(f))$).}
\end{equation*}
The standard example of the macroscopic variables are
hydrodynamic fields (density--velocity--energy density): $\{n,
u, E\}(x):=\sum_i \{1, v_i,  v_i^2/2\} f_i(x)$. But this is not
an obligatory choice. On the other hand, the athermal lattice
Boltzmann models with a shortened list of macroscopic variables
$\{n, n u\}$ are very popular.

The quasiequilibrium is the positive fixed point of the
collision operator for given macroscopic variables $M$. We
assume that this point exists, is unique and depends smoothly
on $M$. For the quasiequilibrium population vector for given
$M$ we use the notation $f^*_M$, or simply $f^*$, if the
correspondent value of $M$ is obvious. We use $\Pi^*$ to denote
the equilibration projection operation of a distribution $f$
into the corresponding quasiequilibrium state:
\begin{equation*}
\Pi^*(f)=f^*_{m(f)}.
\end{equation*}

Usually, collision operators are taken in the form:
\begin{equation}\label{collLin}
F(f) :=\Pi^*(f) + A(\Pi^*(f)-f)\, ,
\end{equation}
where $A$ is a linear operator, whose spectrum belongs to the
interior of the unit circle. A special case of (\ref{collLin})
is very popular, the lattice Bhatnagar--Gross--Krook (LBGK)
model:
\begin{equation}\label{collLBGK}
F(f) :=f + \omega(\Pi^*(f)-f)\, .
\end{equation}
In this brief introduction of LBM we follow the paper
\cite{Brownlee2008}.

The simplest LBGK realization of Fick's law (in 1D) gives the
following system. The discrete velocity set includes two
elements only, $v$ and $-v$. The time step is $\tau$ the
correspondent grid step is $h=v\tau$. The microscopic
variables, the populations, are: $f^-$ for velocity $-v$ and
$f^+$ for $v$. The macroscopic variable, the density, is $\rho=
f^-+f^+$. The corresponding equilibrium is $\Pi^*(f)=f^*$:
$$f^{*+}=f^{*-}=\frac{f^-+f^+}{2}\, .$$
For the non-negative populations, the equilibrium distribution
is the maximizer of the entropy $S=-f^-\ln f^-+f^+ \ln f^+$
under a given value of the  macroscopic variable $\rho$.

Let us take the LBGK collisions (\ref{collLBGK}) with $\omega=1$,
i.e.
\begin{equation}\label{LBGKEhr}
F(f)^{\pm} =\Pi^*(f)^{\pm}=\frac{f^-+f^+}{2}\, .
\end{equation}
This particular case of the LBGK collision integral is an
example of the so-called Ehrenfests' coarse-graining. The idea
of artificial partial equilibration steps was proposed by T.
and P. Ehrenfest for the foundation of statistical physics
\cite{Ehrenfests1911} and further developed to a general
formalism of nonequlibrium thermodynamics
\cite{GKOeTPRE2001,Raz2002,UNIMOLD2004}.

A review and comparative analysis of different approaches to
coarse-graining was published in \cite{GorbanBasic2006}.

The LBGK chain for the collision integral (\ref{LBGKEhr}) has a
very simple form:
\begin{equation}
\begin{split}
&f^+(nh,(m+1)\tau)=f^-(nh,(m+1)\tau)\\ &=\frac{f^+((n-1)h,m\tau)+f^-((n+1)h,m\tau)}{2}
\, .
\end{split}
\end{equation}
Therefore, for the density $\rho$ we get
\begin{equation}
\rho(nh,(m+1)\tau)=\frac{\rho((n-1)h,m\tau)+\rho((n+1)h,m\tau)}{2}
\, .
\end{equation}
This appears to be the one of the most common explicit finite
difference methods for Fick's diffusion equation. The diffusion
coefficient is $D=h^2/(2\tau)=v^2 \tau/2$ and depends
explicitly on the lattice parameters. We can decouple $D$ and
the lattice parameters if we use $\omega \in [1,2]$ in the LBGK
collision integral (\ref{collLBGK}). This lattice-gas scheme
does not coincide with any of the finite difference schemes.
Nevertheless, it also models diffusion and, to the first order
in $\tau$, $D=v^2 \tau \frac{2-\omega}{2\omega}$
\cite{Succi2001,GorbanBasic2006}.

Now, the area of applications of the cellular and lattice
Boltzmann automata is very wide and, in addition to classical
fluid dynamics, includes many areas of chemistry
\cite{Kier2005}, models of phase separation \cite{Rothman1994},
dynamics of macromolecules and many other topics.

We use cellular automata and lattice models of nonlinear
multicomponent diffusion for two purposes:
\begin{itemize}
\item{As a tool for model creation (after that, this model
    could be translated into other languages, such as
    partial differential equations (PDE);}
\item{As a tool for numerical simulation without the stage
    of PDE model.}
\end{itemize}

Elliott and Stuart \cite{ElliottStuart1993} used the cell model
of diffusion to study semilinear parabolic equations. They
proved the existence of absorbing sets, bounded independently
of the mesh size for discrete models. Discrete Lyapunov
functions were constructed. We use the special quasichemical
approach for the generation of the the cell models
\cite{Ocherki1986} that allowed us to construct the Lyapunov
functions for semi-discrete systems and to prove stabilization
of the solution in space and time under proper conditions.

\begin{figure}
\centering
\includegraphics[width=0.4\textwidth]{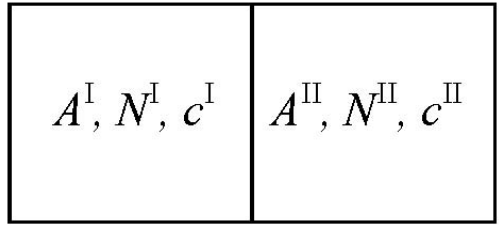}
\caption{Cell Jump Model \label{fig:cell} }
\end{figure}

Let us consider our space divided into cells, a system
represented as a chain of cells of homogeneous composition and
elementary transfer processes between them. It is sufficient
for our purposes to discuss two sells (Fig~\ref{fig:cell}). Let
us numerate these cells by the Roman numbers I and II and mark
all the components and quantities related to them by the upper
index I or II, correspondingly. The lists of the components for
cells are different just by the upper index: $A_1^{\mathrm{I}},
\ldots A_n^{\mathrm{I}}$, $A_1^{\mathrm{II}}, \ldots
A_n^{\mathrm{II}}$.

The mechanism of diffusion is defined as a list of elementary
transitions between cells described by their stoichiometric
equation. Since diffusion is a sort of jumping reaction on the
border, for these jumps the stoichiometric equation is written
as,
\begin{equation}\label{elementaryActDiff}
\sum_{i}\alpha_{ri}^{\mathrm{I}}A_{i}^{\mathrm{I}}+
\sum_{i}\alpha_{ri}^{\mathrm{II}}A_{i}^{\mathrm{II}}\rightarrow
\sum_{i}\beta_{ri}^{\mathrm{I}}A_{i}^{\mathrm{I}}+\sum_{i}\beta_{ri}^{\mathrm{II}}A_{i}^{\mathrm{II}}\,
,
\end{equation}
where $r$ is the number of processes,
$\alpha_{ri}^{\mathrm{I,II}},$ and $
\beta_{ri}^{\mathrm{I,II}},$ are the stoichiometric
coefficients which indicate the number of particles in cells
involved in the process. The direction of changes in the
elementary event (\ref{elementaryActDiff}) is defined by two
stoichiometric vectors
$$\gamma_{ri}^{\mathrm{I}}= \beta_{ri}^{\mathrm{I}}-\alpha_{ri}^{\mathrm{I}}\,
; \gamma_{ri}^{\mathrm{II}}=
\beta_{ri}^{\mathrm{II}}-\alpha_{ri}^{\mathrm{II}}\, .$$

\begin{figure}
\centering \subfigure[\label{fig:SurfElActs1}Simple diffusion: a
particle from the cell I jumps into the cell II and inverse]
{\includegraphics[width=0.6\textwidth]{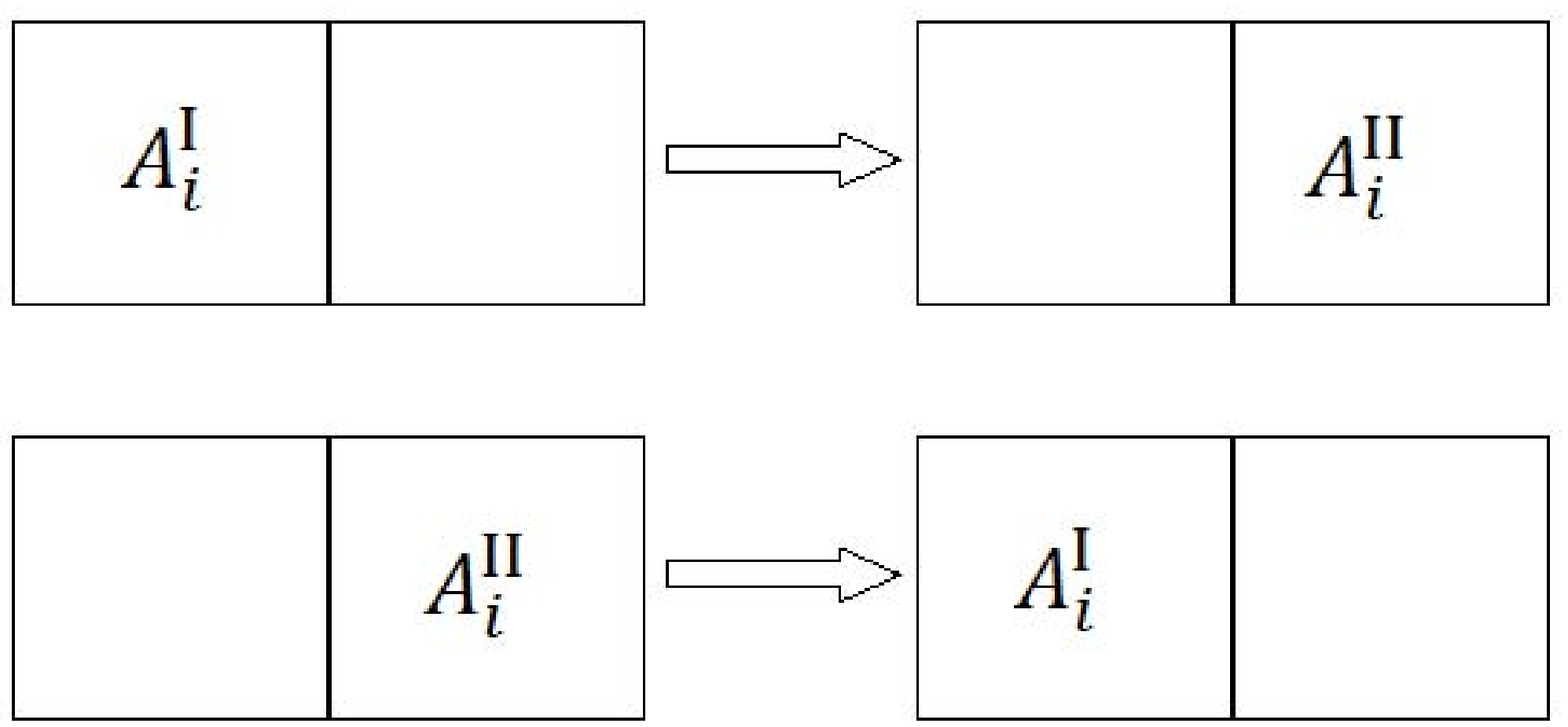}}\hspace{0.7cm}
 \subfigure[\label{fig:SurfElActs2}Jumps to free places: a
particle from the cell I jumps to the free place in cell II
and inverse] {\includegraphics[width=0.6\textwidth]{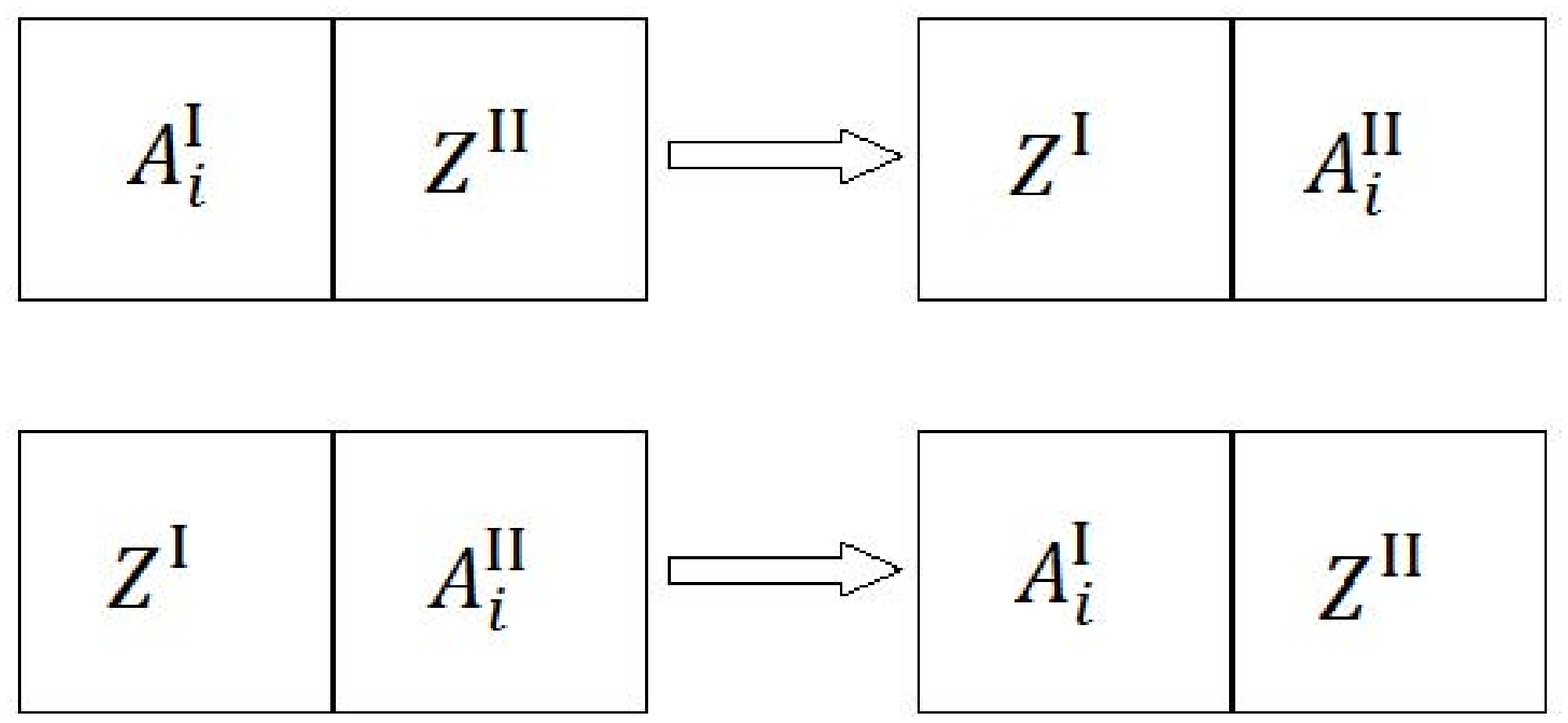}}
\hspace{0.7cm}
 \subfigure[\label{fig:SurfElActs3}Jumps with clustering: two particle attract the third one]
{\includegraphics[width=0.6\textwidth]{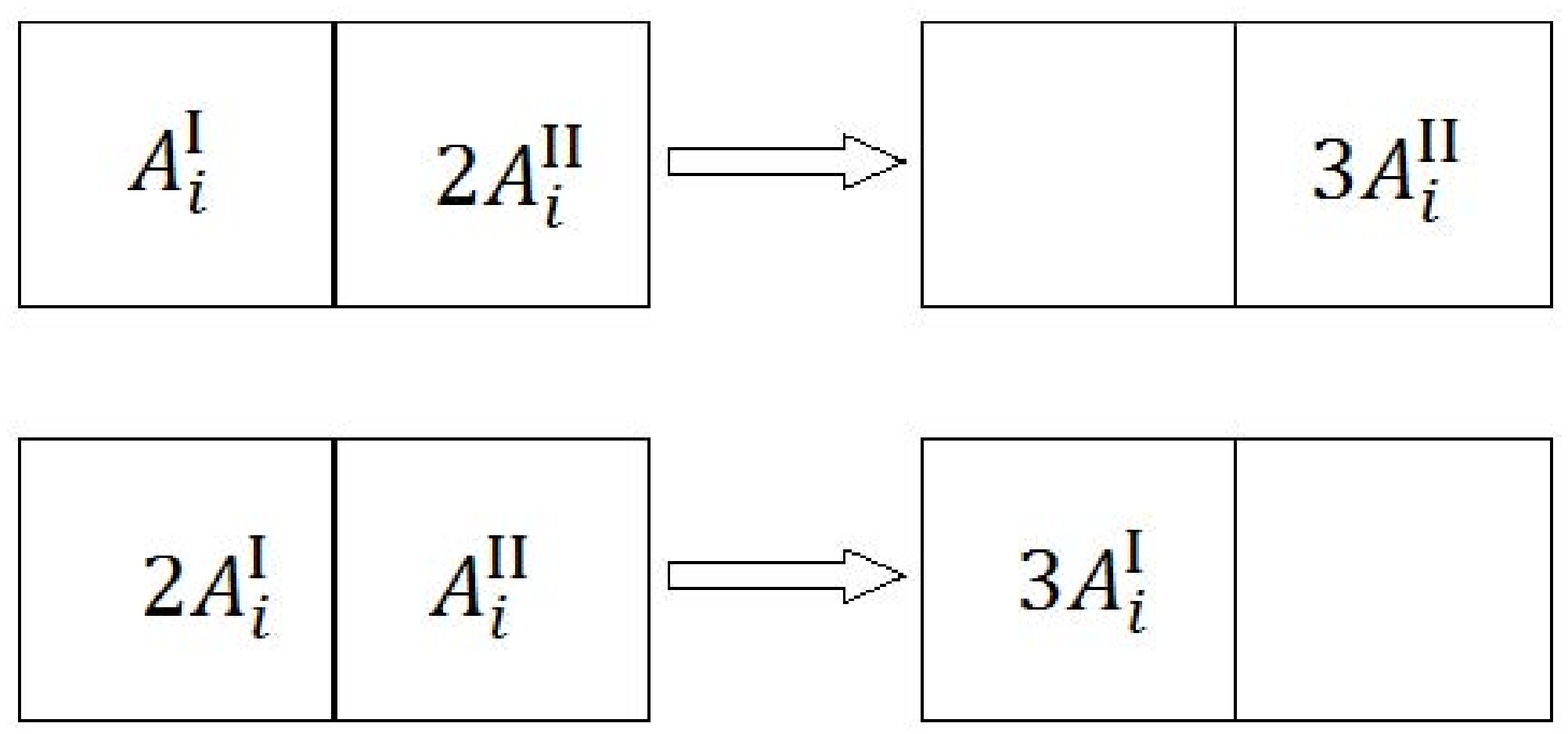}}
\caption{Elementary acts of diffusion,
examples.\label{fig:SurfElActs}}
\end{figure}

Examples of elementary acts are presented in
Fig.~\ref{fig:SurfElActs}.

Elementary events (\ref{elementaryActDiff}) should not include
reactions. Therefore, for each $i$, the amount of $A_i$ in the
system ($A_{i}^{\mathrm{I}}+A_{i}^{\mathrm{II}}$) should not
change. This means exactly that for all $i, r$
$$\gamma_{ri}^{\mathrm{I}}=-\gamma_{ri}^{\mathrm{II}}\, .$$

Let us use the notation
$$\gamma_{ri}=\gamma_{ri}^{\mathrm{I}}=-\gamma_{ri}^{\mathrm{II}}\, .$$

The composition of each cell is a vector $N^{\mathrm{I,II}}$.
The components of this vector, $N_i^{\mathrm{I,II}}$ are the
amounts of $A_i$ in the correspondent cell. We describe the
dynamics of the compositions of two cells by the equations:
\begin{equation}\label{difKinUrGen}
\frac{d N^{\mathrm{I}}}{dt}= - \frac{d N^{\mathrm{II}}}{dt}=  S
\sum_{r} \gamma_{r}w_{r} ,
\end{equation}
where $S$ is the area of the boundary between two cells and
$w_r$ is the rate of the process. For many cells the equations
are the same, but with more pairs of cells interacting, and
therefore there are more terms.

The rates are intensive variables and should be defined as functions
of concentrations or chemical potentials. The crucial question is:
how to describe function $w_r(c^{\mathrm{I}},c^{\mathrm{II}})$,
where $c^{\mathrm{I,II}}$ are concentrations components in cells.

\begin{figure}
\centering
\includegraphics[width=0.6\textwidth]{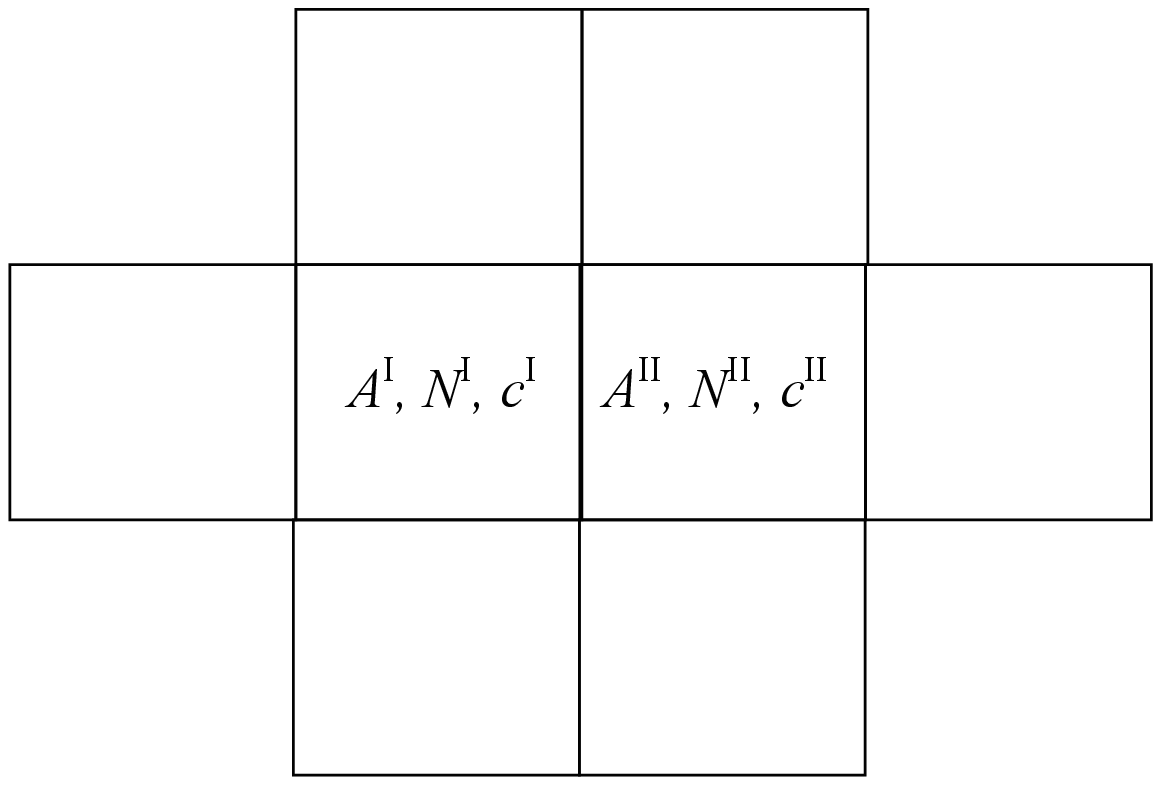}
\caption{Cell Jump Model  with first surroundings. \label{fig:cellSurround} }
\end{figure}

The real physics of diffusion may be more complicated. For
example, the intensity of jumps and the reaction rate
$w_r(c^{\mathrm{I}},c^{\mathrm{II}})$ may depend not only on
$(c^{\mathrm{I}},c^{\mathrm{II}})$ but on the surrounding. For
example, direct simulation of the jumps on the surface
\cite{BykGilGorY1985} demonstrates that the influence of the
surrounding is crucial for structures and critical effects on
the surface.

For each process (\ref{elementaryActDiff}) there is the {\em
space-inverted process} that is defined just by changing I to
II and vice versa. We mark the quantities for the
space-inverted processes by $'$. For example, $\gamma ' = -
\gamma$. The {\em detailed space-inversion symmetry} requires
that the rate functions for them should differ just by the
transposition of the vectors of variables,
$c^{\mathrm{I}},c^{\mathrm{II}}$:
\begin{equation}\label{SpaceSymmetry}
w'_r(c^{\mathrm{I}},c^{\mathrm{II}})=w_r(c^{\mathrm{II}},c^{\mathrm{I}})\,
.
\end{equation}
This requirement of {\em detailed space symmetry} allows us, in
particular, to exclude various types of advection and transport
driven by external force. Diffusion, by its definition, is
driven by the gradients of the concentrations (or, in the
thermodynamical approach, by the gradients of the chemical
potentials). This is not the only way to formulate of pure
diffusion equations without advection. Another possibility
gives us, for example, the diffusion systems with complex
balance (Section~\ref{Sec:CompBalMALDif}).

There are three ways to define the rate functions: from a
phenomenological law (like the mass action law), from
thermodynamics (like the generalized mass action law) or by
direct stochastic simulation of particles jumps in cells (like
in the Gillespie approach \cite{Gillespie1977,Gillespie2007}).

In our research, we focus on the first two approaches.
Therefore, we consider our lattice model as a semi-discrete
model (discrete in space and continuous in time). For this
semi-discrete model, the system of kinetic equations
(\ref{difKinUrGen}) describes diffusion. The continuous limit
of these equations gives us the diffusion PDE. The discrete
scheme by itself can serve as a computational model.

A couple of simple examples can clarify our approach:
\begin{itemize}
\item{Simple diffusion (Fig.~\ref{fig:SurfElActs1}),
    $A_{i}^{\mathrm{I}} \to A_{i}^{\mathrm{II}}$ and
    $A_{i}^{\mathrm{II}} \to A_{i}^{\mathrm{I}}$ with the
    same rate constants. Particles jump into the neighbor
    cells. For perfect mixtures, $w_r = k c^{\mathrm{I}}_i,
    \; w_r' = k c^{\mathrm{II}}_i$ and in the continuous
    limit we get Fick's law (\ref{Fick}) as the first
    Taylor approximation. In this approximation, $D=kl$
    where $l$ is the cell size.}
\item{Jumps to free places (Fig.~\ref{fig:SurfElActs2}),
    $A_{i}^{\mathrm{I}}+Z^{\mathrm{II}}\to
    A_{i}^{\mathrm{II}}+Z^{\mathrm{I}} $ and
    $A_{i}^{\mathrm{II}}+Z^{\mathrm{I}}\to
    A_{i}^{\mathrm{I}}+Z^{\mathrm{II}} $. According to the
    mass action law, $w_r(c^{\mathrm{I}},c^{\mathrm{II}})=k
    c_i^{\mathrm{I}}z^{\mathrm{II}}$,
    $w'_r(c^{\mathrm{I}},c^{\mathrm{II}})=k
    c_i^{\mathrm{II}}z^{\mathrm{I}}$, where $z$ is the
    concentration of free places. In the first Taylor
    approximation, $J=-k l (z \nabla c_i -c_i \nabla z )$
    and we get the model (\ref{ExchangeFlux}),
    (\ref{Excangediff}).}
\end{itemize}
To get the continuous limit, we take $c^{\mathrm{I}}=c(x)$,
$c^{\mathrm{II}}=c(x+l)$ and use the Taylor expansion:
$c(x+l)=c(x)+ l \partial_x c+o(l)$. If we the consider a
sequence of cell representations of diffusion with various $l$
then, for the invariance of the first order, the {\em scaling
rule} should be implemented: $D=k l$ does not change with the
change of size, therefore, the rate constant $k$ depends on
$l$: $k=D/l$.

It is not always possible to keep to first order only. If this
approach gives a negative diffusion coefficient then for
regularity we have to keep the higher derivatives. For example,
let us take the diffusion mechanism with attraction:
\begin{equation}\label{diffusionAttraction}
A_{i}^{\mathrm{I}} + 2A_{i}^{\mathrm{II}} \to 3
A_{i}^{\mathrm{II}}\, .
\end{equation}
 The space-inverted process in this case does not coincide with the
inverse one. If we change the upper indexes (I to II and II to I)
then we obtain
\begin{equation}\label{diffusionAttractionInvert}
2A_{i}^{\mathrm{I}} + A_{i}^{\mathrm{II}} \to 3A_{i}^{\mathrm{I}}\,
.
\end{equation}
This mechanism means that 2 particles attract the third one.
This mechanism is represented in Fig.~\ref{fig:SurfElActs3}.

The reaction rates are: $$w_r=k_r c_{i}^{\mathrm{I}}
(c_{i}^{\mathrm{II}})^2\, , \; w_r'=k_r (c_{i}^{\mathrm{I}})^2
c_{i}^{\mathrm{II}}\, .$$ The flux of $A_i$ from the first cell
to the second one is
$$J=w_r-w_r'=k_rc_{i}^{\mathrm{I}}c_{i}^{\mathrm{II}}
(c_{i}^{\mathrm{II}}-c_{i}^{\mathrm{I}})\, .$$ Therefore, to
first order we have $$J=kl c^2 \nabla c=\frac{1}{3}kl  \nabla
c^3 \, ;$$
 the sign is opposite to standard diffusion. This flux goes
in the direction of gradients. The diffusion equation is
\begin{equation}\label{inverseDiffMAL}
\frac{\partial c}{\partial t}=- kl \mathrm{div} (c^2  \nabla c)=-kl
\frac{1}{3}\Delta c^3 \, .
\end{equation}
Of course, if we take the mechanism ($n>1$)
$$A_{i}^{\mathrm{I}} + nA_{i}^{\mathrm{II}} \to (n+1)
A_{i}^{\mathrm{I}}\, , \; A_{i}^{\mathrm{I}} + A_{i}^{\mathrm{II}}
\to (n+1)A_{i}^{\mathrm{I}}\, ,$$ then we get the equation

$$\frac{\partial c}{\partial t}=- kl(n-1)
\mathrm{div} (c^n  \nabla c)=-kl \frac{n-1}{n+1}\Delta c^{n+1} \,
.$$

This diffusion process has two properties: first, it goes along
gradients and all deviations from the uniform state will increase.
Second, this diffusion is slow for small concentrations (the
diffusion coefficient goes to 0 when $c$ approaches 0) and
accelerates with the concentration growth.

The equation $\partial_t c= - D \Delta c^n$ ($n>1$) admits a
family of self-similar solutions with bounded support, which
collapse in finite time. These solutions have the form
$$c(\tau)=\frac{A}{\rho ^q}\phi\left(\frac{r}{\rho}\right)\,
,$$ where
\begin{itemize}
\item{$\tau$ is the time till collapse;}
\item{$q$ is the dimension of space (usually, $q=$1, 2 or
    3);}
\item{$\rho$ is the radius of the sphere, outside of which
    the solution is zero
$$\rho=B (D \tau)^{\frac{1}{q(n-1)+2}}\, ;$$}
\item{$\phi(\vartheta)=(1-\vartheta^2)^{\frac{1}{n-1}}$ for $\vartheta < 1$ and $\phi(\vartheta)=0$ if $\vartheta\geq 1$;}
\item{The constants $A,B$ depend on $q$, $n$ and the total amount $N=\int c(x) \, \D x$.}
\end{itemize}
This is the so-called Barenblatt solution \cite{Barenblatt1952}
for the equation of porous media $\partial_{\tau} c= + D \Delta
c^n$. Such solutions were used in the analysis of an explosion
which starts from a singularity for equations $\partial_t c= +
D \Delta c^n$ (the classical review of self-similar solutions
was published by Barenblatt and Zeldovich
\cite{BarenZeld1972}).

The cell model of diffusion with attraction
(\ref{diffusionAttraction}) for a finite number of cells of a
given size $l$ is a rather regular system of nonlinear ODE, but
to first order of the Taylor expansion in $l$ the PDE
(\ref{inverseDiffMAL}) produces a singularity in an arbitrarily
short time from smooth initial data. The second order Taylor
approximation adds nothing because the even terms in $l$ cancel
out if we take into account both the left and right neighbors
of the cell. The third order Taylor expansion gives a
regularized equation:
$$J=J=w_r-w_r'=klc^2\frac{\partial }{\partial
x}\left(c+\frac{l^2}{3}\frac{\partial^2 c}{\partial x^2}
\right) +o(l^3)\, ;$$ $$ \frac{\partial c}{\partial
t}=-kl\frac{\partial }{\partial x}c^2 \frac{\partial }{\partial
x}\left(c+\frac{l^2}{3}\frac{\partial^2 c}{\partial x^2}
\right)\, .$$ This is an example of the Cahn--Hilliard type
equation for spinodal decomposition with the regularizing term
$-\mathrm{div} (c^2\mathrm{grad} \Delta c)$. In this equation,
the cell size cannot be eliminated by scaling. The length $l$
is the ``regularization length". All inhomogeneities of size
smaller than $l$ are smoothed by the biharmonic term.

As we can see, the mass action law and the cell representation
of the elementary acts of diffusion give the opportunity to
model the Cahn--Hilliard type phase separation. Nevertheless,
the approach based on the non-perfect thermodynamic potential
(\ref{GinzLandFree}) gives a better representation of the basic
physics and does not require complicated elementary processes.
Just the simplest Fick scheme, $$A_{i}^{\mathrm{I}} \to
A_{i}^{\mathrm{II}}\, , \; A_{i}^{\mathrm{II}} \to
A_{i}^{\mathrm{I}}$$ with the non-perfect Ginzburg--Landau free
energy gives the Cahn--Hilliard equation
(Sec.~\ref{Sec:ContLimNonPerfect}).

The diffusion mechanism with attraction
(\ref{diffusionAttraction}) (Fig.~\ref{fig:SurfElActs3})
differs from the elementary Fick mechanism
(Fig.~\ref{fig:SurfElActs3}) and from the mechanism of jumps to
free places (Fig.~\ref{fig:SurfElActs2}). The dynamical
difference is obvious, the diffusion mechanism with attraction
generates instabilities of the homogeneous state, clustering
and singularities. On the other hand, Fick's law and the
mechanism of jumps to free places (Fig.~\ref{fig:SurfElActs2})
allow a global Lyapunov functional and, in the systems without
external fluxes, lead to homogeneous equilibrium.

These mechanisms have also a very important structural
difference. If we look at the direct and the space-inverted
processes (Fig.~\ref{fig:SurfElActs}) then we find that for the
first two mechanisms, the space-inverted processes coincide
with the inverse processes, which we get just by inversion of
the arrow (or by the exchange $\alpha$ and $\beta$ coefficients
in the stoichiometric equations (\ref{elementaryActDiff})). For
the elementary processes with attractions
(Fig.~\ref{fig:SurfElActs3}) the inverse processes are
processes with repulsion:
\begin{equation}\label{diffusionAttractionTimeInverse}
3 A_{i}^{\mathrm{II}}\to A_{i}^{\mathrm{I}} + 2A_{i}^{\mathrm{II}}
\, , \; 3A_{i}^{\mathrm{I}} \to 2A_{i}^{\mathrm{I}} +
A_{i}^{\mathrm{II}} \, .
\end{equation}

The diffusion processes for which {\em space-inverted
elementary processes coincide with the inverse processes}, have
a fundamental property: The entropy production is positive for
the corresponding mass action law diffusion equations.

Let us consider a complex diffusion process in a bounded domain
with smooth boundary and without external fluxes. We prove the
following theorem in Section~\ref{spaceANDtime}

\vspace{3mm} {\bf Theorem 2.} {\em Let a complex diffusion
process consist of elementary processes, which satisfy the
following property: the space-inverted elementary process
coincides with the inverse process. Then, for the mass action
law equation of diffusion (\ref{difKinUrGenGenN}), the
principle of detailed balance is valid, the global convex
Lyapunov functional exists and the uniform distribution is
asymptotically stable.} \vspace{3mm}

This global Lyapunov functional may be selected in the form of
the (minus) classical entropy, the sum of terms $c\ln c$ for
all cells and components, or, for the continuous limit,
$$\sum_i \int c_i \ln c_i \, \D x \, .$$

A particular case of the dissipation inequality for such
processes is inequality (\ref{ThermodynamicInequalityExc}) for
the diffusion equations (\ref{ExchangeGeneralDiff}) that
describe diffusion by exchange of positions. It is valid
because the exchange mechanism satisfies this fundamental
property: the space-inverted elementary processes coincide with
the inverse processes.

\subsubsection{Thermodynamics and Intermediate Complexes}

Thermodynamics is not always a good leader, but it is always a
good judge. We cannot create nonlinear equations directly from
thermodynamic principles, but we  must always check whether our
equations satisfy thermodynamics. They should satisfy the
thermodynamic restrictions if we do not want to produce a {\em
perpetuum mobile} in our theory.

We also include some other fundamental restrictions like
micro-reversibility in the thermodynamic requirements.

It is not always simple to coordinate the lattice models with
thermodynamics, nevertheless it is possible
\cite{Karlin1998,SucciKarlin2002}.

There are two main approaches for the introduction of
thermodynamics into kinetic models. First, we can start from
general kinetic equations based on the representation of a
complex process as an ensemble of elementary processes with a
given simple kinetic law of elementary processes (for example,
the mass action law). After that, we will find that the rate
constants of the elementary processes are not independent. They
must be coordinated to meet the thermodynamic requirements.
Therefore, not all the possible kinetic systems are allowed
thermodynamically.

Another approach starts from the thermodynamic description of
the system. We should find thermodynamic potentials which
describe the system under given conditions. We have to know
entropy, free energy (Helmholtz energy), or free enthalpy
(Gibbs energy) for the proper set of independent variables
\cite{Callen1985}. After that, we define the rate of elementary
process through the thermodynamic functions but with some
arbitrariness: some constants remain free of thermodynamic
restrictions. These constants are independent for different
elementary processes.

Which way is better? It is not a proper question: both are good
for their purposes. The first approach (we start from kinetics
and then add thermodynamics) is very flexible. In particular,
it can be used when thermodynamic restrictions are not needed.
For example, when we consider subsystems of open systems like
the system of surface components in heterogeneous catalysis,
then the constants of elementary processes include additional
dependencies on some additional concentrations and are not the
``proper" rate constants. Therefore, they do not satisfy the
thermodynamic restrictions, and a subsystem  may demonstrate
non-thermodynamic behavior like non-decaying oscillations or
bifurcations.

The second approach is unavoidable for non-perfect systems. The
kinetic law of elementary processes depends on the
thermodynamic potential. For all perfect systems it is the same
mass action law, but any deviation from the perfect
thermodynamic function requires its own deviation of the
kinetic law from the mass action law \cite{Kudryav2001}. This
deviation may be reformulated as the generalized mass action
law with activities instead of concentrations but the
activities are defined through thermodynamic potentials.

In our work we follow both approaches: First, we formulate the
mass action law for diffusion and study this with and without
the thermodynamic restrictions. Secondly, we introduce the
thermodynamic formalism for diffusion in non-perfect systems.
The ideas for both approaches for diffusion were formulated in
the early 1980s \cite{GorbanSrk1986,Ocherki1986}. The detailed
analysis of the thermodynamic restrictions on chemical kinetics
was performed by Gorban in 1982 \cite{G1}.

Our work was influenced by the works of N.G. Van Kampen
\cite{VKampen1973}, M. Feinberg
\cite{Feinberg1972_a,Feinberg1972} and Horn and Jackson
\cite{HornJackson1972}. In 1973, N.G. Van Kampen proposed a
general formulation for the rates of irreversible processes as
a combination of ``unilateral transfer flows". Each unilateral
flow transfers energy and particles in one direction. Van
Kampen decomposed the total in partial systems, each of which
is in equilibrium and therefore possesses a well-defined
temperature, entropy, and other thermodynamic quantities.
Although the total system Y is not in equilibrium, it is still
possible to attribute an entropy to it. Then Van Kampen studied
the unilateral fluxes between subsystems.

We decompose the Van Kampen unilateral processes further and
represent them as a collection of essentially one-dimensional
elementary processes with the simple kinetic mechanism, the
mass action law or the generalized mass action law.

We start from a similar representation of the total system and
supplement it with the system of {\it stoichiometric equations}
of elementary unilateral processes. To find the rate of the
elementary processes we use an idea of intermediate complex
(compound). This approach is borrowed from the theory of
absolute reaction rates but we do not use the special
idealization of the reaction pass and postulate the more
general microscopic Markov kinetics instead.

If the concentrations of compounds are small and the
equilibrium between intermediates and other components is fast
(both assumptions are important) then we approach the
generalized mass action law, which is very similar to the
Marselin--de Donder kinetics and the generalized mass action
law studied by Feinberg, Horm and Jackson and other authors
\cite{Feinberg1972_a,Feinberg1972,HornJackson1972,BykGOrYab1982}.

This formalism is very convenient for implementation of the
microreversibility consequences in the form of detailed balance
conditions \cite{Mahan1975}. In addition, if there is no
microreversibility then the thermodynamic behavior is also
guaranteed by the special more general relations between
kinetic constants, which follow from the Markov kinetics of
intermediate complexes. First, the idea of such relations was
proposed by Boltzmann as an answer to the Lorentz objections
against Boltzmann's proof of the $H$-theorem. Lorentz stated
nonexistence of inverse collisions for polyatomic molecules.
Boltzmann did not object to this argument but proposed the
``cyclic balance" condition, that means balancing in cycles of
transitions between states $S_1 \to S_2 \to \ldots \to S_n \to
S_1$. Almost 100 years later, Cercignani and Lampis
\cite{CercignaniLamp1981} demonstrated that the Lorenz
arguments are wrong and this Boltzmann new relations are not
needed for the polyatomic molecules under the
microreversibility conditions. The detailed balance conditions
should hold.

Nevertheless, this Boltzmann's idea is very seminal. It was
studied further by Heitler \cite{Heitler1944} and Coester
\cite{Coester1951} and the results are sometimes cited as the
``Heitler-Coestler theorem of semi-detailed balance". In 1952
\cite{Stueckelberg1952} proved these conditions for the
Boltzmann equation. For the micro-description he used the
$S$-matrix representation, which is in this case equivalent for
the Markov microkinetics (see also \cite{Watanabe1955}). Later,
this sort of relation was rediscovered for chemical kinetics
\cite{Feinberg1972,HornJackson1972}. The general proof for
nonlinear nonequlibrium processes was presented recently
\cite{GorbanShahzad2010}. In our analysis of these
Boltzmann--...--Stueckelberg relations we follow the later.

We extend the usual stoichiometric equations by additional
reactions: an input linear combination of reagents forms a
corresponding compound; this compound transforms into another
compound that disintegrates into the corresponding output
linear combination of reagents:
\begin{equation}\label{stoichiometricequationcompaund}
\sum_i\alpha_{\rho i}A_i \rightleftharpoons B_{\rho}^- \to
B_{\rho}^+ \rightleftharpoons \sum_i \beta_{\rho i} A_i \, .
\end{equation}
Here $\rho$ is the elementary reaction number.

\begin{figure}
\centering{
\includegraphics[width=0.6\textwidth]{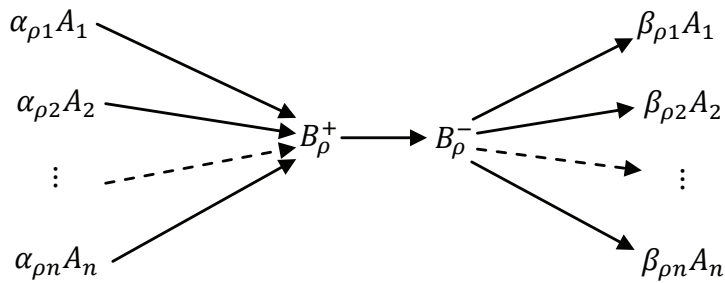}
}\caption{A $2n$-tail scheme of an extended elementary process
(\ref{stoichiometricequationcompaund}). \label{n-tail}}
\end{figure}

It is useful to visualize the reaction scheme. In
Fig.~\ref{n-tail} we represent the $2n$-tail scheme of an
elementary reaction sequence
(\ref{stoichiometricequationcompaund}). This scheme was
proposed in \cite{GorbanShahzad2010}.

We assume that the amount of each compound $B_{\rho}$ is small
enough to apply the perfect entropy formula, and that the
equilibrium between each compound and the corresponding linear
combinations of reagents is fast enough to apply the
quasiequilibrium approximation \cite{GorbanShahzad2010} (the
detailed analysis of this approximation was given in
\cite{GorbanKarIlgOt2001,GorbanKArlinChEngSci2003,GorbanKarlinBook2005}).

The main difference from the Eyring approach
\cite{EyringAtAlDiff1941} is in the model of the hidden
reaction of the ``activated complexes":
\begin{itemize}
\item{Eyring used for each reaction one complex with a
    continuum of energetic states along the ``reaction
    path", whereas we use two compounds (or two states);}
\item{Eyring modeled the reaction of the intermediate
    complex as a classical motion along the additional
    coordinate and even added this degree of freedom with
    classical kinetic energy of this motion to the free
    energy calculation, whereas we follow the Stueckelberg
    approach and model this reaction as a first order
    Markov kinetics, a Markov process with two states.}
\end{itemize}

The difference in the macroscopic consequences of these
approaches seems to be not very large because the result  of
the Eyring approach is one relaxation time approximation for
each reaction. From the dynamical point of view, this result
coincides with the two-state Markov model.

The main differences may arise in the hints which these
approaches give to the microscopic calculation of the
macroscopic quantities. In our work, we concentrate on the
macroscopic dynamics.

\section{Mass Action Law for Diffusion}

\subsection{Mass Action Law \label{Sec:MALKin}}

\subsubsection{Mass Action Law Kinetic Equations }

This is an auxiliary subsection where we collect main
definitions and results about the Mass Action Law (MAL).

To construct a system of kinetic equations by MAL, one needs
the following inputs:
\begin{enumerate}
\item{A list of components;}
\item{A list of elementary reactions represented by their
    stoichiometric equations;}
\item{A set of reaction rate constants.}
\end{enumerate}

The list of components is just a set of symbols (the component
names). We usually assume that this set is finite, $A_1, A_2,
\ldots , A_n$.

Elementary reactions are given by their stoichiometric equations,
\begin{equation}\label{StoiEq}
\sum_i \alpha_{ri} A_i \to \sum_i \beta_{ri} A_i\, ,
\end{equation}
where $r$ is a reaction number,  $\alpha_{ri}$ and $\beta_{ri}$
are nonnegative numbers, the {\em stoichiometric coefficients}.
By default, they are assumed to be integer but, sometimes,
there occurs a need in nonnegative real coefficients.

For each elementary reaction (\ref{StoiEq}), a {\em
stoichiometric vector} is defined, $$\gamma_r: \;
\gamma_{ri}=\beta_{ri}-\alpha_{ri} \, .$$ This is a
``bookkeeping" vector, whose components are ``gain minus loss"
(or ``income minus outcome").

We will also use the  loss and gain vectors of elementary
reactions: $\alpha_r$ (loss) with coordinates $\alpha_{ri}$ and
$\beta_r$ (gain) with coordinates $\beta_{ri}$. Of course,
$$\gamma_r=\beta_r - \alpha_r \, .$$

The stoichiometric matrix $\Gamma$ is the matrix with columns
$\gamma_i$: $\Gamma_{ij}=\gamma_{ji}$, the first index in
$\Gamma_{ij}$ corresponds to component and the second index
corresponds to reaction. Reaction rate constants $k_r$ are
non-negative numbers. They should be defined for all elementary
reactions. For each component $A_i$, a real variable,
concentration $c_i$ is defined. Vector of concentrations $c$
has coordinates $c_i$.

The reaction rate for the elementary reaction (\ref{StoiEq}) is
the following function of $c$
\begin{equation}\label{MALreactrate}
w_r=k_r\prod_{i=1}^n c_i^{\alpha_{ri}}\, .
\end{equation}

 The MAL kinetic equations are
\begin{equation}\label{MALkineticEq}
\frac{\D c}{\D t}=\sum_r \gamma_r w_r \, .
\end{equation}

From the physical point of view, these equations describe
isochoric isothermal processes for perfect systems. For
non-isochoric or non-isothermal processes it is necessary to
introduce the volume (together with pressure) and the enthalpy
(together with temperature) explicitly and describe their
dynamics.

For a given reaction mechanism, a linear {\em stoichiometric
conservation law} is a linear functional $b(c)=\sum_i b_i c_i$
that annihilates all stoichiometric vectors: $$b(\gamma_r)=0
\mbox{ for all } r \, .$$ The stoichiometric conservation law
is strictly positive, if all $b_i>0$. The assumption about
existence of a positive stoichiometric conservation law plays
an important role in the MAL kinetics.

\subsubsection{Existence and Uniqueness of Solutions}

Let in the reaction mechanism all nonzero coordinates of the
loss vectors be not less than 1: $$\alpha_{ri}\geq 1 \, \mbox{
or } \alpha_{ri}=0\, .$$ This assumption is valid, for example,
if all the stoichiometric coefficients are nonnegative
integers. Let us assume also that there exists a strictly
positive stoichiometric conservation law $b$. Then the
following existence and uniqueness theorem for the MAL equation
holds.

\begin{theorem}\label{existposit}For any nonnegative initial data $c(0)$ ($c_i(0)\geq 0$)
there exists a unique solution of (\ref{MALkineticEq}) $c(t)$
for all $t>0$. This solution is nonnegative  ($c_i(t)\geq 0$)
and satisfies the conservation law: $b(c(t))=b(c(0))$.
\end{theorem}

This is a well known result (see, for example,
\cite{VolpertKhudyaev1985}). The proof is quite simple. First,
of all, let us consider a bounded neighborhood $U$ of the
simplex $\Sigma_0$: $c_i\geq 0$, $b(c)=b(c(0))$. The right hand
site of the MAL kinetic equations (\ref{MALkineticEq}) has
continuous first derivatives and these derivatives are bounded
in $U$. Therefore, according to a standard existence and
uniqueness theorem its solution exists on some time interval
$t\in [0,T]$ and this $T$ is the same for a compact set of
initial data $c(0)\in \Sigma_0 \Subset U$. Secondly, let us
mention that if $\gamma_{ri}<0$ then $\alpha_{ri}\geq 1$ and
the reaction rate $w_r$ (\ref{MALreactrate}) includes the
factor $c_i^{\alpha_{ri}}$. Therefore,
$$\sum_{\gamma_{ri}<0}\gamma_{ri} w_r =c_i g(c) \, ,$$
where $g(c)$ is continuous function.

If $c_i \to 0$ then $\sum_{\gamma_{ri}<0}\gamma_{ri} w_r \to
0$. If $c_i=0$ then $\dot{c}_i=\sum_r \gamma_{ri} w_r \geq 0$.
Therefore, the simplex $\Sigma_0$ is positively invariant with
respect to equations (\ref{MALkineticEq}): the existent
solutions $c(t)$ do not leave this simplex for $t>0$ if
$c(0)\in \Sigma_0$. Finally, this implies global existence of
solutions in $\Sigma_0$.

Both conditions of existence of a strictly positive
stoichiometric conservation law $b$ and of coordinates of the
loss vectors, $\alpha_i=0$ or $\alpha_i\geq 1$ are significant.
Just for example, we can consider mechanisms that violate these
conditions: $2A \to 3A$ and $\frac{1}{2}A\to A$. For the first
mechanism, $\dot{c}=kc^2$ and there is no global existence, for
the second system, $\dot{c}=k\sqrt{c}$ and there is no
uniqueness of solution.

\subsubsection{Detailed Balance \label{sec:D3etBal}}

The expected behavior of a system of physical or chemical
kinetics is simple in the absence of external fluxes:
everything goes to equilibrium. A Lyapunov function for this
relaxation is the corresponding thermodynamic potential.

The MAL kinetics (\ref{MALkineticEq}) do not assume any
thermodynamic properties ``from scratch". Moreover, this class
of kinetic equations is so rich that it is dense in the class
of all smooth semi-dynamical systems in $\Sigma_0$ for a given
conservation law $b$ \cite{Ocherki1986}. Additional assumptions
are needed to guarantee the thermodynamic behavior.

The most celebrated sufficient condition for the thermodynamic
behavior of the MAL kinetics is the {\em principle of detailed
balance}. This principle, as a realization of {\em
microreversibility} was known for the Boltzmann equation
\cite{Boltzmann} (since his proof of the $H$-theorem in 1872) long
before Onsager's reciprocal relations \cite{Onsager1,Onsager2}.
A.~Einstein used this principle for the linear kinetics of emission
and absorption of radiation \cite{Einstein1916}. In 1901, R.
Wegscheider published analysis of detailed balance for chemical
kinetics \cite{Wegscheider1901}.

To formulate the principle of detailed balance, it is
convenient to join pairs of direct and inverse elementary
reactions in (\ref{StoiEq}) and write
\begin{equation}\label{StoiEqCoupled}
\sum_i \alpha_{ri} A_i \rightleftharpoons \sum_i \beta_{ri} A_i\, .
\end{equation}
If the inverse reaction does not exist in the original mechanism, we
formally add it but assume that its rate constant is zero. We mark
quantities for the direct and inverse reactions by the upper indexes
$^+$ and $^-$ and write the MAL reaction rate:
\begin{equation}\label{MALrevers}
\begin{split}
&w_r=w_r^+-w_r^-\, , \\
&w_r^+=k_r^+\prod_{i=1}^n c_i^{\alpha_{ri}} \, , \;
w_r^-=k_r^-\prod_{i=1}^n c_i^{\beta_{ri}} \, .
\end{split}
\end{equation}

For the MAL kinetics the principle of detailed balance is:
there exists a strictly positive point of detailed balance that
is such vector of concentration $c^*$ that $c^*_i>0$ and
\begin{equation}\label{detailed balance}
w_r^+(c^*)=w_r^-(c^*)\, (=w_r^*>0) \, .
\end{equation}
This means that at least at one positive point the direct elementary
processes are equilibrated by the inverse elementary processes.

Existence of one such point implies that all equilibria are also
points of detailed balance (\ref{detailed balance}) and, moreover,
there exists a global Lyapunov function that has the form of
relative entropy. This is, precisely, the analogue of the Boltzmann
$H$-theorem for the MAL kinetics.

For the formulation, use and proof of this theorem, it is convenient
to rewrite the formulas for the direct and inverse reaction rates
(\ref{MALrevers}) using $c^*$, $w_r^*$ and the detailed balance
relations (\ref{detailed balance}):
\begin{equation}\label{MALreversDB}
\begin{split}
&w_r=w_r^+-w_r^-\, , \\
&w_r^+=w_r^*\prod_{i=1}^n
\left(\frac{c_i}{c_i^*}\right)^{\alpha_{ri}} \, , \;
w_r^-=w_r^*\prod_{i=1}^n \left(\frac{c_i}{c_i^*}\right)^{\beta_{ri}}
\, .
\end{split}
\end{equation}

The Lyapunov function is:
\begin{equation}\label{Gfunction}
G=\sum_i c_i \left(\ln\left(\frac{c_i}{c_i^*}\right)-1\right)+\sum_i
c_i^* \, .
\end{equation}
Here, the last constant term is added to satisfy $G(c^*)=0$.

The partial derivatives of $G$ (the analogs of chemical potentials)
are
\begin{equation}\label{Gpotentials}
\frac{\partial G}{\partial c_i}=\ln\left(\frac{c_i}{c_i^*}\right) \,
.
\end{equation}
Therefore, we have one more form for the MAL kinetic law with
detailed balance (\ref{MALreversDB}):
\begin{equation}\label{MALreversDB+}
\begin{split}
&w_r=w_r^+-w_r^-\, , \\
&w_r^+=w_r^* \exp \left(\sum_i \alpha_{ri} \frac{\partial
G}{\partial c_i}\right)=w_r^* \exp (\alpha_r,\nabla_c G)\, , \\
&w_r^-=w_r^* \exp \left(\sum_i \beta_{ri} \frac{\partial G}{\partial
c_i}\right)=w_r^* \exp (\beta_r,\nabla_c G)\, ,
\end{split}
\end{equation}
It is worth mentioning that
$$\frac{w_r^+}{w_r^-}=\exp[-(\gamma_r,\nabla_c G)]\, .$$

For the time derivative of $G$ due to the MAL kinetics
(\ref{MALkineticEq}) with the detailed balance, simple algebra gives
the dissipation inequality:
\begin{equation}\label{dissIneqMALDB}
\frac{\D G}{\D t}=\sum_r w_r (\gamma_r,\nabla_c G)=-\sum_r
(w_r^+-w_r^-)\ln\left(\frac{w_r^+}{w_r^-}\right)\leq 0 \, .
\end{equation}
The last inequality holds because $\ln x$ is a strictly
monotone function and $\ln x - \ln y$ has the same sign as
$x-y$ has. Obviously, $\dot{G}|_c=0$ if and only if $c$ is a
point of detailed balance. This equilibrium point may be
different from the point $c^*$, which was used for the
definition. All the positive points of detailed balance for the
MAL system (\ref{MALreversDB}) form a smooth manifold with
dimension $$n-\mathrm{rank}\{\gamma_1, \gamma_2, \ldots \}\,
,$$ where $n$ is the number of components and
$\mathrm{rank}\{\gamma_1, \gamma_2, \ldots \}$ is the rank of
the system of the stoichiometric vectors for the given reaction
mechanism.

If we fix values of all stoichiometric linear conservation laws
then the strictly positive point of detailed balance is unique
for the given values. Indeed, the dissipation inequality is
valid for every pair of mutually inverse reactions and all the
terms $w_r(\gamma_r,\nabla_c G)$ in $\dot{G}$
(\ref{dissIneqMALDB}) are non-positive. Therefore, at any
strictly positive equilibrium point $c^{\mathrm{eq}}$,
$(\gamma_r,\nabla_c G)=0$ for all $r$. This means that
$c^{\mathrm{eq}}$ is a critical point of $G$ in $c^{
\mathrm{eq}}+\mathrm{span}\{\gamma_1,\gamma_2, \ldots \}$. The
function $G$ is strictly convex at any positive point: its
Hessian is positive definite,
$$\frac{\partial^2 G}{\partial c_i \partial
c_j}=\frac{1}{c_i}\delta_{ij}\, ,$$ where $\delta_{ij}$ is the
Kronecker delta. Therefore, there may exist only one positive
critical point of $G$ on a linear manifold.

Due to the logarithmic singularity of the gradients at the
boundary of $\mathbb{R}_+$ (where some of $c_i=0$), $G$
achieves its global minimum in
$$(c^{ \mathrm{eq}}+\mathrm{span}\{\gamma_1,\gamma_2, \ldots
\})\bigcap \mathbb{R}_+^n$$ at a positive point. This point is a positive point
of detailed balance.

For any positive vector, $c^0$,  the polyhedron
$$\mathcal{V}=(c^0+\mathrm{span}\{\gamma_1,\gamma_2, \ldots
\})\bigcap \mathbb{R}_+^n\, ,$$ is positively invariant with
respect to (\ref{MALkineticEq}). For systems with detailed
balance it includes one and only one positive point of detailed
balance. This was first demonstrated by Zeldovich in 1938
(reprinted in 1996 \cite{Zeldovich1938}). In our analysis we
mainly follow \cite{VolpertKhudyaev1985}.

\subsubsection{Complex Balance \label{Sec:ComplBal}}

In this subsection we consider the direct and inverse reactions
separately as we did it before in (\ref{StoiEq}),
(\ref{MALreactrate}). Detailed balance is s sufficient but not
necessary condition of the thermodynamic behavior. The simple
example of thermodynamic behavior gives any monomolecular
(linear) reaction mechanism, which consists of reactions $A_i
\to A_j$. Let us use notation $k_{ji}$  for this reaction rate
constant.

The MAL equations for a monomolecular reaction mechanism are
\begin{equation}\label{monomolKin}
\frac{\D c_i}{\D t}=\sum_{j, \, j\neq i} (k_{ij}c_j-k_{ji}c_i) \, .
\end{equation}
Let $c^0$ be a strictly positive steady state for these
equations (not necessarily a point of detailed balance):
$$\sum_{j, \, j\neq i} k_{ij}c_j^0= c_i^0 \sum_{j, \, j\neq
i}k_{ji}\, .$$ With this $c^0$ we can rewrite the second term
in (\ref{monomolKin}):
$$\sum_{j, \, j\neq i}k_{ji}=\sum_{j, \, j\neq i}
k_{ij}\frac{c_j^0}{c_i^0}\, ;$$
$$\sum_{j, \, j\neq i}k_{ji}c_i=\sum_{j, \, j\neq i}
k_{ij}c_j^0\frac{c_i}{c_i^0}\, .$$ Therefore, the kinetic
equations (\ref{monomolKin}) have the equivalent form for given
$c^0$:
\begin{equation}\label{monomolKinEquiv}
\frac{\D c_i}{\D t}=\sum_{j, \, j\neq i} k_{ij}c_j^0 \left(\frac{c_j}{c^0_j}-\frac{c_i}{c^0_i}\right) \, .
\end{equation}
Then we can define
\begin{equation}\label{Gcomplex}
G=\sum_i c_i \left(\ln\left(\frac{c_i}{c^0_i}\right) -1
\right)\, .
\end{equation}
After simple transformations, we find that due to
(\ref{monomolKinEquiv})
\begin{equation}\label{linEntProd}
\begin{split}
\frac{\D G}{\D t}=\sum_{ij\, i\neq j}k_{ij}c_j^0
&\left[\frac{c_i}{c_i^0}\left(\ln\left(\frac{c_i}{c_i^0}\right)-1\right)
-\frac{c_j}{c_j^0}\left(\ln\left(\frac{c_j}{c_j^0}\right)-1\right) \right.\\ & \left. +
\ln\left(\frac{c_i}{c_i^0}\right)\left(\frac{c_j}{c_j^0}-\frac{c_i}{c_i^0}\right)\right]\leq 0\, .
\end{split}
\end{equation}
To prove this formula, it is worth mentioning that for any $n$
numbers $f_i$, $$\sum_{ij\, i\neq j}k_{ij}c_j^0(f_i-f_j)=0\,
.$$ This gives us the first two terms in the square brackets
with
$$f_i=\frac{c_i}{c_i^0}\left(\ln\left(\frac{c_i}{c_i^0}\right)-1\right)\, .$$
The last term,
$$\ln\left(\frac{c_i}{c_i^0}\right)\left(\frac{c_j}{c_j^0}-\frac{c_i}{c_i^0}\right)\, ,$$
appears in the straightforward computation of the time
derivative of $G$ due to kinetic equations
(\ref{monomolKinEquiv}).

The expressions in square brackets in (\ref{linEntProd}) have
the form $$f(a)-f(b)+f'(a)(b-a)$$ for the convex function
$f(x)=x(\ln x -1)$. This expression is always non-positive
because of Jensen's inequality.

Linear MAL kinetics can obviously violate the principle of
detailed balance. For example, an irreversible cycle $$A_1 \to
A_2 \to \ldots \to A_n \to A_1$$ has always a positive steady
state but never has a positive point of detailed balance if
$n>2$.

There is a nice generalization of the dissipation inequality
(\ref{linEntProd}) for the nonlinear MAL equations under some
algebraic conditions on the kinetic constants. These conditions
are strictly weaker than the principle of detailed balance.
They were discovered for the Boltzmann equation by Stueckelberg
\cite{Stueckelberg1952} in 1952 and called later the ``complex
balancing condition" for the general MAL
\cite{HornJackson1972,Feinberg1972}.

To formulate this condition for the MAL kinetics, let us start
from the function $G$ (\ref{Gcomplex}) and look for conditions
that guarantee the inequality $\dot{G}\leq 0$.

For a given $c^0$, we can rewrite the MAL reaction rate in the
form
\begin{equation}
w_r=\varphi_r \exp(\alpha_r, \nabla_c G)=\varphi_r \prod_i \left(\frac{c_i}{c_i^0}\right)^{\alpha_{ri}} \, ,
\end{equation}
where $\varphi_r=k_r\prod_i (c_i^0)^{\alpha_{ri}}=w_r(c^0)$.

It is convenient to express $\dot{G}$ using an auxiliary
function $\theta$ of an auxiliary variable $\lambda$: for any
concentration vector,
\begin{equation}\label{auxtheta}
\theta(\lambda)=\sum_r \varphi_r \exp[\lambda (\alpha_r, \nabla_c G)
+(1-\lambda)(\beta_r, \nabla_c G)]\, .
\end{equation}
This function is convenient because
\begin{equation}\label{auxthetaderiv}
\frac{\D \theta(\lambda)}{\D \lambda}=-\sum_r \varphi_r (\mu, (\beta_r-\alpha_r))  \exp[\lambda (\alpha_r, \nabla_c G)
+(1-\lambda)(\beta_r, \nabla_c G)]\, .
\end{equation}
In this notation, $$\dot{G}=-\theta'(1)\, .$$

The function $\theta(\lambda)$ is a sum of exponents. It is
convex ($\theta''(\lambda)\geq 0$). Therefore, if
$\theta(0)=\theta(1)$ then $\theta'(1) \geq 0$.

This condition, $\theta(0)=\theta(1)$ (for all positive $c$) is
called the {\em complex balancing condition} and it is
sufficient for the dissipation inequality: $$\dot{G}=-\theta'(1)
\leq 0\, .$$

The principle of detailed balance for the MAL equations has a
form of existence of a special equilibrium point, a point of
detailed balance. This existence implies important dynamical
properties in the whole of $\mathbb{R}_+^n$ because of the very
``rigid" monomial structure of MAL.

The complex balancing condition also can be formulated as
existence of a special ``point of complex balance". Let us
reformulate it in this way.

Some vectors  $\alpha_r$, $\beta_r$ for a given reaction
mechanism may coincide. Let us denote by $\{y_1, \ldots ,
y_q\}$ the set of all different vectors $\alpha_r$, $\beta_r$.
For each $y_i$ we define $R_i^+=\{r\, | \, \alpha_r=y_i\}$,
$R_i^-=\{r\, | \, \beta_r=y_i\}$. In this notation,

\begin{equation}
\begin{split}
\theta(1)=\sum_i \left(\sum_{r\in R_i^+} w_r(c^0)\right) \exp(y_i, \nabla_c G)\, , \\
\theta(0)=\sum_i \left(\sum_{r\in R_i^-} w_r(c^0)\right) \exp(y_i, \nabla_c G)\, .
\end{split}
\end{equation}

For any finite set of (different) vectors $\{y_1, \ldots ,
y_q\}$ the correspondent functions $\exp(y_i, \nabla_c G)$ of
$c \in \mathbb{R}_+^n$ are linearly independent because the
Hessian of $G$ is strictly positive definite. Therefore, the
condition $\theta(0)=\theta(1)$ (for all positive $c$) is
equivalent to
\begin{equation}\label{complex balance}
\sum_{r\in R_i^+} w_r(c^0)=\sum_{r\in R_i^-} w_r(c^0)\, .
\end{equation}
The point $c^0$ that satisfies (\ref{complex balance}) is
called the point of {\em complex balance} and the complex
balancing condition means that there exists such a strictly
positive point of complex balance. In this case, the
dissipation inequality, $\dot{G}\leq 0$, with $G$ defined by
(\ref{Gcomplex}) holds for all positive points.

Of course, a point of detailed balance is a point of complex
balance as well. The reverse statement is not valid: for
example, all the positive steady states of linear MAL kinetics
(\ref{monomolKin}) are the points of complex balance but they
are not necessarily the points of detailed balance.

The microscopic background for  detailed balance is
microreversibility, i.e. invariance of the microscopic
classical or quantum equations with respect to the time
inversion.

The microscopic backgrounds for complex balance were formulated
by Stueckelberg as unitarity of the $S$-matrix
\cite{Stueckelberg1952}. It is necessary to add that the
validity of the scattering (or Markov) model of elementary
reactions is also needed: see \cite{GorbanShahzad2010} and
discussion in Section~\ref{sec:GenMAL}

\subsection{Mass Action Cell-Jump Formalism}

\subsubsection{Stoichiometry of Diffusion Jumps}

We represent the physical space as a network of compartments.
Each compartment is modeled as a cubic cell with an edge size
$l$. The stoichiometric equations of diffusion describe
interaction of two neighboring cells. To distinguish the
quantities related to these two cells we use the upper indexes
I and II (Fig.~\ref{fig:cell}).

The general stoichiometric equation for an elementary event of
diffusion is (\ref{elementaryActDiff})
\begin{equation}\label{elementaryActDiffChap}
\sum_{i}\alpha_{ri}^{\mathrm{I}}A_{i}^{\mathrm{I}}+
\sum_{i}\alpha_{ri}^{\mathrm{II}}A_{i}^{\mathrm{II}}\rightarrow
\sum_{i}\beta_{ri}^{\mathrm{I}}A_{i}^{\mathrm{I}}+\sum_{i}\beta_{ri}^{\mathrm{II}}A_{i}^{\mathrm{II}}\,
.
\end{equation}

Coefficients $\alpha_{ri}^{\mathrm{I,II}}$,
$\beta_{ri}^{\mathrm{I,II}}$ are nonnegative. Usually, we assume
that they are integers but in some situations real numbers are
needed.

Elementary events (\ref{elementaryActDiffChap}) describe diffusion
and do not include the transformation of components (reactions).
Therefore, the total amounts of each component $A_i$ coincide  in
the left and the right hand sides of (\ref{elementaryActDiffChap}):
\begin{equation}\label{DiffCond}
\alpha_{ri}^{\mathrm{I}}+\alpha_{ri}^{\mathrm{I}}=
\beta_{ri}^{\mathrm{I}}+\beta_{ri}^{\mathrm{II}}\, .
\end{equation}

Each elementary process (\ref{elementaryActDiffChap}) has two
loss vectors, $\alpha_{r}^{\mathrm{I,II}}$ with coordinates
$\alpha_{ri}^{\mathrm{I,II}}$ and two output vectors,
$\beta_{r}^{\mathrm{I,II}}$ with coordinates
$\beta_{ri}^{\mathrm{I,II}}$. Because of the conservation of
particles of all types (\ref{DiffCond}), the stoichiometric
vectors of processes for the cells differ just by the sign of
coordinates:
$$\gamma_r=\gamma_r^\mathrm{I}=\gamma_r^\mathrm{II}=
\beta_{r}^{\mathrm{I}}-\alpha_{r}^{\mathrm{I}}\, .$$

We define here a mechanism of diffusion as a system of
stoichiometric equations for elementary events. The simple and basic
examples are:
\begin{itemize}
\item{Fick's diffusion, $A_i^{\mathrm{I}} \to
    A_i^{\mathrm{II}}$, $A_i^{\mathrm{II}} \to
    A_i^{\mathrm{I}}$;}
\item{The exchange of particles, $A_i^{\mathrm{I}}+A_j^{\mathrm{II}} \to
A_i^{\mathrm{II}}+A_j^{\mathrm{I}}$;}
\item{Clustering (diffusion with attraction), $A_i^{\mathrm{I}}+s A_i^{\mathrm{II}} \to
(s+1)A_i^{\mathrm{II}}$, $sA_i^{\mathrm{I}}+ A_i^{\mathrm{II}} \to
(s+1)A_i^{\mathrm{I}}$,  $s>1$;}
\item{Diffusion with repulsion, $(s+1)A_i^{\mathrm{I}}\to sA_i^{\mathrm{I}}+ A_i^{\mathrm{II}}$,
$(s+1)A_i^{\mathrm{I}}\to A_i^{\mathrm{I}}+ sA_i^{\mathrm{II}}$,
($s>0$).}
\end{itemize}
Formally, diffusion with repulsion is the time-inverted process
of diffusion with attraction (the porous medium model) but for
$s=1$ diffusion with attraction has no sense (the exactly
uniform state cannot produce the nonuniform distribution).
Therefore, the restrictions on $s$ are different.

\subsubsection{MAL Equations for Diffusion}

Let us consider the system of stoichiometric equations
(\ref{elementaryActDiffChap}) as a reaction mechanism for MAL
(\ref{StoiEq}). If we apply MAL then the rate of the elementary
diffusion process is
\begin{equation}
w_r(c^{\mathrm{I}},c^{\mathrm{II}})=k_r\prod_i
(c_i^{\mathrm{I}})^{\alpha^{\mathrm{I}}_{ri}} \prod_i
(c_i^{\mathrm{II}})^{\alpha^{\mathrm{II}}_{ri}} \, .
\end{equation}

For example, for Fick's diffusion, we have two elementary
processes, $A_i^{\mathrm{I}} \to A_i^{\mathrm{II}}$ and
$A_i^{\mathrm{II}} \to A_i^{\mathrm{I}}$. The corresponding
reaction rates are $k_1 c_i^{\mathrm{I}}$ and $k_2
c_i^{\mathrm{II}}$.

The composition of each cell is vector $N^{\mathrm{I,II}}$.
Components of this vector,
$N_i^{\mathrm{I,II}}=V^{\mathrm{I,II}}c_i^{\mathrm{I,II}} $ are
amounts of $A_i$ in the corresponding cell and
$V^{\mathrm{I,II}}$ are volumes of cells. We describe the
dynamics of the compositions of two cells by the equations:
\begin{equation}\label{difKinUrGenGen}
\frac{d N^{\mathrm{I}}}{dt}= - \frac{d N^{\mathrm{II}}}{dt}=
S^{\mathrm{I,II}} \sum_{r}
\gamma_{r}w_{r}(c^{\mathrm{I}},c^{\mathrm{II}})\,  ,
\end{equation}
where $S^{\mathrm{I,II}}$ is the area of the boundary between cells
I and II. If there are many cells then
\begin{equation}\label{difKinUrGenGenN}
\frac{d N^{\mathrm{I}}}{dt}= \sum_{\mathrm{J}}S^{\mathrm{I, J}}
\sum_{r} \gamma_{r}w_{r}(c^{\mathrm{I}},c^{\mathrm{J}})\,  ,
\end{equation}
with summation through all interacting pairs (I,J).

For example, for Fick's diffusion, we have two elementary
processes, $A_i^{\mathrm{I}} \to A_i^{\mathrm{II}}$ and
$A_i^{\mathrm{II}} \to A_i^{\mathrm{I}}$. The corresponding
reaction rates are $k_1 c_i^{\mathrm{I}}$ and $k_2
c_i^{\mathrm{II}}$. Equations (\ref{difKinUrGenGen}) give
 $$\frac{d N_i^{\mathrm{I}}}{dt}=-S^{\mathrm{I,II}}
 k_1c_i^{\mathrm{I}}+ S^{\mathrm{I,II}} k_2 c_i^{\mathrm{II}}\, .$$

For several pairs, let us mention the symmetry in pairs:
$k_1=k_2=k$ (we discuss this symmetry in more detail  in the
next subsection):
$$\frac{d N_i^{\mathrm{I}}}{dt}=\sum_J k
S^{\mathrm{I,J}}(c_i^{\mathrm{J}}-c_i^{\mathrm{I}}) \, .$$ For
example, on a straight line (two neighbors), this equation gives
$$\frac{d N_i^{\mathrm{I}}}{dt}=
k
S^{\mathrm{I,J}}(c_i^{\mathrm{I+1}}+c_i^{\mathrm{I-1}}-2c_i^{\mathrm{I}})
\, .$$ From this expression, the proper scaling of $k$ with the cell
size is obvious:
$$\frac{d c_i^{\mathrm{I}}}{dt}=
k \frac{S^{\mathrm{I,J}}l^2}{V}
\frac{c_i^{\mathrm{I+1}}+c_i^{\mathrm{I-1}}-2c_i^{\mathrm{I}}}{l^2}
\, ,$$ the fraction
${c_i^{\mathrm{I+1}}+c_i^{\mathrm{I-1}}-2c_i^{\mathrm{I}}}/{l^2}$
approximates the second derivative and, hence,
${kS^{\mathrm{I,J}}l^2}/{V}=\mathrm{const}$.

For a cubic cell, $V=Sl$ and $kl= \mathrm{const}$.

\subsubsection{Space Symmetry and Time Symmetry \label{spaceANDtime}}

The system of elementary events should be symmetric with
respect to space-inversion. For each elementary process
(\ref{elementaryActDiffChap}) a space-inverted process is
defined just by changing I to II and vice versa. We mark the
quantities for the space-inverted processes by $'$. For
example, $$\gamma' = - \gamma \, .$$ Space inversion is an
involution: if we apply it two times then we return to the
original process.

The key condition is: the rate functions for the space-inverted
processes should differ just by transposition by vectors of
variables, $c^{\mathrm{I}},c^{\mathrm{II}}$ (\ref{SpaceSymmetry}):
\begin{equation}\label{SpaceSymmetry1`}
w'_r(c^{\mathrm{I}},c^{\mathrm{II}})=w_r(c^{\mathrm{II}},c^{\mathrm{I}})\,
.
\end{equation}
For MAL this means that $k_r=k_r'$.

The requirement of space symmetry distinguishes diffusion from
various types of advection and transport driven by external
force. This condition is necessary for existence of diffusion
equations when the cell size $l \to 0$ (see the next
subsection).

Inversion in time differs, in general, from the inversion in space.
For example, for the elementary process $3A^{\mathrm{I}} \to
2A^{\mathrm{I}}+A^{\mathrm{II}}$ the space inversion gives
$2A^{\mathrm{II}} \to A^{\mathrm{I}}+A^{\mathrm{II}}$ (we exchange
the upper indexes, I$\to$II and II$\to$I), and the inversion of
elementary events ($T$-transformation) gives
$2A^{\mathrm{I}}+A^{\mathrm{II}} \to 3A^{\mathrm{I}}$ (here, we
change direction of arrow). We have to stress that inversion in time
assumes micro-reversion. At the macroscopic level it does not mean
change $t$ to $-t$ in the kinetic equations but the transformation
of the direct processes into reverse ones (inversion of collisions,
for example).

Time symmetry (microreversibility) means that the principle of
detailed balance is valid. In this case, all the consequences of the
principle of detailed balance are applicable
(Section~\ref{sec:D3etBal}), a global Lyapunov functional exists,
and every positive equilibrium is a point of detailed balance.

Microreversibility and symmetry of space are independent
properties of the system. Nevertheless, for some elementary
processes, the space-inverted process coincides with the
reverse (the time-inverse) process. If a diffusion mechanism is
constructed from such processes then the symmetry in space is
equivalent to symmetry in time (to the principle of detailed
balance).

This specific class of diffusion mechanisms includes such
mechanisms as Fick's diffusion or the diffusion by exchange of
particle positions.

Indeed, for Fick's diffusion, when we exchange the upper
indexes in an elementary process $A^{\mathrm{I}} \to
A^{\mathrm{II}}$ (inversion in space) then we get the reverse
process as well, $A^{\mathrm{II}} \to A^{\mathrm{I}}$
(inversion of arrows). Analogously, for
$A^{\mathrm{I}}+B^{\mathrm{II}} \to
A^{\mathrm{II}}+B^{\mathrm{I}}$ inversion of space gives the
reverse process as well: $A^{\mathrm{II}}+B^{\mathrm{I}} \to
A^{\mathrm{I}}+B^{\mathrm{II}}$.

This fundamental property is formulated in the following theorem.

\begin{theorem}\label{space=time} Let a complex diffusion process consist
of elementary processes, which satisfy the following property:
the space-inverted elementary process coincides with the
inverse process. Then, for the mass action law equation of
diffusion (\ref{difKinUrGenGenN}), the principle of detailed
balance is valid, the global convex Lyapunov functional exists
and the uniform distribution is asymptotically stable.
\end{theorem}

Indeed, due to space symmetry, a uniform distribution is an
equilibrium and each process is equilibrated at this state by
its space-inverted process. At the same time, this distribution
is a point of detailed balance. Therefore, the results about
the principle of detailed balance are applicable.

\subsubsection{Arrested Diffusion and Boundary Equilibria}

Existence of a uniform distribution, which is a point of
detailed balance, existence of the global Lyapunov function and
asymptotic stability of the uniform equilibrium distributions
do not mean that there exist no nonuniform equilibria.

The phenomenon of boundary equilibria is well known. For
example, an autocatalytic reversible reaction
$A+B\rightleftharpoons 2 A$ has two equilibria for a given
value of a stoichiometric conservation law, $b=c_A+c_B=$const.
One equilibrium is strictly positive, with positive
concentrations of $A$ and $B$. Another is a boundary
equilibrium, $c_A=0$, $c_B=b$. The positive equilibrium is
asymptotically stable, the boundary equilibrium is unstable but
if the initial state is near the boundary ($c_A$ is close to
zero) then slow relaxation occurs \cite{GorbanSlorelax2004},
and the motion may be arrested for a long time near this state.

There are well known effects of arrested diffusion caused by
changing of temperature. The solid solutions show the effects
of diffusion, which has been arrested by chilling below a
threshold temperature in a very short time \cite{Lyon1959}.

Kinetic effects of arrested diffusion are also possible. For
example, let us consider the diffusion mechanism by jumps to
free places (Fig.~\ref{fig:SurfElActs2}): any distribution of
components $A_i$ for zero concentrations of free places $Z$ is
stationary, and for a small concentration of free places
diffusion is slow.

For effects of arrested diffusion, the average concentration of
free places should not be small. There may be, for example, a
layer of particles with low mobility between a dense island of
particles with high mobility and an island of free places.
Formally, it is possible to construct many such situations. All
of them may be characterized as follows: either a small change
of concentrations or a small change of constants (or both) lead
the system to a nonuniform equilibrium. At this nonuniform
equilibrium some of the concentrations in some cells take zero
values and, therefore, some of fluxes are also zero. Because of
the appearance of these zero concentrations, such an
equilibrium is called a {\em boundary equilibrium}.

\subsection{Continuous Diffusion Equation}

\subsubsection{MAL Diffusion Flux}

Let us consider an elementary process together with its
space-inverted process
\begin{equation}\label{elementaryActDiffChapII}
\begin{split}
\sum_{i}\alpha_{ri}^{\mathrm{I}}A_{i}^{\mathrm{I}}+
\sum_{i}\alpha_{ri}^{\mathrm{II}}A_{i}^{\mathrm{II}}\rightarrow
\sum_{i}\beta_{ri}^{\mathrm{I}}A_{i}^{\mathrm{I}}+\sum_{i}\beta_{ri}^{\mathrm{II}}A_{i}^{\mathrm{II}}\,
, \\
\sum_{i}\alpha_{ri}^{\mathrm{I}}A_{i}^{\mathrm{II}}+
\sum_{i}\alpha_{ri}^{\mathrm{II}}A_{i}^{\mathrm{I}}\rightarrow
\sum_{i}\beta_{ri}^{\mathrm{I}}A_{i}^{\mathrm{II}}+\sum_{i}\beta_{ri}^{\mathrm{II}}A_{i}^{\mathrm{I}}\,
.
\end{split}
\end{equation}
The reaction rates are
\begin{equation}
\begin{split}
&w_r(c^{\mathrm{I}},c^{\mathrm{II}})=k_r\prod_i
(c_i^{\mathrm{I}})^{\alpha^{\mathrm{I}}_{ri}} \prod_i
(c_i^{\mathrm{II}})^{\alpha^{\mathrm{II}}_{ri}} \, ,\\
&w_r'(c^{\mathrm{I}},c^{\mathrm{II}})=w_r(c^{\mathrm{II}},c^{\mathrm{I}})=k_r\prod_i
(c_i^{\mathrm{II}})^{\alpha^{\mathrm{I}}_{ri}} \prod_i
(c_i^{\mathrm{I}})^{\alpha^{\mathrm{II}}_{ri}} \, ,
\end{split}
\end{equation}
where we take $k'_r=k_r$ due to the symmetry in space.

To first order in $l$, the flux vector for $A_i$ in this
process is
\begin{equation}\label{FluxCOnt}
\begin{split}
J_{ri}&=-\gamma_{ri}[w_r (c(x),c(x+l))-w_r(c(x+l),c(x))]\\&=
-l\gamma_{ri}\sum_j\left(\left.\frac{\partial w_r
(c^{\mathrm{I}},c^{\mathrm{II}})}{\partial
c_j^{\mathrm{II}}}\right|_{c^{\mathrm{I}}=c^{\mathrm{II}}=c(x)}-\left.\frac{\partial
w_r (c^{\mathrm{I}},c^{\mathrm{II}})}{\partial
c_j^{\mathrm{I}}}\right|_{c^{\mathrm{I}}=c^{\mathrm{II}}=c(x)}\right)\nabla
c_j(x) \\&
 =-l\gamma_{ri}
 w_r(c(x),c(x))\sum_j\frac{\alpha^{\mathrm{II}}_{rj}-\alpha^{\mathrm{I}}_{rj}}{c_j}\nabla
c_j(x)\\&
 =-lk\gamma_{ri} \left(\prod_q
 c_q^{\alpha^{\mathrm{I}}_{rq}+\alpha^{\mathrm{II}}_{rq}}\right)
 \sum_j \frac{\alpha^{\mathrm{II}}_{rj}-\alpha^{\mathrm{I}}_{rj}}{c_j}\nabla
 c_j(x) \, .
\end{split}
\end{equation}
Here,
$\gamma_{ri}=\beta_{ri}^{\mathrm{I}}-\alpha_{ri}^{\mathrm{I}}$
(input minus output in the first cell); the minus in front of
the formula appears because the direction of flux from cell I
to cell II (from $x$ to $(x+l)$) is positive.

The factor $1/c_j$ never leads to a singularity in the flux
because $c_j$ enters in the monomial $\prod_q
c_q^{\alpha^{\mathrm{I}}_{rq} +\alpha^{\mathrm{II}}_{rq}}$ with
the power $\alpha^{\mathrm{I}}_{rj}
+\alpha^{\mathrm{II}}_{rj}$. This power is strictly positive if
the coefficient
$({\alpha^{\mathrm{II}}_{rj}-\alpha^{\mathrm{I}}_{rj}})$ is not
zero.

The proper scaling of $k$ for grid refinement or coarsening is
$kl=d=const$ in order not to change the first order expression for
flux (\ref{FluxCOnt}).

According to (\ref{FluxCOnt}), the matrix of diffusion coefficients
for the elementary process (\ref{elementaryActDiffChapII}) (together
with its space-inverted process) is
\begin{equation}\label{DiffCoeff}
D_{r\, ij}(c)=d \left(\prod_q
 c_q^{\alpha^{\mathrm{I}}_{rq}+\alpha^{\mathrm{II}}_{rq}}\right)
 \frac{\gamma_{ri}(\alpha^{\mathrm{II}}_{rj}-\alpha^{\mathrm{I}}_{rj})}{c_j}\,
 ,
\end{equation}
where $d=\mathrm{const}(=kl)$.

The corresponding diffusion equations have the divergent form:
\begin{equation}
\frac{\partial c}{\partial t}=\mathrm{div} (D(c) \nabla c) \, ,
\end{equation}
where $c$ is the vector of concentrations and $D$ is the matrix
of diffusion coefficients (\ref{DiffCoeff}).

It might be useful to represent the flux  (\ref{FluxCOnt})
similarly to the Teorell formula (\ref{Teorell}). For this
purpose, let us collect under $\nabla$ the terms which
represent the chemical potential in perfect media: $\mu=RT\ln c
+ \mu_0$. We assume that $T$ and $\mu_0$ are constant in space.
With these conditions,
\begin{equation}\label{MALTeorell}
 J_{ri}=-\frac{lk}{RT}\gamma_{ri} \left(\prod_q
 c_q^{\alpha^{\mathrm{I}}_{rq}+\alpha^{\mathrm{II}}_{rq}}\right)
 \sum_j (\alpha^{\mathrm{II}}_{rj}-\alpha^{\mathrm{I}}_{rj})\nabla
 \mu_j(x) \, .
\end{equation}

\subsubsection{Examples}

Let us illustrate application of formula (\ref{FluxCOnt}) by several
elementary examples.

First of all, Fick's law: $A^{\mathrm{I}} \to A^{\mathrm{II}}$
and $A^{\mathrm{II}} \to A^{\mathrm{I}}$. For this system,
$\alpha^{\mathrm{I}}=1$, $\alpha^{\mathrm{II}}=0$,
$\beta^{\mathrm{I}}=0$, $\beta^{\mathrm{II}}=1$ and
$\gamma=-1$.

Formula (\ref{FluxCOnt}) gives
$$J=lk c \frac{-1}{c} \nabla c= -lk \nabla c\, .$$
This is exactly the standard Fick law. The diffusion equation is
$\partial_t c=d \Delta c$. Here and further in this subsection we
use $d$ for $lk$.

Exchange of positions: $A^{\mathrm{I}} + B^{\mathrm{II}}\to
A^{\mathrm{II}}+B^{\mathrm{I}}$ together with the space-inverted
process $A^{\mathrm{II}}+B^{\mathrm{I}} \to A^{\mathrm{I}} +
B^{\mathrm{II}}$, which is the same as the reverse process. For this
case, $\alpha_A^{\mathrm{I}}=1$, $\alpha_B^{\mathrm{I}}=0$,
$\alpha_A^{\mathrm{II}}=0$, $\alpha_B^{\mathrm{II}}=1$,
$\beta_A^{\mathrm{I}}=0$, $\beta_B^{\mathrm{I}}=1$,
$\beta_A^{\mathrm{II}}=1$, $\beta_B^{\mathrm{II}}=0$,  $\gamma_A =
\beta_A^{\mathrm{I}}-\alpha_A^{\mathrm{I}}=-1$, and $\gamma_B =
\beta_B^{\mathrm{I}}-\alpha_B^{\mathrm{I}}=1$. Due to
(\ref{FluxCOnt}),
\begin{equation}
\begin{split}
&J_A=-d (-1) c_A c_B \left[ \frac{-1}{c_A}\nabla c_A +\frac{1}{c_B}
\nabla c_B\right]= -d (c_B \nabla c_A - c_A\nabla c_B) \, , \\
&J_B=-d c_A c_B \left[ \frac{-1}{c_A}\nabla c_A +\frac{1}{c_B}
\nabla c_B\right]= d (c_B \nabla c_A - c_A\nabla c_B)\, .
\end{split}
\end{equation}
The diffusion equations are
$$\partial_t c_A=d (c_B \Delta c_A - c_A \Delta c_B) \, , \;
\partial_t c_B=d (c_A \Delta c_B- c_B \Delta
c_A)
\, . $$

Repulsion of components $A$ and $B$ ($A$ is mobile). The mechanism
is $A^{\mathrm{I}} + B^{\mathrm{I}}\to
A^{\mathrm{II}}+B^{\mathrm{I}}$ together with the space-inverted
process $A^{\mathrm{II}}+B^{\mathrm{II}} \to A^{\mathrm{I}} +
B^{\mathrm{II}}$, which does not coincide with the reverse process.
For this mechanism,  $\alpha_A^{\mathrm{I}}=1$,
$\alpha_B^{\mathrm{I}}=1$, $\alpha_A^{\mathrm{II}}=0$,
$\alpha_B^{\mathrm{II}}=0$, $\beta_A^{\mathrm{I}}=0$,
$\beta_B^{\mathrm{I}}=1$, $\beta_A^{\mathrm{II}}=1$,
$\beta_B^{\mathrm{II}}=0$,  $\gamma_A =
\beta_A^{\mathrm{I}}-\alpha_A^{\mathrm{I}}=-1$, and $\gamma_B =
\beta_B^{\mathrm{I}}-\alpha_B^{\mathrm{I}}=0$. Formula
(\ref{FluxCOnt}) gives:
\begin{equation}
\begin{split}
J_A&=-d (-1) c_A c_B \left[\frac{-1}{c_A}\nabla c_A +
\frac{-1}{c_B}\nabla c_B\right]\\ &=-d(c_B \nabla c_A + c_A \nabla
c_B)=-d \nabla (c_A c_B)\, , \\ J_B&=0\, .
\end{split}
\end{equation}
We can see that $B$ activates diffusion of $A$ (the term $c_B \nabla
c_A$) and, at the same time, pushes $A$ in the area with lower
concentration of $B$ (the term $c_A \nabla c_B$).

The diffusion equation is
$$\partial_t c_A= d\Delta (c_Ac_B)\,  .$$

The no-flux steady states of these diffusion equations are
given by the condition: $c_Ac_B=\mathrm{const}$.

If we assume that $B$ is also mobile by a similar mechanism then we
get a system of equations (with two different diffusion
coefficients):
\begin{equation}
J_A=-d_A\nabla (c_Ac_B)\, , \; J_B=-d_B \nabla (c_Ac_B)\, ,
\end{equation}

\begin{equation}
\partial_t c_A=
d_A\Delta (c_Ac_B)\, , \; \partial_t c_B= d_B\Delta (c_Ac_B)\, .
\end{equation}

Let us change variables: $$c_+=\frac{c_A}{d_A}+\frac{c_B}{d_B}, \;
c_-=\frac{c_A}{d_A}-\frac{c_B}{d_B} \, .$$ In these variables,
$c_Ac_B=\frac{d_Ad_B}{4}(c_+^2-c_-^2)$,  $\partial_t c_-=0$ and for
$c_+$ we have the porous media equation:
$$\partial_t c_+=\frac{d_Ad_B}{2}\Delta c_+^2 \, .$$

\subsection{Principle of Detailed Balance and Dissipation Inequality}

\subsubsection{Detailed balance and Coupling of Direct and Reverse
Processes}

In this subsection, we formulate the principle of detailed
balance for MAL diffusion. Physically, it follows from
microreversibility.

For every elementary process,
\begin{equation}\label{elementaryActDiffChapIII}
\sum_{i}\alpha_{ri}^{\mathrm{I}}A_{i}^{\mathrm{I}}+
\sum_{i}\alpha_{ri}^{\mathrm{II}}A_{i}^{\mathrm{II}}\rightarrow
\sum_{i}\beta_{ri}^{\mathrm{I}}A_{i}^{\mathrm{I}}+\sum_{i}\beta_{ri}^{\mathrm{II}}A_{i}^{\mathrm{II}}\,
,
\end{equation}
 the reverse process is (just invert
arrows):
\begin{equation}\label{elementaryActDiffReverse}
\sum_{i}\beta_{ri}^{\mathrm{I}}A_{i}^{\mathrm{I}}+\sum_{i}\beta_{ri}^{\mathrm{II}}A_{i}^{\mathrm{II}}
\to \sum_{i}\alpha_{ri}^{\mathrm{I}}A_{i}^{\mathrm{I}}+
\sum_{i}\alpha_{ri}^{\mathrm{II}}A_{i}^{\mathrm{II}} \, .
\end{equation}
Let us distinguish the quantities for the reverse and direct
processes by the upper indices $\pm$. The simple algebraic relations
hold:
\begin{equation}\label{pmDBrelations}
\alpha_{ri}^{\mathrm{I,II}\mp}=\beta_{ri}^{\mathrm{I,II}\pm}\,
\mbox{ and }\gamma_{ri}^{\mathrm{I,II}\mp}=
-\gamma_{ri}^{\mathrm{I,II}\pm}\, .$$ Therefore,
$$\alpha_{ri}^{\mathrm{I+}}+\alpha_{ri}^{\mathrm{II+}}=\beta_{ri}^{\mathrm{I-}}+\beta_{ri}^{\mathrm{II-}}=
\alpha_{ri}^{\mathrm{I-}}+\alpha_{ri}^{\mathrm{II-}}=\beta_{ri}^{\mathrm{I+}}+\beta_{ri}^{\mathrm{II+}}\,
.
\end{equation}

The reaction rates are
\begin{equation}
\begin{split}
&w_r^+(c^{\mathrm{I}},c^{\mathrm{II}})=k_r^+\prod_i
(c_i^{\mathrm{I}})^{\alpha^{\mathrm{I}}_{ri}} \prod_i
(c_i^{\mathrm{II}})^{\alpha^{\mathrm{II}}_{ri}} \, ,\\
&w_r^-(c^{\mathrm{I}},c^{\mathrm{II}})=k_r^-\prod_i
(c_i^{\mathrm{I}})^{\beta^{\mathrm{I}}_{ri}} \prod_i
(c_i^{\mathrm{II}})^{\beta^{\mathrm{II}}_{ri}} \, ,
\end{split}
\end{equation}

Let there exist a uniform strictly positive point  of detailed
balance: such a strictly positive vector $c^*$ that for all $r$
$$w_r^+(c^*,c^*)=w_r^-(c^*,c^*)\, .$$

This means that
$$k_r^+\prod_i
(c_i^*)^{\alpha^{\mathrm{I}}_{ri}+\alpha^{\mathrm{II}}_{ri}}=
k_r^-\prod_i
(c^*_i)^{\beta^{\mathrm{I}}_{ri}+\beta^{\mathrm{II}}_{ri}}\, .$$

Let us use the relations
$\alpha^{\mathrm{I}}_{ri}+\alpha^{\mathrm{II}}_{ri}=
\beta^{\mathrm{I}}_{ri}+\beta^{\mathrm{II}}_{ri}$. Therefore, the
principle of detailed balance for MAL diffusion can be reformulated:
for all $r$
$$k_r^+=k_r^- \, .$$

Let us join processes (\ref{elementaryActDiffChapIII}),
(\ref{elementaryActDiffReverse}) and write for them the diffusion
flux analogously to (\ref{FluxCOnt}):
\begin{equation}\label{FluxCOntDetBal}
\begin{split}
J_{ri}&=-\gamma_{ri}[w_r^+ (c(x),c(x+l))-w_r^-(c(x+l),c(x))]\\&=
-l\gamma_{ri}\sum_j\left(\left.\frac{\partial w_r^+
(c^{\mathrm{I}},c^{\mathrm{II}})}{\partial
c_j^{\mathrm{II}}}\right|_{c^{\mathrm{I}}=c^{\mathrm{II}}=c(x)}-\left.\frac{\partial
w_r^- (c^{\mathrm{I}},c^{\mathrm{II}})}{\partial
c_j^{\mathrm{I}}}\right|_{c^{\mathrm{I}}=c^{\mathrm{II}}=c(x)}\right)\nabla
c_j(x) \\&
 =-l\gamma_{ri}
 w_r(c(x),c(x))\sum_j\frac{\gamma_{rj}}{c_j}\nabla
c_j(x)\\&
 =-lk\gamma_{ri} \left(\prod_q
 c_q^{\alpha^{\mathrm{I}}_{rq}+\alpha^{\mathrm{II}}_{rq}}\right)
 \sum_j \frac{\gamma_{rj}}{c_j}\nabla
 c_j(x) \, .
\end{split}
\end{equation}

According to this expression for the diffusion flux,
(\ref{FluxCOntDetBal}), the matrix of diffusion coefficients for the
elementary process (\ref{elementaryActDiffChapII}) (together with
its space-inverted process) is
\begin{equation}\label{DiffCoeffDB}
D^r_{ij}(c)=d_r \left(\prod_q
 c_q^{\alpha^{\mathrm{I}}_{rq}+\alpha^{\mathrm{II}}_{rq}}\right)
 \frac{\gamma_{ri}\gamma_{rj}}{c_j}\,  ,
\end{equation}
where $d=\mathrm{const}(=k_rl)$.

This matrix is symmetric with respect to the inner product
\begin{equation}\label{entropicprodC}
\langle a , b\rangle_c=\sum_i \frac{a_ib_i}{c_i} \, .
\end{equation}
This means that $$\langle D a ,  b\rangle_c=\langle a , D
b\rangle_c$$ or in coordinates
$$\sum_{ij} \frac{D_{ij} a_j b_i}{c_i}=\sum_{ij} \frac{a_i D_{ij}
b_j}{c_i} \, .$$
 Indeed, let $\widetilde{w}_r=d_r \left(\prod_q
 c_q^{\alpha^{\mathrm{I}}_{rq}+\alpha^{\mathrm{II}}_{rq}}\right)$.
 For two scalar products, $\langle D a, b\rangle_c$ and $\langle a , D
b\rangle_c$ we get
$$\langle D a, b\rangle_c=\sum_{ij} \frac{D_{ij} a_j
b_i}{c_i}=\widetilde{w}_r
 \sum_{ij}\frac{\gamma_{ri}\gamma_{rj}a_j b_i}{c_jc_i}\, $$
 and
$$\langle a , D
b\rangle_c=\sum_{ij} \frac{a_i D_{ij}b_j}{c_i}=\widetilde{w}_r
 \sum_{ij}\frac{a_i \gamma_{ri}\gamma_{rj} b_j}{c_ic_j}\, ,$$
 These two expressions differ only in the notation of dummy indexes,
 $i$ and $j$. Therefore, $D$ is symmetric with respect to the inner
 product (\ref{entropicprodC}).

$D$ is also positive semi-definite. Indeed, for any vector $\xi$,
$$\langle \xi, D \xi \rangle =\sum_i \frac{\xi_i D^r_{ij} \xi_j}{c_i}=
\widetilde{w}_r \langle\xi,\gamma)^2\geq 0 \rangle\, .$$ This
expression may be zero at a positive state ($c_i>0$) if and
only if the vector $\xi$ is orthogonal to the vector $\gamma_r$
in the inner product (\ref{entropicprodC}). If we summarize
diffusion coefficients for all pairs of mutually inverse
elementary processes then we get
\begin{equation}\label{DiffCoeffDBfull}
D_{ij}=\sum_r D^r_{ij}(c)=\sum_r \widetilde{w}_r
\frac{\gamma_{ri}\gamma_{rj}}{c_j}\,  .
\end{equation}

For this $D_{ij}$, $$\sum \frac{\xi_i D_{ij} \xi_j}{c_i} \geq
0$$ and is zero if and only if vector $\xi$ is orthogonal to
all vectors $\gamma_r$ in the inner product
(\ref{entropicprodC}).

The corresponding diffusion equations have the divergent form:
\begin{equation}
\frac{\partial c}{\partial t}=\mathrm{div} (D(c) \nabla c) \, .
\end{equation}

It might be useful to represent the flux (\ref{FluxCOntDetBal})
similarly to the Teorell formula (\ref{Teorell}). For this
purpose, let us collect under $\nabla$ the terms which
represent the chemical potential in perfect media: $\mu=RT\ln c
+ \mu_0$. We assume that $T$ and $\mu_0$ are constant in space.
With these conditions,
\begin{equation}\label{MALTeorellDB}
J_{ri} =-\frac{lk}{RT}\gamma_{ri} \left(\prod_q
 c_q^{\alpha^{\mathrm{I}}_{rq}+\alpha^{\mathrm{II}}_{rq}}\right)
 \sum_j \gamma_{rj}\nabla
 \mu_j(x) \, .
\end{equation}

The difference from the MAL Teorell formula (\ref{MALTeorell})
is obvious: for the systems with detailed balance
(\ref{MALTeorellDB}) the matrix of the coefficients in the
Teorell formula is symmetric for each  elementary process
together with its reverse process because it is a product of a
number (in square brackets) and the symmetric matrix
$\gamma_{ri}\gamma_{rj}$:
\begin{equation}
\left[\frac{lk}{RT} \left(\prod_q
 c_q^{\alpha^{\mathrm{I}}_{rq}+\alpha^{\mathrm{II}}_{rq}}\right)\right]
\gamma_{ri}\gamma_{rj} \, .
\end{equation}

For the general MAL diffusion (with detailed symmetry in space)
the matrix these coefficients for the elementary process
together with its space-inverted process is not symmetric in
general: it is a scalar multiple of the matrix
$$\gamma_{ri}(\alpha^{\mathrm{II}}_{rj}-\alpha^{\mathrm{I}}_{rj})\, .$$
Its symmetry may be guaranteed if the space inversion of the
elementary process coincides with its reversion in time (i.e.
$\alpha^{\mathrm{II}}_{r}$ coincides with
$\beta^{\mathrm{I}}_{r}$).

For MAL chemical kinetics, there is a sufficient algebraic
condition for detailed balance that is independent of the
microreversibility and follows just from the stoichiometric
equations. Indeed, let us assume that all the reactions are
reversible and all the stoichiometric vectors $\gamma_r$ in
(\ref{MALreversDB}) are linearly independent. Then, at the
equilibrium, from the condition $\dot{c}=\sum_r w_r \gamma_r=0$
we get $w_r=0$ for all $r$. That is the detailed balance
condition.

For MAL diffusion we have also a specific (``diffusion") algebraic
condition: if for all elementary processes the space-inverted
reaction is the reverse reaction then the principle of detailed
balance follows from the space symmetry condition. For such systems,
the coupling ``process--space inverted process" coincides with the
coupling ``process--reverse process" and equations
(\ref{FluxCOntDetBal}) coincide with (\ref{FluxCOnt}).

For an elementary process (\ref{elementaryActDiffChapIII}), let
the space-inverted process not coincide with the reverse
process. With the condition of detailed balance, it is
straightforward to check that for the pair of elementary
processes (\ref{elementaryActDiffChapIII}),
(\ref{elementaryActDiffReverse}), the space-inverted process to
(\ref{elementaryActDiffChapIII}) and the reverse process to
this space-inverted one  generate the same diffusion equations
and the diffusion coefficient as the original pair does. These
two couples together produce the flux that is twice as large as
(\ref{FluxCOntDetBal}).

\subsubsection{The Dissipation Inequality and Detailed Balance}

In this subsection, we construct the Lyapunov functional for
the general MAL diffusion models and calculate its time
derivative due to the diffusion equations.

Let us select any strictly positive reference concentration vector
$c^*$ and take (\ref{Gfunction})
$$G=\sum_i c_i \left(\ln\left(\frac{c_i}{c_i^*}\right)-1\right)+\sum_i
c_i^* \, .$$

Let us consider this system in a bounded domain $V$ with smooth
boundary and with zero fluxes through its boundary: $(n,J)=0$ at any
point of $\partial V$ at any time ($n$ is the vector of the outer
normal). Due to the definition of flux (\ref{FluxCOnt}), it is
sufficient that $(n,\mathrm{grad}c_j)=0$ on $\partial V$ for all $j$
(but it is not necessary).

The Lyapunov functional is

\begin{equation}\label{GfunctionINT}
\mathbf{G}=\sum_i \int_V
\left[c_i\left(\ln\left(\frac{c_i}{c_i^*}\right)-1\right)+\sum_i
c_i^* \right] \, \D x \, .
\end{equation}

Due to the boundary conditions and the Gauss--Ostrogradskii theorem
\begin{equation}\label{GPRODUCTIONDIFDB}
\frac{\D \mathbf{G}}{\D t}=-\sum_i \int_V
\ln\left(\frac{c_i}{c_i^*}\right)
 \mathrm{div} J_i \, \D x  = \sum_i \int_V
 \left(\nabla_x \left(\ln\frac{c_i}{c_i^*}\right),
  J_i \right)  \, \D x \, .
\end{equation}

Let us assume the principle of detailed balance and calculate
$\dot{\mathbf{G}}$ (\ref{GPRODUCTIONDIFDB}) due to diffusion
equation with the matrix of diffusion coefficients
(\ref{DiffCoeffDBfull}):

\begin{equation}\label{GPRODUCTIONDIFDB2}
\begin{split}
\frac{\D \mathbf{G}}{\D t}&= \sum_i \int_V
 \left(\nabla_x \left(\ln\frac{c_i}{c_i^*}\right),
  J_i \right)  \, \D x \,  \\
&=-\sum_r  \int_V \widetilde{w}_r \sum_{ij}  \left(\nabla_x
(\ln{c_i}),
  \gamma_{ri}\gamma_{rj} \nabla_x (\ln{c_j}) \right)  \, \D x \,  \\
&=-\sum_r  \int_V \widetilde{w}_r \left(\nabla_x\left(
\sum_i\ln(\gamma_{ri}c_i)\right),
   \nabla_x \left( \sum_j\ln(\gamma_{rj}c_j)\right) \right)  \, \D x \\
&=-\sum_r  \int_V \widetilde{w}_r (\nabla_x (\gamma_r,\nabla_c G))^2
\, \D x \leq 0 \, .
 \end{split}
\end{equation}
Here, $(\ , \ )$ is the standard Euclidean inner product both in
space and in the concentration space.

As we can see from this dissipation inequality, the only positive
equilibria for diffusion equations with detailed balance conditions
satisfy the conditions $(\gamma_r,\nabla_c G)=\mathrm{const}$ for
all $r$. Another form of these conditions is
$$\prod_i c_i^{\gamma_{ri}}=\mathrm{const} \, .$$

Inequality (\ref{GPRODUCTIONDIFDB2}) demonstrates a significant
difference between the two classes of diffusion mechanism. If
the stoichiometric vectors $\gamma_r$ form a basis in the
concentration space then all the equilibria are uniform because
in this case the condition $(\gamma_r,\nabla_c
G)=\mathrm{const}$ (for all $r$) implies $\nabla_c
G=\mathrm{const}$, that is, $\ln c_i = \mathrm{const}$ for all
$i$, hence, $c_i=\mathrm{const}$ for all $i$. Let us call this
mechanism the {\em mechanisms with mixing}. The first example
is Fick's mechanism: if the diffusion constant is not zero for
all components then all the equilibria of the system are
uniform.

If $\mathrm{span} \{\gamma_r\} $ does not coincide with the
concentration space then there exist the invariants of
diffusion. They are given by the linear functionals that
annihilate all the  vectors $\gamma_r$.  For example, in
diffusion by jumps on the free places
(Fig.~\ref{fig:SurfElActs2}) the value of the sum $c_Z(x)
+\sum_i c_i(x)$ is locally conserved. In the mechanism,
$A^{\mathrm{I}} + B^{\mathrm{I}}\to
A^{\mathrm{II}}+B^{\mathrm{I}}$,
$A^{\mathrm{II}}+B^{\mathrm{II}} \to A^{\mathrm{I}} +
B^{\mathrm{II}}$,  concentration of $B$ is conserved. These
locally conserved quantities together with the condition of
positivity $c_i\geq 0$ define a convex body where the vector of
concentrations may be situated at a given point. This body
depends on the values of the conserved quantities, differs for
different points, but does not change in time.

\subsection{Complex Balance in MAL Diffusion
\label{Sec:CompBalMALDif}}

\subsubsection{Complex Balance Conditions for MAL Diffusion}

The complex balance condition does not assume any space or time
symmetry. The only microscopic assumption is the Markov fast
microscopic kinetic with relatively small amount of active
intermediate complexes \cite{Stueckelberg1952,GorbanShahzad2010}.

We discussed this condition for the MAL kinetics in
Section~\ref{Sec:ComplBal}, and now let us transform it into a
condition for the MAL diffusion equation. First of all, we
should abandon the symmetry conditions $k'=k$ (space symmetry)
and $k^+=k^-$ (microreversibility). Without these conditions,
the zero-order terms in the expression for fluxes will not be
annihilated by balance between the elementary processes with
given pair of vectors
$(\alpha_r^{\mathrm{I}},\alpha_r^{\mathrm{II}})$ and the
processes with the same pair
$(\beta_r^{\mathrm{I}},\beta_r^{\mathrm{II}})$ (\ref{complex
balance}).

Let the stoichiometric mechanism of the diffusion be given:
\begin{equation}\label{elementaryActDiffChap4}
\sum_{i}\alpha_{ri}^{\mathrm{I}}A_{i}^{\mathrm{I}}+
\sum_{i}\alpha_{ri}^{\mathrm{II}}A_{i}^{\mathrm{II}}\rightarrow
\sum_{i}\beta_{ri}^{\mathrm{I}}A_{i}^{\mathrm{I}}+\sum_{i}\beta_{ri}^{\mathrm{II}}A_{i}^{\mathrm{II}}\,
,
\end{equation}
where all the elementary processes have different numbers $r$
and the space-inverted and reverse processes are represented
separately. Let us consider all pairs of vectors,
$(\alpha_r^{\mathrm{I}},\alpha_r^{\mathrm{II}})$  and
$(\beta_r^{\mathrm{I}},\beta_r^{\mathrm{II}})$. Let us
enumerate all the different pairs: $y_1,y_2,\ldots$,
$y_q=(y_q^{\mathrm{I}},y_q^{\mathrm{II}})$. For each $y_q$
there are two sets of reactions, $R_q^+$, $R_q^-$:

$$R_q^+=\{r\, | \, (\alpha_r^{\mathrm{I}},\alpha_r^{\mathrm{II}})=y_q\}\, , \;
R_q^-=\{r\, | \, (\beta_r^{\mathrm{I}},\beta_r^{\mathrm{II}})=y_q\}
\, .$$

The complex balance condition is: there exists a strictly
positive vector $c^*$ such that
\begin{equation}\label{complex balanceDIF}
\sum_{r\in R_q^+} w_r(c^*,c^*)=\sum_{r \in R_q^-} w_r(c^*,c^*)\,
\mbox{ for all } q \, .
\end{equation}
For MAL this means
\begin{equation}\label{ComBalPol}
\sum_{r\in R_q^+} k_r \prod_i
(c^*_i)^{\alpha_r^{\mathrm{I}}+\alpha_r^{\mathrm{II}}}=\sum_{r\in
R_l^-}k_q \prod_i
(c^*_i)^{\alpha_q^{\mathrm{I}}+\alpha_q^{\mathrm{II}}}\, \mbox{ for
all } l \, .
\end{equation}

For $r\in R_q^+$,
$$\alpha_r^{\mathrm{I}}+\alpha_r^{\mathrm{II}}=y_q^{\mathrm{I}}+y_q^{\mathrm{II}}\, ,$$
and for $r\in R_l^-$
$$\beta_r^{\mathrm{I}}+\beta_r^{\mathrm{II}}=y_q^{\mathrm{I}}+y_q^{\mathrm{II}}\, .$$
Therefore, all the monomials coincide in the right and the left
hand sides of (\ref{ComBalPol}) for the given $q$ and we can
write the complex balance condition  for diffusion in the
following form:
\begin{equation}\label{complexBalance constDif}
\sum_{r\in R_q^+} k_r =\sum_{r\in R_q^-}k_r \, \mbox{ for all } q \,
.
\end{equation}

Let us calculate the flux
$$J=\sum_r \gamma_r w_r(c(x),c(x+l))$$ at
first order in $l$. We group terms in this sum and each group
will corresponds to a pair $y_q$. For each elementary process
with number $r$ there are two $q$, $q_{\alpha}(r)$ and
$q_{\beta}(r)$: ${r\in R_{q_{\alpha}(r)}^+}$ and ${r\in
R_{q_{\beta}(r)}^-}$. We split the term $\gamma_r
w_r(c(x),c(x+l))$ in two terms:
$$\gamma_r
w_r(c(x),c(x+l))= \beta_r w_r(c(x),c(x+l))- \alpha_r
w_r(c(x),c(x+l))$$ We associate the first of them with
$y_{q_{\beta}(r)}$ and the second with $y_{q_{\alpha}(r)}$. The
result is represented below:
\begin{equation}\label{ComplBalFluxMAL}
\begin{split}
J&=-\sum_r \gamma_r w_r(c(x),c(x+l)) \\
&=\sum_q \left(\sum_{r\in R_q^+} \alpha_r^{\mathrm{I}} w_r(c(x),c(x+l)) -
\sum_{r\in R_q^-} \beta_r^{\mathrm{I}} w_r(c(x),c(x+l))\right) \\
&=\sum_q y_q^{\mathrm{I}} \left(\sum_{r\in R_q^+} w_r(c(x),c(x+l)) -
\sum_{r\in R_q^-} w_r(c(x),c(x+l))\right) \\
&=\sum_q y_q^{\mathrm{I}} \prod_i c_i^{y_q^{\mathrm{I}}+y_q^{\mathrm{II}}} \left(\sum_{r\in R_q^+} k_r   -
\sum_{r\in R_q^-}k_r \right) \\
&\quad + l \sum_q y_q^{\mathrm{I}} \prod_i c_i^{y_q^{\mathrm{I}}+y_q^{\mathrm{II}}}
\left(\sum_{r\in R_q^+} k_r \sum_j \frac{\alpha^{\mathrm{II}}_{rj}}{c_j} \nabla c_j  -
\sum_{r\in R_q^-}k_r \sum_j \frac{\alpha^{\mathrm{II}}_{rj}}{c_j} \nabla c_j\right)+o(l) \, .
\end{split}
\end{equation}
The zero-order term is zero because of the complex balance
condition (\ref{complexBalance constDif}). Let us take into
account that both for ${r\in R_q^+}$ and ${r\in R_q^-}$
$$\alpha^{\mathrm{I}}+\alpha^{\mathrm{II}}=y_q^{\mathrm{I}}+y_q^{\mathrm{II}}\,
.$$ Therefore, if for two $y_q,y_s$ $$R_q^+ \bigcap R_s^- \neq \emptyset$$ then
$$y_q^{\mathrm{I}}+y_q^{\mathrm{II}}=y_q^{\mathrm{I}}+y_q^{\mathrm{II}}\, .$$
For the set of all vectors $y_q^{\mathrm{I}}+y_q^{\mathrm{II}}$
we use notation $Z$. For each $z \in Z$
$$Y_z=\{y_q \ | \ y_q^{\mathrm{I}}+y_q^{\mathrm{II}} = z \}$$

Let us group the terms with the same vectors
$z=y_q^{\mathrm{I}}+y_q^{\mathrm{II}}\in Z$ together and write
the expression for the flux for the first order:
\begin{equation}\label{firstOrdFluxComBal}
\begin{split}
&J=l \sum_{z \in Z} \prod_i c_i^{z_i} J^z\, , \\
&J^z= \sum_{y_q\in Y_z} y_q^{\mathrm{I}}
\sum_j\left(\sum_{r\in R_q^+} k_r
\frac{\alpha^{\mathrm{II}}_{rj}}{c_j}  - \sum_{r\in R_q^-}k_r
\frac{\alpha^{\mathrm{II}}_{rj}}{c_j} \right) \nabla c_j \, .
\end{split}
\end{equation}
Let us study the expression for $J^z$ for given $z$. First of
all,
$$\sum_{y_q\in Y_z} \sum_j\left(\sum_{r\in
R_q^+} k_r \frac{\alpha^{\mathrm{II}}_{rj}}{c_j}  - \sum_{r\in
R_q^-}k_r \frac{\alpha^{\mathrm{II}}_{rj}}{c_j} \right) \nabla
c_j=0 \, $$ because each $k_r$ from this sum enters it twice,
with the opposite signs, one time as an element of $ R_q^+$ for
some $y_q \in Y_z$ (with $+$) and the second time (with $-$) as
an element of $R_s^-$ for another $y_s\in Y_z$ with the same
sum
$$y_s^{\mathrm{I}}+y_s^{\mathrm{II}}=y_q^{\mathrm{I}}+y_q^{\mathrm{II}}=z\, .$$

We note that $y_q^{\mathrm{I}}=z-y_q^{\mathrm{II}}$ for $y_q
\in Y_z$ in the sum (\ref{firstOrdFluxComBal}). This allows us
to rewrite the expression for $J^z$:
\begin{equation}\label{firstOrdFluxComBal2}
J^z=- \sum_{y_q \in Y_z} y_q^{\mathrm{II}}
\sum_j\left(\sum_{r\in R_q^+} k_r
\frac{\alpha^{\mathrm{II}}_{rj}}{c_j}  - \sum_{r\in R_q^-}k_r
\frac{\alpha^{\mathrm{II}}_{rj}}{c_j} \right) \nabla c_j
\, .
\end{equation}

\subsubsection{The Dissipation Inequality and Complex Balance for MAL
Diffusion}

We would like to demonstrate some similarity of the expression
for $J^z$ and some formulas from the theory of Markov chains.
Vectors $y_q\in Y_z $  numerate the states. Elementary
processes correspond to transitions between states. Each
nonzero constant $k_r$ corresponds to two vectors $y_q,y_s\in
Y_z$: ${r\in R_q^+}$ and ${r\in R_s^-}$. We substitute the
index $r$ by two indexes $q,s$ and use notation $k_{sq}$ (or
even $k_{s \leftarrow q}$). If there is no nonzero constant for
this pair $q,s$ then we take $k_{s \leftarrow q}=0$. In
particular, $k_{qq}=0$. The complex balance condition
(\ref{complexBalance constDif}) reads:
\begin{equation}\label{MarkovStSt}
\sum_q k_{sq}= \sum_q k_{qs} \, .
\end{equation}

This means that the constants $k_{sq}$ describe a continuous time
Markov chain with the Master equation
\begin{equation}\label{MAsterUni}
\dot{\pi}_q=\sum_s(k_{qs}\pi_s - k_{sq}\pi_q)\,
\end{equation}
and equidistribution in equilibrium. Here $\pi_q$ is the probability
to find the system in the state $q$ and $k_{qs}\pi_s$ is the
probability flux from the state $s$ to the state $q$. We can use the
steady state condition (\ref{MarkovStSt}) and rewrite the Master
equation (\ref{MAsterUni}):
\begin{equation}\label{MAsterUni1}
\dot{\pi}_q=\sum_s k_{qs}(\pi_s - \pi_q)\, .
\end{equation}
From this form, it is easy to see that the functional
$$H=\frac{1}{2}\sum_{q} \pi_q^2$$
monotonically decreases due to the system dynamics:
\begin{equation}\label{MasterLyapunovSimple}
\frac{\D H}{\D t}=\sum_{sq} \pi_q k_{qs}(\pi_s-\pi_q) \leq 0 \, .
\end{equation}

To prove (\ref{MasterLyapunovSimple}), let us use the identity
\begin{equation}\label{fullProb}
\sum_q \dot{\pi}_q =\sum_s(k_{qs}\pi_s - k_{sq}\pi_q)=\sum_s
k_{qs}(\pi_s - \pi_q)=0 \, .
\end{equation}
This condition holds for all values of numbers $\pi_q$ (this is
obvious for the Master equation in the form (\ref{MAsterUni})).

In particular, $$\sum_s k_{qs}\left(\frac{1}{2}\pi_q^2 -
\frac{1}{2}\pi_s^2\right)=0 \, .$$ Let us add this expression to the
right hand side of (\ref{MasterLyapunovSimple}):
\begin{equation}\label{MasterLyapunovSimple1}
\frac{\D H}{\D t}=\sum_{sq} k_{qs} \left(\frac{1}{2}\pi_q^2 -
\frac{1}{2}\pi_s^2 + \pi_q (\pi_s-\pi_q)\right) \leq 0
\end{equation}
because $k_{qs}\geq 0$ and the expression in the parentheses is
non-positive:
$$\frac{1}{2}\pi_q^2 - \frac{1}{2}\pi_s^2 + \pi_q
(\pi_s-\pi_q)=-\frac{1}{2}(\pi_q- \pi_s)^2\leq 0 \, . $$

Therefore, we have proved the inequality for our set of coefficients
$k_{qs}$:

\begin{equation}\label{IneqMark2}
\sum_{sq} \pi_q (k_{qs}\pi_s - k_{sq}\pi_q)= \sum_{sq} \pi_q
k_{qs}(\pi_s-\pi_q) \leq 0 \,
\end{equation}
for any set of numbers $\pi_q$. This inequality means that $\D
H/\D t \leq 0$ (\ref{MasterLyapunovSimple}). It is zero if all
$\pi_q$ coincide ($\pi_s=\pi_q$ for all $s$, $q$).

For our purposes, it is important to know when the zero time
derivative of $H$ ($\D H/\D t = 0$) is equivalent to the
equidistribution ($\pi_s=\pi_q$ for all $s$, $q$). They are
equivalent if the Markov chain (\ref{MAsterUni}) is {\em
ergodic}. The conditions of ergodicity are well known
\cite{Seneta1981,VanMieghem2006}: the chain (\ref{MAsterUni})
is ergodic if for any two $s$, $ q$ ($s \neq q$) there exists a
oriented path from $s$ to $ q$ in the graph of the network,
that is such a sequence
$$r_0, r_1, \ldots ,r_g \mbox{ that } s=r_0, \, q=r_g \mbox{ and }
k_{r_{j+1}r_j}>0 \mbox{ fo all } j=0,\ldots, g \, .$$ This
means that the graph of of transitions of the Markov chain
(\ref{MAsterUni}) is {\em strongly connected}.

To apply this inequality to  the proof of the dissipation
inequality, we have to rewrite the expression for $J^z$
(\ref{firstOrdFluxComBal2}) using these notations, $k_{qs}$
instead of $k_r$:

\begin{equation}\label{firstOrdFluxComBal3}
J^z= \sum_{y_q \in Y_z}
y_q^{\mathrm{II}} \sum_j \sum_s (k_{qs} {y_{sj}^{\mathrm{II}}}-
k_{sq} {y_{qj}^{\mathrm{II}}} ) \nabla_x \ln c_j  \, .
\end{equation}

Now we are in the position to prove the dissipation inequality
for MAL diffusion equation with complex balance. In a bounded
domain $V$ with smooth boundary and without fluxes through
boundary we have to estimate $\dot{\mathbf{G}}$
(\ref{GfunctionINT}), (\ref{GPRODUCTIONDIFDB}).

\begin{equation}\label{GPRODUCTIONDIFCB}
\begin{split}
\frac{\D \mathbf{G}}{\D t}&=-\int_V \sum_j
 \ln\left(\frac{c_j}{c_j^*}\right) \mathrm{div}
  J_j   \, \D x \\
&=\int_V \sum_j
 (\nabla_x \ln c_j,
  J_j )   \, \D x \\
&=l \sum_{z \in Z} \int_V   \prod_i c_i^{z_i} \sum_j (\nabla_x
\ln c_j,
  J_{j}^z)   \, \D x \, .
\end{split}
\end{equation}
Due to the representation of $J^z$ (\ref{firstOrdFluxComBal3}),
\begin{equation}
\sum_j (\nabla_x \ln c_j,
  J_{j}^z ) =
\sum_{q \, s} (\pi_q, (k_{qs} \pi_s- k_{sq} \pi_q )) \leq 0 \, ,
\end{equation}
where $\pi_q$ is a space vector: $\pi_q=\sum_i
y_{qi}^{\mathrm{II}} \nabla_x \ln c_i$ and for $y_q$, $y_s$
$$y_{q,s}^{\mathrm{I}}+y_{q,s}^{\mathrm{II}}=z\, .$$ This
expression has exactly the form (\ref{IneqMark2}) for each
space coordinate. Finally,
\begin{equation}
\sum_j (\nabla_x \ln c_j,  J_{j}^z) \leq 0
\end{equation}
and it is zero if all $\pi_q=\sum_i y_{qi}^{\mathrm{II}}
\nabla_x \ln c_i$ coincide.

The reverse statement, $$\mbox{all } \pi_q=\sum_i
y_{qi}^{\mathrm{II}} \nabla_x \ln c_i \mbox{ coincide if }
\sum_j (\nabla_x \ln c_j,  J_{j}^z) = 0\, ,$$ is true if the
auxiliary Markov chain is ergodic for given $z$ (i.e. the graph
of transitions is strongly connected). Let us assume this
ergodicity. For every $z$ we can define a linear subspace $E_z$
in the concentration space given by the system of equation
$$E_z=\{e \ | \ (y_{qi}^{\mathrm{II}}, e) \mbox{ coincide for
all } y_q \in Y_z \} \, . $$

If $$\bigcap_{z\in Z} E_z = \{0\}$$ then all the equilibria for
this mechanism of diffusion are uniform. In particular, they
are uniform if at least one $y_{qi}^{\mathrm{II}}=0$.

\subsection{Intermediate Summary}

We presented the formal language of the stoichiometric
mechanism for description of nonlinear diffusion
(\ref{elementaryActDiffChap}), (\ref{elementaryActDiffChapII}),
(\ref{elementaryActDiffChapIII}),
(\ref{elementaryActDiffReverse}). The general construction of
the diffusion equations under various conditions is given:
\begin{itemize}
\item{For systems with symmetry with respect to inversion
    in space ($k_r=k_r'$) (\ref{FluxCOnt}),
    (\ref{DiffCoeff});}
\item{For systems with microreversibility ($k_r^+=k_r-$)
    (\ref{FluxCOntDetBal}), (\ref{DiffCoeffDB});}
\item{For general systems with Markov microkinetics, which
    satisfy the complex balance conditions
    (\ref{complexBalance constDif}),
    (\ref{firstOrdFluxComBal}).}
\end{itemize}

For systems with detailed balance (microreversibility) and with
complex balance (Markov microkinetics) the explicit formula for
the free energy--type Lyapunov functional $\mathbf{G}$
(\ref{GfunctionINT}) is
$$\mathbf{G}=\sum_i \int_V
\left[c_i\left(\ln\left(\frac{c_i}{c_i^*}\right)-1\right)+\sum_i
c_i^* \right] \, \D x \, .$$

We found that $\mathbf{\dot{G}}\leq 0$ for the systems with
detailed or complex balance (\ref{GPRODUCTIONDIFDB}),
(\ref{GPRODUCTIONDIFCB}). These inequalities guarantee the
thermodynamic behavior of diffusion. The detailed symmetry with
respect to the inversion in space is insufficient for such an
inequality and the diffusion collapse for them is possible.

The Mass Action Law by itself does not imply thermodynamics. In
that sense, it is too flexible and needs additional
requirements to respect the basic physics. This redundancy of
MAL allows, at the same time, to use them in many other areas.
The famous Lotka and Volterra models in Mathematical ecology
\cite{Lotka1925,Volterra1926} are implementations of MAL for
description of surviving and reproduction of animals, that is
far from the initial area of MAL applications. Creation of
mathematical genetics \cite{Fisher1930} and analysis of
dynamical aspects of biological evolution \cite{Gause1934} also
use MAL in their backgrounds. Combination of MAL kinetics with
nonlinear diffusion is very important for analysis of
biological invasions and other phenomena in ecology
\cite{Gurney1976,Hengeveld1989,ShigesadaKawasaki1997,Petrovskii2006}.
The MAL for diffusion can also generate equations that have no
direct physical sense but may be used for modeling of some
phenomena of non-physical nature.

On the other hand, in this approach, we did not take into account some basic physical properties,
namely,  the momentum and the center of mass conservation. The diffusion transport should be
coupled with the viscous transport or elastic deformation (or both)
because two reasons:
\begin{itemize}
\item{The mass average velocity of diffusion $$u=\frac{\sum_i m_i J_i}{\sum_i
m_i}\, ,$$ where $m_i$ is the molecular mass of the $A_i$ particles,
is, in general, not zero;}
\item{The change of the mixture composition implies the change  of
pressure and, hence, the viscous flux or the elastic deformation (or
both).}
\end{itemize}
The careful analysis of these effects should give, for example a
theory of the Kirkendall effect. In 1942, Kirkendall demonstrated
experimentally that different atoms can migrate at different rates
in an alloy, and this diffusion is accompanied by measurable local
volume change and displacement of interfaces
\cite{Kirkendall1942,Nakajima1977}. The kinetic theory of this
effect is still under development and there remain open problems and
new ideas are needed \cite{Narasimhan2007}.

We return to the problem of correct description of the general
transport equation in next Section.

The MAL formalism for diffusion is a flexible and effective
tool for modeling. The semi-discrete MAL model may be used for
numerical modeling directly as a sort of finite elements. Their
coarse-graining and refinement should follow the main rule
$kl=d$ where $l$ is the cell size, $k$ is a kinetic constant
for the finite elements, $d$ is the invariant diffusion
coefficient. For unstable processes, these provide a biharmonic
regularization. These {\em kinetic finite elements} respect the
basic physical properties like positivity of concentration,
conservation laws and the second law of thermodynamics (under
the relevant conditions of detailed or complex balance).

\section{Generalized Mass Action Law for Diffusion \label{sec:GenMAL}}

\subsection{Free Energy, Free Entropy, Chemical Potentials, \\ Activities, and Generalized Mass Action Law}

\subsubsection{Thermodynamic Potentials}

In this Subsection, we present the thermodynamic approach to the
generalized MAL. Exactly as in Section~\ref{Sec:MALKin} we start
from the chemical kinetic equations and then extend our approach to
the transport processes.

In the thermodynamic approach, the kinetic description of the
multicomponent system requires the following inputs

\begin{enumerate}
\item{A list of components;}
\item{A thermodynamic potential;}
\item{A list of elementary reactions represented by their
    stoichiometric equations;}
\item{A set of reaction rate constants.}
\end{enumerate}

Exactly as it was in Section~\ref{Sec:MALKin}, the list of
components is just a set of symbols (the component names). We
usually assume that this set is finite, $A_1, A_2, \ldots ,
A_n$. The definitions of stoichiometric equation and the
corresponding vectors  $\alpha_{ri}$, $\beta_{ri}$ are also the
same.

There are many thermodynamic potentials and they form two series:
energy and free energies and, on another hand, entropy and free
entropies (the Massieu--Planck functions). Each of them has its own
``natural variables" and if one of them is given in the natural
variables then all other thermodynamic functions can be produced.

The names and standard notations of variables are:
\begin{itemize}
\item{Internal energy, $U$;}
\item{Entropy, $S$;}
\item{Enthalpy, $H$;}
\item{Temperature, $T$;}
\item{Volume, $V$;}
\item{Pressure, $P$;}
\item{Number of particles (or moles) composing the $i$th
component $A_i$, $N_i$ ($N$ is vector with coordinates $N_i$, the
vector of composition);}
\item{Chemical potential of the  $i$th component $A_i$,
    $\mu_i$.}
\end{itemize}

The first potential in the energetic series is the internal energy
$U(S,V,N)$ in the natural coordinates $S$, $V$, $N$ and
$$\D U= T \D S- P\D V+ \sum_i \mu_i \D N_i\, .$$
The enthalpy, $H(S,P,N)=U+PV$ has the natural coordinates $S$, $P$,
$N$ and
$$\D H= T \D S+ V \D P+ \sum_i \mu_i \D N_i\, .$$
The free energy (the Helmholtz energy), $F(T,V,N)=U-TS$ and
$$\D F=- S \D T - P \D V+ \sum_i \mu_i \D N_i\, .$$
The free enthalpy (the Gibbs energy), $G(T,P,N)=H-TS$ and
$$\D G=- S \D T+ V \D P+ \sum_i \mu_i \D N_i\, .$$
The grand potential, $\Omega(T,V,\mu)=U-TS-\sum_i \mu_i N_i$ and
$$\D \Omega=- S \D T - P \D V - \sum_i N_i \D \mu_i \, .$$

The entropic series starts from the entropy $S(U,V,N)$ and
$$\D S= \frac{1}{T} \D U + \frac{P}{T} \D V - \sum_i
\frac{\mu_i}{T} \D N_i \, .$$ Therefore, the main set of the {\em
intensive variables} for the entropic series is
$$\frac{1}{T}=\frac{\partial S(U,V,N)}{\partial U}\, , \frac{
P}{T}=\frac{\partial S(U,V,N)}{\partial V}\, ,
-\frac{\mu_i}{T}=\frac{\partial S(U,V,N)}{\partial N_i}\, .$$

The Massieu function, $\Phi(T^{-1}, V, N)=S-T^{-1}U \ (=-F/T)$ and
$$\D \Phi= -U \D\left(\frac{1}{T}\right) + \frac{P}{T} \D V  -\sum_i
\frac{\mu_i}{T} \D N_i \, .$$ The Planck function,
$\Xi(T^{-1},T^{-1}P , N)=S-T^{-1}U-T^{-1}PV \ (=-G/T)$ and
$$\D \Phi= -U \D\left(\frac{1}{T}\right)- V \D\left(\frac{P}{T}\right)-\sum_i
\frac{\mu_i}{T}\D N_i \, .$$

All these functions are used  for the definition of
equilibrium. The main definition for an isolated system follows
the  R. Clausius two main laws formulated in 1865
\cite{Clausius1865}:
\begin{enumerate}
\item{The energy of the Universe is constant.}
\item{The entropy of the Universe tends to a maximum.}
\end{enumerate}
More precisely, the entropy of the isolated system tends to a
maximum under given $U,V$ and values of other conservation
laws. Let the  conservation laws be given:
$$B_j=\sum_i b_{ji}N_j \, ,$$
then the equilibrium is the maximizer of the entropy under given
values of $U$, $V$ and $B_j$:
\begin{equation}\label{eqSmax}
 S(U,V,N) \to \max \mbox{ subject to
given } U, V \mbox{ and } \{B_j\} \, .
\end{equation}
Gibbs \cite{Gibbs1875} paid much attention to the dual formulation
of this condition:
\begin{equation}\label{eqUmin}
U(S,V,N) \to \min \mbox{ subject to given } S, V \mbox{ and }
\{B_j\} \, .
\end{equation}
Other definitions of {\em the same} equilibrium are available
through the free energies and entropies. For the free energies this
is condition of minimum. Analogously to (\ref{eqUmin}) we get
\begin{equation}\label{eqHFGmin}
\begin{split}
&H(S,P,N)  \to \min \mbox{ subject to given } S, P \mbox{ and }
\{B_j\} \, , \\
&F(T,V,N) \to \min \mbox{ subject to given } T, V \mbox{ and }
\{B_j\} \, , \\
&G(T,P,N)  \to \min \mbox{ subject to given } T, P \mbox{ and }
\{B_j\} \, .
\end{split}
\end{equation}

For the free entropies the equilibrium should be the maximizer: in
addition to (\ref{eqSmax})
\begin{equation}\label{eqPhiXimax}
\begin{split}
&\Phi\left(\frac{1}{T}, V, N\right)  \to \max \mbox{ subject to
given } \frac{1}{T}, V \mbox{ and }
\{B_j\} \, , \\
&\Xi\left(\frac{1}{T}, \frac{P}{T}, N\right)  \to \max \mbox{
subject to given } \frac{1}{T}, \frac{P}{T} \mbox{ and } \{B_j\} \,
.
\end{split}
\end{equation}

In {\em extensive thermodynamics}, $V$, $U$,  $S$, and $N$  are
the {\em extensive variables}, that is they are directly
proportional to the system volume if we just join several
copies of the system with the proportional increase of the
volume. Therefore, $U(S,V,N)$ is the homogeneous function of
the first order, and the equation for $\D U$ can be easily
integrated (this is the Euler theorem):

$$U=T S - P V + \sum_i \mu_i N_i \, .$$

In this case,
$$H=U+PV=T S+ \sum_i \mu_i N_i\, ,$$
$$F=U-TS=-PV + \sum_i \mu_i N_i\, ,$$
$$G=H-TS=\sum_i \mu_i N_i\, ,$$
$$\Omega=U-TS-\sum_i \mu_i N_i=-PV\, ,$$

Free energies have a very important physical and technical sense.
They measure the {\em available work} under given conditions.

Free entropies coincide (up to some constant additions) with the
entropies of the minimal isolated system, which includes the system
under consideration. This statement was analyzed in detail in
\cite{G1} but is still not very well known. Therefore, let us prove
it.

Physically, when we consider a system under isothermal condition,
this means that the system is in contact with a large thermal bath.
The state of the thermal bath is characterized by two variables,
$U_{\mathrm{T}}$ and $V_{\mathrm{T}}$. The entropy of a thermal bath
is $S_{\mathrm{T}}(U_{\mathrm{T}},V_{\mathrm{T}})$. The total
entropy of the isolated system ``a system + the thermal bath" is
$$\widetilde{S}(U,V,N,U_{\mathrm{T}},V_{\mathrm{T}})=S(U,V,N)+S_{\mathrm{T}}(U_{\mathrm{T}},V_{\mathrm{T}})\, .$$

The equilibrium is the maximizer of the total entropy
$\widetilde{S}$ for given total energy
$\widetilde{U}=U+U_{\mathrm{T}}$, values of volumes $V$,
$V_{\mathrm{T}}$ and linear conservation laws $\{B_j\}$. In
particular,
\begin{equation}
\begin{split}
&\frac{\partial
[S(U,V,N)+S_{\mathrm{T}}(\widetilde{U}-U,V_{\mathrm{T}})]}{\partial
U}=0\, , \\
&\frac{\partial S(U,V,N)}{\partial U}=\left.\frac{\partial
S_{\mathrm{T}}(U_{\mathrm{T}},V_{\mathrm{T}})}{\partial
U}\right|_{U_{\mathrm{T}}=\widetilde{U}-U}\, .
\end{split}
\end{equation}
This means that the temperatures of the thermal bath and the system
are equal, $T=T_{\mathrm{T}}$.

Let us take the conditional maximum function (under conditions
$U+U_{\mathrm{T}}=\widetilde{U}$, $T=T_{\mathrm{T}}$:
\begin{equation}
\begin{split}
S_{\Phi}(\widetilde{U},V,N,V_{\mathrm{T}})&=S(U,V,N)+S_{\mathrm{T}}(\widetilde{U}-U,V_{\mathrm{T}})
\\
&=S(U,V,N)+V_{\mathrm{T}}S_{\mathrm{T}}\left(\frac{\widetilde{U}-U}{V_{\mathrm{T}}},1\right)
\\
&=S(U,V,N)+V_{\mathrm{T}}S_{\mathrm{T}}\left(\frac{\widetilde{U}}{V_{\mathrm{T}}},1\right)-\frac{U}{T}\\
&=\Phi+V_{\mathrm{T}}S_{\mathrm{T}}\left(\frac{\widetilde{U}}{V_{\mathrm{T}}},1\right)
+O(V_{\mathrm{T}}^{-1})
\end{split}
\end{equation}
This function differs from the free entropy $\Phi$ by a constant
$S_{\mathrm{T}}(\widetilde{U},V_{\mathrm{T}})$ and an infinitesimal
$O(V_{\mathrm{T}}^{-1})$, which goes to zero when the bath
increases.

Therefore, the free entropy $\Psi$ (the Massieu function) is equal
to the entropy of the minimal isolated system, which includes the
system under consideration and the large thermal bath (up to a large
constant and an infinitesimal additions).

For the Planck function $\Xi$ we have to consider a system under a
constant pressure in the contact with the same large thermal bath.
The only difference is that instead of the internal energy of our
system we have to take the enthalpy. The enthalpy is the energy of
the system plus the device, which keeps the pressure constant
(potential energy of a heavy piston of the given weight). So, the
total energy is the energy of the minimal isolated system
$\widetilde{U}=H+U_{\mathrm{T}}$ and everything else is the same as
for $\Phi$: the free entropy is the entropy of the minimal isolated
system, which includes the system under consideration, under
condition of the partial equilibrium with auxiliary systems and up
to a constant summand.

For perfect systems, by definition,
\begin{equation}\label{perfectDefin}
\begin{split}
&U=\sum_i N_i u_i(T)\, ,\\
&PV=RT\sum_i N_i \, ,
\end{split}
\end{equation}
where $u_i(T)$ is the energy of one mole of $A_i$ at the
temperature $T$.

Under this assumption, the entropy $S$ is defined up to an
arbitrary uniform function of first order $S_0(N)$:
\begin{equation}\label{perfectEntropy}
S=RS_0(N)+\sum_i N_i\left[-R\ln c_i +
\int_{T_0}^T \frac{1}{\tau}\frac{\D u_i(\tau)}{\D \tau} \, \D \tau \right] \, .
\end{equation}

If we assume that $S_0(N)$ is a linear function,
$$S_0(N)=\sum_i \delta_i N_i$$
then
\begin{equation}\label{perfectEntropy1}
S=\sum_i N_i\left[-R(\ln c_i-\delta_i) +
\int_{T_0}^T \frac{1}{\tau}\frac{\D u_i(\tau)}{\D \tau} \, \D \tau \right] \, ,
\end{equation}
where $c_i=N_i/V$ is the concentration, $T_0>0$ is a reference
temperature (we assume that on the interval $[T_0,T]$ the
system is perfect (\ref{perfectDefin})).

It is necessary to stress that linearity of $S_0(N)$ does not
follow from the assumption (\ref{perfectDefin}) and is an
additional hypothesis. For a general perfect system
(\ref{perfectDefin}) we can state that $S_0(N)$ is a uniform
function of the first order only.

Formulas (\ref{perfectDefin}), (\ref{perfectEntropy1}) allow us
to express the free energy in the proper variables, $F(T,V,N)$:
\begin{equation}\label{FEperfect}
\begin{split}
F(T,V,N)&=U-TS \\
&=\sum_i N_i u_i(T)+ RT\sum_i N_i \left[\ln
\left(\frac{N_i}{V}\right) - \delta_i -\int_{T_0}^T
\frac{1}{R\tau}\frac{\D u_i(\tau)}{\D \tau} \, \D \tau\right] \, .
\end{split}
\end{equation}

From this formula for $F(T,V,N)$, all other thermodynamic
functions can be expressed locally:
$$S=\frac{\partial F}{\partial T}\, , \; U=F-T \frac{\partial F}{\partial
T}\,, \; P=-\frac{\partial F}{\partial V} \,, \; \mu_i=\frac{\partial F}{\partial N_i}\,, \ldots$$

Generalizations of the free energy for non-perfect systems are often
produced by transformations of (\ref{FEperfect}). The first
generalization describes a system of small admixtures to a general
system. Let the ``background" system have the extensive state
variables $M$. Interaction of small admixtures is negligible and the
formula $PV=RT\sum N_i$ describes the osmotic pressure of the
admixtures.. Then we can write, analogously to (\ref{FEperfect}):
\begin{equation}\label{FEAlmostPerfect}
\begin{split}
F(T,V,N,M)=&F_0(T,V,M)+ \sum_i N_i u_i\left(T,\frac{M}{V}\right)
\\&+ RT\sum_i N_i \left[\ln \left(\frac{N_i}{V}\right) -
\delta_i\left(\frac{M}{V}\right) -\int_{T_0}^T
\frac{1}{R\tau}\frac{\D u_i\left(\tau,\frac{M}{V}\right)}{\D \tau}
\, \D \tau\right] \, .
\end{split}
\end{equation}
Here, the energies $u_i$ and parameters $\delta_i$ are functions of
densities $M/V$. For each given value of these densities, this
formula coincides with (\ref{FEperfect}).

The first model of non-perfect gases is the van der Waals gas. To
write the free energy for this type of gas (or gas of admixtures) we
have to take into account two effects: the excluded volume per mole
of particles $A_i$, $v_i$, and the energy of attraction for
particles $A_i$, $A_j$ with the energy density $$\epsilon_{ij}= -
a_{ij}c_i c_j\, $$ (minus because this is attraction).

For the free energy these effects give:
\begin{equation}\label{FEvdW}
\begin{split}
F(T,V,N)=&\sum_i N_i u_i(T)- V
\sum_{ij}a_{ij}\frac{N_i}{V}\frac{N_j}{V} \\ &+RT\sum_i N_i
\left[\ln \left(\frac{N_i}{V-\sum_i v_i N_i}\right) - \delta_i
-\int_{T_0}^T \frac{1}{R\tau}\frac{\D u_i(\tau)}{\D \tau} \, \D
\tau\right] \, .
\end{split}
\end{equation}

For adsorbed particles on a surface, the model of an {\em ideal
adsorbed layer} implies a lattice of places, and a multicomponent
gas with components $A_0=Z, A_1,\ldots ,A_n$ where $Z$ ia a free
place and $A_i$ are adsorbed particles on their places (each
adsorbed particle occupies a place). There is a conservation law:
$$\sum_{i=0}^n c_i=\theta=\mathrm{const}\, ,$$
therefore, $c_0=\theta-\sum_{i=1}^n c_i$ and the free energy has the
form of the energy of the Fermi-gas:

\begin{equation}\label{FEperfectFermi}
\begin{split}
F(T,\sigma,N)=&F_0(T)\sigma+\sum_{i=1}^n N_i u_i(T)\\&+
RT\sum_{i=1}^n N_i \left[\ln \left(\frac{N_i}{\sigma}\right) -
\delta_i -\int_{T_0}^T \frac{1}{R\tau}\frac{\D u_i(\tau)}{\D \tau}
\, \D \tau\right]\\&+\left(\sigma \theta - \sum_{i=1}^n
N_i\right)\left(\ln \left( \theta - \sum_{i=1}^n
\frac{N_i}{\sigma}\right) - \delta_0\right) \, ,
\end{split}
\end{equation}
where  $\sigma$ is the surface area, $F_0(T)\sigma$ is the free
energy of empty surface.

For the systems, distributed in space, the density of free energy
may be expressed through the concentrations, for example, for the
perfect system (\ref{FEperfect})

\begin{equation}\label{FEperfectDens}
\begin{split}
f(T,c)&=\frac{F(T,V,cV)}{V} \\
&=\sum_i c_i u_i(T)+ RT\sum_i c_i \left[\ln c_i
 - \delta_i -\int_{T_0}^T
\frac{1}{R\tau}\frac{\D u_i(\tau)}{\D \tau} \, \D \tau\right] \, .
\end{split}
\end{equation}
Therefore, for non-uniform system
\begin{equation}\label{integrFE}
 F=\int_V f(T(x),c(x)) \, \D
x \, . \end{equation}

This formula is applicable if the space gradients are not too
sharp. If $f(T,c)$ is a convex function of $c$ then the
minimizer of $F$ (\ref{integrFE}) under given $T$, $V$ and
$N=\int_V c(x \, \D x$ (i.e. the equilibrium) is a uniform
distribution and we should not expect spontaneous appearance of
singularities on the way to the equilibrium. If $f(T,c)$ is not
a convex function of $c$ then the minimizer of $F$ may be
nonuniform, non-smooth and non-unique: phase transitions are
possible. In that case, the simple integral of the density $f$
should be regularized by addition terms. Now, the standard
approach gives the Ginzburg--Landau free energy:
\begin{equation}\label{GLFE}
\begin{split}
&F=\int_V \psi(T(x),c(x),\nabla c(x)) \, \D x \, , \\
&\psi(T(x),c(x),\nabla c(x))=f(T(x),c(x))+\frac{1}{2}\sum_i \lambda_i (\nabla c_i)^2
\end{split}
\end{equation}
More general dependencies of $\nabla c$ are also under consideration \cite{Gurtin1996}.
The chemical potentials $\mu_i$ for the free energy (\ref{GLFE}) are defined as variational derivatives of
this functional,
\begin{equation}\label{ChemPotGinzLand}
\mu_i=\frac{\delta F}{\delta c_i}=\frac{\partial f(T,c)}{\partial c_i}- \lambda_i  \Delta c_i \, .
\end{equation}

Special analysis of the most general form of diffusion equations which provide the proper decrease  of
the Ginzburg--Landau free energy (\ref{GLFE}) was provided by Gurtin \cite{Gurtin1996}. He
introduced general nonlinear mobility matrices.

The kinetic laws should satisfy the thermodynamic restrictions. That is, the dissipation should be positive.
There are two physical forms of this law: (i) the available work should decrease and (ii) the entropy
of the minimal isolated system, which includes the system under consideration, should increase. These two
equivalent formulations correspond to two series of thermodynamic potentials: energetic or entropic series.
There are several approaches to a general formalisms, which pretend to describe {\em all systems} that satisfy these monotonicity
conditions \cite{GENERIC,GENERIC1}. Such approaches form the special discipline, nonequilibrium
thermodynamics \cite{De Groot1962,Gyarmati1970,Lebon2008} or
beyond equilibrium thermodynamics \cite{Ottinger2005}.

Our goal is different: we construct a method for assembling of a complex transport process from
a mechanism combined by simple elementary processes. This approach for physics should satisfy
the basic thermodynamic requirements.

\subsubsection{Markovian Microkinetics and Generalized Mass Action
Law}

To satisfy the thermodynamic restrictions the kinetic law of the elementary
reactions should have a special form, and the reaction
rate constants for different elementary reactions should be harmonized.

Let us start from a reaction mechanism given by the
stoichiometric equations (\ref{StoiEq}):
\begin{equation}\label{StoiEq2}
\sum_i \alpha_{ri} A_i \to \sum_i \beta_{ri} A_i\, ,
\end{equation}
where $r$ is a reaction number,  $\alpha_{ri}$ and $\beta_{ri}$
are nonnegative numbers, the stoichiometric coefficients.

In this Subsection, we return to some ideas of Michaelis and Menten \cite{MichMen1913} and to the Stueckelberg analysis of
the Boltzmann equation \cite{Stueckelberg1952} and represent the general
kinetic law for elementary processes. Detailed analysis was provided recently in \cite{GorbanShahzad2010}.

This law could be proved under the following assumptions:
\begin{enumerate}
\item{The elementary processes go through intermediate
    states (complexes or compounds)
    \ref{stoichiometricequationcompaund}
    (Fig.~\ref{n-tail}):
\begin{equation}\label{stoichcomp2}
    \sum_i\alpha_{r i}A_i \rightleftharpoons B_{r}^-
    \to B_{r}^+ \rightleftharpoons \sum_i \beta_{r i}
    A_i \, ,
\end{equation}
    where $\rho$ is the elementary process number;}
\item{The amount of each compound $B_{\rho}$ is small
    enough to apply the perfect free energy formula
    (\ref{FEAlmostPerfect})  for them;}
\item{The equilibrium between each compound and the
    corresponding linear combinations of reagents is fast
    enough to apply the quasiequilibrium approximation
    \cite{GorbanShahzad2010};}
\item{The transitions between compounds could be described by
a continuous time Markov chain (the Master equation or the monomolecular kinetics).}
\end{enumerate}

The first three items of these assumptions correspond exactly
to the celebrated Michaelis--Menten work \cite{MichMen1913}.
Later, Briggs and Haldane \cite{BriggsHald1925} abandoned the
assumption of fast equilibria and produced the so-called
Michaelis and Menten kinetics approximation. Original approach
of Michaelis and Menten (fast equilibria with intermediate
compounds + small amounts of compounds) was discovered again by
Stueckelberg \cite{Stueckelberg1952} almost 30 years later. It
was applied not to the kinetics of  catalytic reactions but to
the collision in the gas kinetics, as an alternative background
of the Boltzmann equation. Gorban and Shahzad
\cite{GorbanShahzad2010} provided the detailed analysis of this
approach to the general kinetic equation.

Let the concentration of the intermediate compound $B_{\rho}$
(\ref{stoichiometricequationcompaund}) be $\varsigma_{\rho}$.
The free energy (\ref{FEAlmostPerfect}) for the small admixture
of compounds $B_{\rho}$ to the components $A_i$  may be
represented in the form
\begin{equation}\label{FreeEnPerf2}
F=Vf(c,T)+VRT \sum_{{\rho}=1}^q \varsigma_{\rho} \left(\ln
\left(\frac{\varsigma_{\rho}}{\varsigma_{\rho}^*(c,T)}\right)-1\right) \, ,
\end{equation}
where $\varsigma_{\rho}$ is  the {\em standard equilibrium}
concentrations of $B_{\rho}$.

We assume that the standard equilibrium concentrations
$\varsigma_{\rho}^*(c,T)$ as well as the current concentrations
$\varsigma_{\rho}$ are much smaller than the concentrations of
$A_i$.

To formulate the results of
Michaelis--Menten--Stueckelberg--Gorban--\-Sha\-hzad ({\em
MMSGS kinetics}) we have to introduce the basic notions in more
detail. First of all, some of the formal linear combinations
$$(\alpha_r,A) =  \sum_i\alpha_{r i}A_i \, , \; (\beta_r,A)=
\sum_i \beta_{r i} A_i \, $$ may coincide. The same combination
may be, simultaneously, the input combination of several
reactions and the output combination of several other
reactions.

We assume that a fast intermediate compound $B_{\cdots}$
corresponds not to a reaction but to a formal complex of the
form $(\alpha_r,A)$ or $(\beta_r,A)$ and this compound is the
same for all reactions which include this complex.

Let us call the formal linear combinations of the form
$(\alpha_r,A)$ or $(\beta_r,A)$ the complexes and enumerate
them: $\Theta_1$, $\Theta_2$, ... , $\Theta_q$. For each
complex $\Theta_j$, the corresponding vector of coefficients
($\alpha_r$ or $\beta_r$) is $y_j$: $\Theta_j=(y_j,A)$.

The reaction mechanism (\ref{StoiEq2}) may be represented in
the form $\Theta_j \to \Theta_s$ for some pairs $(j,s)$.

The additional component, the fast compound $B_j$, corresponds
to each complex $\Theta_j$ and the reaction mechanism $\Theta_j
\to \Theta_s$ (for some pairs $(j,s)$) (\ref{StoiEq2}) is
extended to
\begin{equation}
\Theta_j \rightleftharpoons B_j \to B_s \rightleftharpoons \Theta_s \mbox{ for some pairs }(j,s)\, .
\end{equation}
The fast equilibrium $\Theta_j \rightleftharpoons B_j$ gives
\begin{equation}\label{equilibrationEqMUMU}
\vartheta_j=\sum_i y_{ji} \frac{\mu_i(c,T)}{RT}\, ,
\end{equation}
or
\begin{equation}\label{equilibrationEq}
\varsigma_j=\varsigma^*_j(c,T)\exp\left(\frac{\sum_i y_{ji} \mu_i(c,T)}{RT}\right) \, ,
\end{equation}
where $$\mu_i=\frac{\partial f(c,T)}{\partial c_i}$$ is the
chemical potential of $A_i$ and
$$\vartheta_j=\ln\left(\frac{\varsigma_j}{\varsigma^*_j}\right)$$
($RT\vartheta_j=\frac{1}{V}\frac{\partial F}{\partial
\varsigma_j}$ is the chemical potential of $B_j$).

For the systems with fixed volume, the  stoichiometric
conservation laws of the monomolecular system of reaction
between compounds are sums of the concentrations of $B_j$ which
belong to various connected components of the reaction graph.
Under the hypothesis of weak reversibility there is no other
conservation law.

Let the graph of reactions $B_j \to B_l$ have $d$ connected
components $C_s$ and let $V_s$ be the set of indexes of those
$B_j$ which belong to $C_s$: $B_j \in C_s$ if and only if $j
\in V_s$. For each $C_s$ there exists a stoichiometric
conservation law
\begin{equation}\label{LinearConservation}
\beta_s=\sum_{j\in V_s}\varsigma_j \, .
\end{equation}

For any set of positive values of $\beta_s$ ($s=1, \ldots , d$)
and given $c,T$ there exists a unique conditional maximizer
$\varsigma^{\mathrm{eq}}_j$ of the free energy
(\ref{FreeEnPerf2}): for the compound $B_j$ from the $s$th
connected component ($j \in V_s$) this equilibrium
concentration is
\begin{equation}\label{compoundEq}
\varsigma^{\mathrm{eq}}_j=\beta_s\frac{\varsigma_j^*(c,T)}{\sum_{l\in V_s
}\varsigma_j^*(c,T)}
\end{equation}

Inversely, positive values of concentrations $\varsigma_j$ are
the equilibrium concentrations (\ref{compoundEq}) for some
values of $\beta_s$ if and only if for any $s=1, \ldots , d$
\begin{equation}\label{equilibriumcompoundTheta}
\vartheta_j=\vartheta_l \;\;{\mathrm{if}} \;\; j,l\in V_s \,
\end{equation}
($\vartheta_j=\ln(\varsigma_j/\varsigma_j^*)$). This system of
equations together with the equilibrium conditions
(\ref{equilibrationEq}) constitute the equilibrium of the
systems. All the equilibria form a linear subspace in the space
with coordinates $\mu_i/RT$ ($i=1, \ldots, n$) and
$\vartheta_j$ ($j=1, \ldots , q$).

Our expression for the free energy does not assume anything
special about the free energy of the mixture of $A_i$. For the
compounds $B_i$, we assume that this is a very small addition
to the mixture of $A_i$, neglect all quadratic terms in
concentrations of $B_i$ and use the free energy of the perfect
systems for this small admixture.

The Master Equation for the concentration of $B_j$ gives:
\begin{equation}\label{MasterEqcomp}
\frac{\D \varsigma_j}{\D t}=\sum_{l, \, l\neq j} \left(\kappa_{jl}
\varsigma_l-\kappa_{lj}\varsigma_j\right)\, .
\end{equation}

This kinetic equation should respect thermodynamics. For the
Master equation this means the equilibrium condition

\begin{equation}\label{MarkovEqEquilibriumCond}
\sum_{l, \, l\neq j} \kappa_{jl} \varsigma_l^*=\sum_{l, \, l\neq j}
\kappa_{lj}\varsigma_j^* \, .
\end{equation}

Under this condition, the  Master Equation (\ref{MasterEqcomp})
has the equivalent form:
\begin{equation}
\frac{\D \varsigma_j}{\D t}=\sum_{l, \, l\neq j} \kappa_{jl}
\varsigma_l^*\left(\frac{\varsigma_l}{\varsigma_l^*}-\frac{\varsigma_j}{\varsigma_j^*}\right)\,
.
\end{equation}
In this form, it is obvious that $\varsigma_j^*$ is equilibrium
for the kinetic equation.

All these expressions for concentrations of compounds and the
Markov reaction rates result in the following kinetic law:
\begin{enumerate}
\item{The reaction rate of the reaction $\Theta_j \to
    \Theta_s$ is
\begin{equation}\label{MMSGSkinlaw}
w_{sj}=\phi_{sj} \exp\left(\frac{\sum_i y_{ji} \mu_i(c,T)}{RT}\right) \, ,
\end{equation}
where the quantity $\phi_{sj}\geq 0 $ is a {\em kinetic
factor}, in the Markov model it corresponds to $\kappa_{sj}
\varsigma_j^*$;}
\item{The kinetic factors $\phi_{sj}$ satisfy the complex
    balance condition:
\begin{equation}\label{MMSGScompbal}
\sum_s \phi_{js} =  \sum_s \phi_{sj} \mbox{ for all } j\, ,
\end{equation}
    in the Markov model this identity corresponds to the equilibrium condition
    (\ref{MarkovEqEquilibriumCond}).}
\end{enumerate}

This is the macroscopic MMSGS kinetics.

For this kinetics, the free energy decreases in time. To
demonstrate this dissipation inequality, let us formulate the
macroscopic MMSGS kinetic equations in the original notations
for the reaction mechanism (\ref{StoiEq2}). Formula (\ref{MMSGSkinlaw}) for the reaction rate gives
\begin{equation}\label{MMSGSkinlawALPHA}
w_r=\phi_r \exp\left(\frac{\sum_i \alpha_{ri} \mu_i(c,T)}{RT}\right) \, ,
\end{equation}

The kinetic equation under the isochoric conditions (the
constant volume) are
\begin{equation}\label{KInUr2}
\frac{\D c}{\D t}=\sum_r \gamma_r w_r \, ,
\end{equation}
where the stoichiometric vector $\gamma_r=\beta_r-\alpha_r$.
According to this equation, the dissipation rate is
\begin{equation}\label{KInUr2dissipation}
\frac{\D F}{\D t}=V\sum_r (\mu, \gamma_r) w_r=
V \sum_r \phi_r (\mu, (\beta_r-\alpha_r))  \exp\left(\frac{(\alpha_r, \mu)}{RT}\right)  \, ,
\end{equation}
where $\mu_i=\partial F /\partial N_i$ is the chemical
potential.

Let us introduce an auxiliary function like we did it for MAL
(\ref{auxtheta}):

\begin{equation}\label{auxtheta2}
\theta(\lambda)=\sum_r \phi_r \exp\left[\frac{\lambda (\alpha_r, \mu)
+(1-\lambda)(\beta_r, \mu)}{RT}\right]\, .
\end{equation}

This function is convenient because

\begin{equation}\label{auxtheta2deriv}
\frac{\D \theta(\lambda)}{\D \lambda}=-\sum_r \phi_r (\mu, (\beta_r-\alpha_r))  \exp\left[\frac{\lambda (\alpha_r, \mu)
+(1-\lambda)(\beta_r, \mu)}{RT}\right]\, .
\end{equation}
Therefore, for the dissipation rate we get
\begin{equation}\label{KInUr2dissipationTheta}
\frac{\D F}{\D t}=-V \theta'(1)
\end{equation}

The function $\theta(\lambda)$ is a sum of exponents. It is
convex ($\theta''(\lambda)\geq 0$). Therefore, if
$\theta(0)=\theta(1)$ then $\theta'(1) \geq 0$. This means that the identity
\begin{equation}\label{ComBalIdent}
\theta(0)\equiv \theta(1)
\end{equation}

is a sufficient condition for the dissipation inequality
$$\frac{\D F}{\D t}\leq 0\, .$$

Some of vectors $\alpha_r$, $\beta_r$ may coincide. Let there be $q$ different vectors among them.
We denote them $y_1, \ldots, y_q$. For
each $y_i$ we define $R_i^+=\{r\, | \, \alpha_r=y_i\}$,
$R_i^-=\{r\, | \, \beta_r=y_i\}$. The sufficient condition for the identity (\ref{ComBalIdent})
is
\begin{equation}\label{CompBalIdent}
\sum_{r\in R_i^+} \phi_r=\sum_{r\in R_i^-} \phi_r \mbox{ for all }  i \, .
\end{equation}
This condition is also necessary if we can vary $\phi_r$ and
$\mu_i$ independently and the Jacobian $|\partial \mu_i /
\partial c_j|$ has the full rank. This condition is the {\em
complex balance} condition.

Of course, the important particular case of the complex balance conditions is the detailed balance condition:
\begin{equation}\label{DetBalIdent}
\phi_r^+=\phi_r^- \, ,
\end{equation}
where $\phi_r^+$, $\phi_r^-$ are the kinetic factors of the
mutually reverse reactions: $\phi_r^+$ for $(\alpha_r,A)\to
(\beta_r,A)$ and $\phi_r^-$ for $(\beta_r,A) \to (\alpha_r,A)$.

\subsection{From Cell-Jump Models to Continuous Diffusion
Equations \\ for Generalized Mass Action Law
\label{Sec:ContLimNonPerfect}}

Let us start again from the stoichiometric mechanism of the diffusion:
\begin{equation}\label{elementaryActDiff5}
\sum_{i}\alpha_{ri}^{\mathrm{I}}A_{i}^{\mathrm{I}}+
\sum_{i}\alpha_{ri}^{\mathrm{II}}A_{i}^{\mathrm{II}}\rightarrow
\sum_{i}\beta_{ri}^{\mathrm{I}}A_{i}^{\mathrm{I}}+\sum_{i}\beta_{ri}^{\mathrm{II}}A_{i}^{\mathrm{II}}\,
.
\end{equation}
All the elementary processes have different numbers $r$ and the
space-inverted and reverse processes are represented
separately. Let us consider all pairs of vectors,
$(\alpha_r^{\mathrm{I}},\alpha_r^{\mathrm{II}})$  and
$(\beta_r^{\mathrm{I}},\beta_r^{\mathrm{II}})$ and numerate all
the different pairs: $y_1,y_2,\ldots$, where
$y_q=(y_q^{\mathrm{I}},y_q^{\mathrm{II}})$. For each $y_q$
there are two sets of reactions, $R_q^+$, $R_q^-$:
$$R_q^+=\{r\, | \, (\alpha_r^{\mathrm{I}},\alpha_r^{\mathrm{II}})=y_q\}\, , \;
R_q^-=\{r\, | \, (\beta_r^{\mathrm{I}},\beta_r^{\mathrm{II}})=y_q\}
\, .$$

The reaction rate for the elementary process (\ref{elementaryActDiff5}) is (\ref{MMSGSkinlawALPHA}):
\begin{equation}\label{MMSGSkinlawALPHAIII}
w_r=\phi_r \exp\left(\frac{ (\alpha_r^{\mathrm{I}}, \mu(c^{\mathrm{I}},T))+(\alpha_r^{\mathrm{II}}, \mu(c^{\mathrm{II}},T))}{RT}\right) \, ,
\end{equation}
The kinetic factors are functions of $c^{\mathrm{I}}$,
$c^{\mathrm{II}}$ and $T$. They should satisfy the identity of
complex balance (\ref{CompBalIdent}).

The cell--jump model gives us
\begin{equation}\label{cellJump3}
\begin{split}
J&=-\sum_r \gamma_r w_r(c(x),c(x+l)) \\
&=\sum_q \left(\sum_{r\in R_q^+} \alpha_r^{\mathrm{I}} w_r(c(x),c(x+l)) -
\sum_{r\in R_q^-} \beta_r^{\mathrm{I}} w_r(c(x),c(x+l))\right) \\
&=\sum_q y_q^{\mathrm{I}} \left(\sum_{r\in R_q^+} w_r(c(x),c(x+l)) -
\sum_{r\in R_q^-} w_r(c(x),c(x+l))\right)
\end{split}
\end{equation}

The zero order term in $l$ is
$$\sum_q y_q^{\mathrm{I}} \exp\left(\frac{(y_q^{\mathrm{I}}+y_q^{\mathrm{II}},\mu)}{RT} \right) \left(\sum_{r\in R_q^+} \phi_r   -
\sum_{r\in R_q^-}\phi_r \right)=0\, .$$
It vanishes because of the complex balance condition.

The first order term gives
\begin{equation}\label{fluxCBGEN}
\begin{split}
J=l \sum_q y_q^{\mathrm{I}} \exp\left(\frac{(y_q^{\mathrm{I}}+y_q^{\mathrm{II}},\mu)}{RT} \right)
&\left(\sum_{r\in R_q^+} \phi_r \left(\alpha^{\mathrm{II}}_{r}, \nabla \left(\frac{\mu}{RT}\right)\right)\right. \\& - \left.
\sum_{r\in R_q^-}\phi_r \sum_j \left(\alpha^{\mathrm{II}}_{r}, \nabla \left(\frac{\mu}{RT}\right)\right)\right)\, .
\end{split}
\end{equation}

Each term in this sum consists of the positive scalar
pre-factor
$$l\exp\left(\frac{(y_q^{\mathrm{I}}+y_q^{\mathrm{II}},\mu)}{RT}
\right)\, $$ and the matrix
$$y_{qi}^{\mathrm{I}} \left(
\sum_{r\in R_q^+} \phi_r \alpha^{\mathrm{II}}_{rj} -
\sum_{r\in R_q^-}\phi_r \alpha^{\mathrm{II}}_{rj}
\right)\, $$ multiplied by
$$\nabla\left(\frac{\mu_j}{RT}\right)\, .$$

This structure of the formula for the flux is very similar to
(\ref{ComplBalFluxMAL}). Let us follow the same logic as in
Subsection~\ref{Sec:CompBalMALDif} to find the more convenient
form of the expression for the flux (\ref{fluxCBGEN}). First of
all, let us group all terms with the same pre-factor.

For the set of all vectors $y_q^{\mathrm{I}}+y_q^{\mathrm{II}}$
we use notation $Z$. For each $z \in Z$
$$Y_z=\{y_q \ | \ y_q^{\mathrm{I}}+y_q^{\mathrm{II}} = z \} \, .$$
Analogously to (\ref{firstOrdFluxComBal}) we get

\begin{equation}\label{FluxComBalGenZ}
\begin{split}
&J=l \sum_{z \in Z} l\exp\left(\frac{(y_q^{\mathrm{I}}+y_q^{\mathrm{II}},\mu)}{RT}
\right) J^z\, , \\
&J^z= \sum_{y_q\in Y_z} y_q^{\mathrm{I}}
\sum_j\left(
\sum_{r\in R_q^+} \phi_r \alpha^{\mathrm{II}}_{rj} -
\sum_{r\in R_q^-}\phi_r \alpha^{\mathrm{II}}_{rj}
\right)
\nabla\left(\frac{\mu_j}{RT}\right) \, .
\end{split}
\end{equation}

Let us analyze $J^z$ for given $z$.

First of all,
$$\sum_{y_q\in Y_z}
\sum_j \left( \sum_{r\in R_q^+} \phi_r
\alpha^{\mathrm{II}}_{rj} - \sum_{r\in R_q^-}\phi_r
\alpha^{\mathrm{II}}_{rj} \right)
\nabla\left(\frac{\mu_j}{RT}\right)=0 \, $$ because each
$\phi_r$ from this sum enters it twice, with the opposite
signs, one time as an element of the sum over $ R_q^+$ for some
$y_q \in Y_z$ (with $+$) and the second time (with $-$) as an
element of  the sum over $R_s^-$ for another $y_s\in Y_z$ with
the same sum
$$y_s^{\mathrm{I}}+y_s^{\mathrm{II}}=y_q^{\mathrm{I}}+y_q^{\mathrm{II}}=z\, .$$
(Let us recall that
$$\alpha_s^{\mathrm{I}}+\alpha_s^{\mathrm{II}}=\beta_s^{\mathrm{I}}+\beta_s^{\mathrm{II}}$$
for all $s$ and, hence, if $r\in R_q^+$ and
$\alpha_r^{\mathrm{I}}+\alpha_r^{\mathrm{II}}=z$ then
$\beta_s^{\mathrm{I}}+\beta_s^{\mathrm{II}}=z$ and
$(\beta_s^{\mathrm{I}},\beta_s^{\mathrm{II}})\in Y_z$.)

Let us mention that $y_q^{\mathrm{I}}=z-y_q^{\mathrm{II}}$.
Therefore,
\begin{equation}\label{JzGenMAL}
J^z=- \sum_{y_q\in Y_z} y_q^{\mathrm{II}} \sum_j\left( \sum_{r\in
R_q^+} \phi_r \alpha^{\mathrm{II}}_{rj} - \sum_{r\in
R_q^-}\phi_r \alpha^{\mathrm{II}}_{rj} \right)
\nabla\left(\frac{\mu_j}{RT}\right) \, .
\end{equation}
In this formula, all the stoichiometric vectors are for the
same cell, for the second one.

Let us demonstrate similarity of the expression
(\ref{JzGenMAL}) to some formulas from the theory of Markov
chains. Let us follow Subsection~\ref{Sec:CompBalMALDif} and
numerate the states of the auxiliary Markov chain by vectors
$y_q\in Y_z $. Elementary processes correspond to transitions
between states. Each nonzero kinetic factor $\phi_r$
corresponds to two vectors $y_q,y_s\in Y_z$: ${r\in R_q^+}$ and
${r\in R_s^-}$.

Let us substitute the index $r$ by two indexes $q,s$ and use
notation $\phi_{sq}$ for transitions $y_q\to y_s$ (i.e. for the
case ${r\in R_q^+}$ and ${r\in R_s^-}$).

The complex balance condition (\ref{CompBalIdent}) reads:
\begin{equation}\label{MarkovStStvarphi}
\sum_q \phi_{sq}= \sum_q \phi_{qs} \, .
\end{equation}
This is precisely the steady state condition for the Markov
chain.

Inequality (\ref{IneqMark2}) holds for the kinetic factors:
\begin{equation}\label{IneqMark2varphi}
\sum_{sq} \pi_q (\phi_{qs}\pi_s - \phi_{sq}\pi_q)= \sum_{sq} \pi_q
\phi_{qs}(\pi_s-\pi_q) \leq 0 \,
\end{equation}
for any set of numbers $\pi_q$.

For the proof of the dissipation inequality it is convenient to
rewrite $J^z$ in these notations:

\begin{equation}\label{firstOrdFluxGenMAL4}
J^z= \sum_{y_q \in Y_z}
y_q^{\mathrm{II}} \sum_j \sum_s (\phi_{qs} {y_{sj}^{\mathrm{II}}}-
\phi_{sq} {y_{qj}^{\mathrm{II}}} ) \nabla\left(\frac{\mu_j}{RT}\right) \, .
\end{equation}

Free energy in a domain $V$ is
$$\mathbf{F}=\int_V f(c,T) \, \D x \, .$$

Let us estimate $\dot{\mathbf{F}}$ in a bounded domain $V$ with
smooth boundary and without fluxes through boundary for
isothermal conditions.

\begin{equation}\label{DissipationDIFCBGenMAL}
\begin{split}
\frac{\D \mathbf{F}}{\D t}&= RT \int_V
 \sum_j \left(\nabla_x \left(\frac{\mu_j}{RT}\right),
  J_j \right)  \, \D x \\
&=l \sum_{z \in Z} \int_V   \exp\left(\frac{(y_q^{\mathrm{I}}+y_q^{\mathrm{II}},\mu)}{RT}
\right)
 \sum_j \left(\nabla_x \left(\frac{\mu_j}{RT}\right),
  J_{j}^z \right)  \, \D x \, .
\end{split}
\end{equation}

Due to the representation of $J^z$ (\ref{firstOrdFluxGenMAL4}),
\begin{equation}
\sum_j \left(\nabla_x \left(\frac{\mu_j}{RT}\right),
  J_{j}^z \right) =
\sum_{q \, s} (\pi_q, (\phi_{qs} \pi_s- \phi_{sq} \pi_q )) \leq 0 \, ,
\end{equation}
where $\pi_q$ is a space vector:$\pi_q=\sum_i
y_{qi}^{\mathrm{II}} \nabla_x \left(\frac{\mu_j}{RT}\right)$
and for $y_q$, $y_s$
$$y_{q,s}^{\mathrm{I}}+y_{q,s}^{\mathrm{II}}=z\, .$$ This
expression has exactly the form (\ref{IneqMark2varphi}) for
each space coordinate. Finally,
\begin{equation}
\sum_j \left(\nabla_x \left(\frac{\mu_j}{RT}\right),
  J_{j}^z \right) \leq 0
\end{equation}
and it is zero if all $\pi_q=\sum_i y_{qi}^{\mathrm{II}} \nabla
\ln c_i$ coincide. (The reverse statement is correct if the
auxiliary Markov chain with the transition coefficients
$\phi_{qs}$ is ergodic for given $z$.)

Therefore,
$$\mathbf{\dot{F}}\leq 0 \, $$
for the generalized MAL (\ref{MMSGSkinlaw}) with the complex
balance conditions (\ref{MMSGScompbal}).

\subsection{Detailed Balance for the Generalized Mass Action Law \\ and the Dissipation Inequality}

\subsubsection{Isothermal conditions}

Systems with detailed balance form an important subclass of the
generalized MAL systems.

Let us join each elementary process with its reverse process
and represent the mechanism of diffusion by the pairs of
mutually reverse processes:
\begin{equation}\label{elementaryActDiff5DB}
\sum_{i}\alpha_{ri}^{\mathrm{I}}A_{i}^{\mathrm{I}}+
\sum_{i}\alpha_{ri}^{\mathrm{II}}A_{i}^{\mathrm{II}}\rightleftharpoons
\sum_{i}\beta_{ri}^{\mathrm{I}}A_{i}^{\mathrm{I}}+
\sum_{i}\beta_{ri}^{\mathrm{II}}A_{i}^{\mathrm{II}}\, .
\end{equation}
All the quantities for the direct process we mark by the upper
index $^+$ and for the reverse process by the upper index $^-$.
The simple algebraic relations hold (see also
(\ref{pmDBrelations}):
\begin{equation}\label{pmDBrelations2}
\alpha_{ri}^{\mathrm{I,II}\mp}=\beta_{ri}^{\mathrm{I,II}\pm}\,
\mbox{ and }\gamma_{ri}^{\mathrm{I,II}\mp}=
-\gamma_{ri}^{\mathrm{I,II}\pm}\, .
\end{equation}
Therefore,
\begin{equation}
\alpha_{ri}^{\mathrm{I+}}+\alpha_{ri}^{\mathrm{II+}}=\beta_{ri}^{\mathrm{I-}}+\beta_{ri}^{\mathrm{II-}}=
\alpha_{ri}^{\mathrm{I-}}+\alpha_{ri}^{\mathrm{II-}}=\beta_{ri}^{\mathrm{I+}}+\beta_{ri}^{\mathrm{II+}}\,
.
\end{equation}

Due to the principle of detailed balance,
$$\phi_r^+=\phi_r^-=\phi_r \, .$$

The reaction rates are
\begin{equation}
\begin{split}
&w_r^+(c^{\mathrm{I}},c^{\mathrm{II}})=\phi_r
\exp\left(\sum_i{\alpha^{\mathrm{I}}_{ri}} \frac{\mu_i^{\mathrm{I}}}{RT}
+\sum_i{\alpha^{\mathrm{II}}_{ri}}\frac{\mu_i^{\mathrm{II}}}{RT}\right) \, ,\\
&w_r^+(c^{\mathrm{I}},c^{\mathrm{II}})=\phi_r
\exp\left(\sum_i{\beta^{\mathrm{I}}_{ri}} \frac{\mu_i^{\mathrm{I}}}{RT}
+\sum_i{\beta^{\mathrm{II}}_{ri}}\frac{\mu_i^{\mathrm{II}}}{RT}\right) \, ,\\
\end{split}
\end{equation}

The cell--jump model gives us (\ref{cellJump3})
\begin{equation}\label{cellJump4}
\begin{split}
J&=\sum_r J_r=-\sum_r \gamma_r (w_r^+(c^{\mathrm{I}},c^{\mathrm{II}})-w_r^-(c^{\mathrm{I}},c^{\mathrm{II}})) \\
&=\sum_r \gamma_r  \phi_r \left(
\exp\left(\sum_i{\alpha^{\mathrm{I}}_{ri}} \frac{\mu_i^{\mathrm{I}}}{RT}
+\sum_i{\alpha^{\mathrm{II}}_{ri}}\frac{\mu_i^{\mathrm{II}}}{RT}\right)\right.\\
&\qquad \qquad\left.-\exp\left(\sum_i{\beta^{\mathrm{I}}_{ri}} \frac{\mu_i^{\mathrm{I}}}{RT}
+\sum_i{\beta^{\mathrm{II}}_{ri}}\frac{\mu_i^{\mathrm{II}}}{RT}\right)
\right) \, .
\end{split}
\end{equation}
In the first order approximation, we get analogously to
(\ref{FluxCOntDetBal}):
\begin{equation}\label{FluxCOntDetBal4}
\begin{split}
&J_{ri} =-l\phi_r\exp\left(\sum_j
 (\alpha^{\mathrm{I}}_{rj}+\alpha^{\mathrm{II}}_{rj})\frac{\mu_j^{\mathrm{II}}}{RT}\right)
 \sum_j \gamma_{ri} \gamma_{rj}\nabla \left(\frac{\mu_j^{\mathrm{II}}}{RT}\right) \, , \\
&J_r=- l\phi_r \exp\left(\frac{(\alpha^{\mathrm{I}}_r+\alpha^{\mathrm{II}}_r,\mu)}{RT}\right)
\gamma_r \left(\gamma_r,\nabla\left(\frac{\mu}{RT}\right) \right)\, .
\end{split}
\end{equation}
This expression for $J_r$ has the form of the multicomponent
Teorell formula with the symmetric matrix of coefficients. This symmetry
for {\em nonlinear diffusion} gives us the generalization of the Onsager reciprocal symmetry
\cite{Onsager1,Onsager2}. We represented the nonlinear multicomponent diffusion a sum of
elementary processes. For each elementary process
\begin{equation}\label{SymmetryRelDiffDB}
J_{ri} =-d_r  \sum_j \gamma_{ri} \gamma_{rj}\nabla \left(\frac{\mu_j^{\mathrm{II}}}{RT}\right) \, ,
\end{equation}
where the scalar coefficient
$$d_r=l\phi_r\exp\left(\sum_j
 (\alpha^{\mathrm{I}}_{rj}+\alpha^{\mathrm{II}}_{rj})\frac{\mu_j^{\mathrm{II}}}{RT}\right)>0$$
is, from the thermodynamic point of view, an almost arbitrary
positive quantity (because it includes the kinetic factor $\phi_r$). ``Almost" here means that some conditions
of zero values (and the orders of these zeros) at the boundary when some of $c_i=0$ are
prescribed by the factor
$$\exp\left(\sum_j
 (\alpha^{\mathrm{I}}_{rj}+\alpha^{\mathrm{II}}_{rj})\frac{\mu_j^{\mathrm{II}}}{RT}\right)$$
and the logarithmic singularity of $\mu_i$ when $c_i\to 0$.

The internal symmetry of this formula makes the dissipation
inequality obvious: in a bounded domain $V$ with smooth
boundary and without fluxes through boundary for isothermal
conditions
\begin{equation}\label{DissipationDIFDBGenMAL}
\begin{split}
\frac{\D \mathbf{F}}{\D t}&= RT \int_V
 \sum_j \left(\nabla_x \left(\frac{\mu_j}{RT}\right),
  J_j \right)  \, \D x \\
&=-l \sum_r \phi_r \exp\left(\frac{(\alpha^{\mathrm{I}}_r+\alpha^{\mathrm{II}}_r,\mu)}{RT}\right)
 \left(\gamma_r,\nabla\left(\frac{\mu}{RT}\right) \right)^2 \leq 0 \, .
\end{split}
\end{equation}

\subsubsection{Generalization: Non-isothermal Processes}

Extension of the generalized MAL (\ref{MMSGSkinlaw}) on the
non-isothermal processes is quite simple. Let us follow the
paper \cite{BykGOrYab1982} and include the ``energetic
component" $A_U$ in the list of components. Instead of the
stoichiometric equations (\ref{StoiEq}), (\ref{StoiEq2}) we
get:
\begin{equation}\label{StoiEq3}
\sum_i \alpha_{ri} A_i +\alpha_Q A_U \to \sum_i \beta_{ri} A_i+\beta_Q A_U\, ,
\end{equation}
The corresponding macroscopic extensive variable for $A_U$ is
the internal energy $U$ with the density (``concentration")
$u$. To consider energy as the additional extensive variable,
we should take the main thermodynamic potential for the
isolated system from the entropic series. This is the entropy
$S$:
$$\D S= \frac{1}{T} \D U + \frac{P}{T} \D V - \sum_i
\frac{\mu_i}{T} \D N_i \, .$$

Let us extend the formulas for the generalized kinetics by
additional component and take $-\frac{1}{RT}$ instead of
$\frac{\mu_i}{RT}$. All the formulas including the dissipation
inequalities remain the same.

In isolated (isochoric) systems, $\dot{U}=0$ and
$$\dot{N}_i=\sum_r \gamma_{ri} w_r\, ,$$
where
$$w_r=\phi_r \exp\left[\left(\alpha_r,
\frac{\mu}{RT}\right)-\frac{\alpha_Q}{RT}\right]\, .$$

 For  transport processes, conservation of energy gives the
following relations:
$$\alpha_Q^{\mathrm{I}}+\alpha_Q^{\mathrm{II}}=\beta_Q^{\mathrm{I}}+\beta_Q^{\mathrm{II}}\, .$$

The space gradient of $-\frac{1}{RT}$ enters the multicomponent
Teorell formula as an additional force and the gradients of
$\frac{\mu_i}{RT}$ also enter the formulas for the heat flux.

In particular, the simplest mechanism of transport,
$$\alpha_Q A_U^{\mathrm{I}}\rightleftharpoons \alpha_Q A_U^{\mathrm{II}}\, ,$$
generates the Fourier law:
$$J_Q=- l \alpha_Q \phi \exp\left(-\frac{\alpha_Q}{RT}\right) \nabla \left(-\frac{1}{RT(x)}\right)=-\lambda(T)\nabla T \, .$$

The thermodynamic consideration cannot produce the temperature dependence of
the thermal conductivity $\lambda(T)>0$. From the thermodynamic point of view,
$\phi$ here is an arbitrary positive quantity. The problem of temperature dependence of
$\lambda$ and its relations with other constants like diffusivity is widely discussed from the kinetic point of
view \cite{McLaughlin1964}. For computing thermal conductivity various methods were developed including
direct simulation and the Green--Kubo approach \cite{Hoover1991}. These methods were compared in
\cite{Schellingatal2002}.

Thermodynamics may give the relations between different coefficients. For example,
the principle of detailed balance produces the multicomponent Teorell formula with the
symmetric matrix of coefficients (\ref{FluxCOntDetBal4}). The nonlinear reciprocal relations (\ref{SymmetryRelDiffDB})
could be automatically  extended to the heat flux: just use the heat flux $J_Q$ as additional flux
and $-1/RT$ instead $\mu_i/RT$ for the component $A_U$.

More relations we get for the specific mechanisms of transport. For example, for the activation mechanism of diffusion
\begin{equation}\label{activationMech}
A^{\mathrm{I}}+ \alpha_Q A_U^{\mathrm{I}}\rightleftharpoons A^{\mathrm{II}}+ \alpha_Q A_U^{\mathrm{II}}
\end{equation}
the fluxes are:
\begin{equation}
\begin{split}
&J_A=- l \phi \exp\left(\frac{\mu-\alpha_Q}{RT}\right) \left[\nabla \left(\frac{\mu}{RT}\right)+ \alpha_Q \nabla \left(-\frac{1}{RT}\right)\right]\, ,\\
&J_Q=- l \phi \exp\left(\frac{\mu-\alpha_Q}{RT}\right) \left[\alpha_Q \nabla \left(\frac{\mu}{RT}\right)+ \alpha_Q^2 \nabla \left(-\frac{1}{RT}\right)\right]\, ,
\end{split}
\end{equation}
or in the matrix form
\begin{equation}
\left(\begin{array}{l}
J_A \\
J_Q
\end{array}\right)=- l \phi \exp\left(\frac{\mu-\alpha_Q}{RT}\right)
\left(\begin{array}{cc}
1 & \alpha_Q \\
\alpha_Q  & \alpha_Q^2
\end{array}\right)
\left(\begin{array}{l}
\nabla \left(\frac{\mu}{RT}\right) \\
\nabla\left(-\frac{1}{RT}\right)
\end{array}\right) \, .
\end{equation}

For the mechanism (\ref{activationMech}), the heat flux $J_Q$ is proportional to the diffusion flux $J_A$ with
the coefficient $\alpha_Q$, that is the activation heat. In this activation mechanism, the
activation heat travels with the particle from the cell I to the cell II. We can, for example, assume
different behavior of the activation heat: let $\beta_Q$ distribute symmetrically after the diffusion jump:
\begin{equation}\label{activationMech2}
A^{\mathrm{I}}+ \alpha_Q A_U^{\mathrm{I}}\rightleftharpoons A^{\mathrm{II}}
+ \frac{1}{2}\alpha_Q A_U^{\mathrm{I}}+ \frac{1}{2}\alpha_Q A_U^{\mathrm{II}} \, .
\end{equation}
For this mechanism, $\gamma_U=-\frac{1}{2}\alpha_U$ and
\begin{equation}
\left(\begin{array}{l}
J_A \\
J_Q
\end{array}\right)=- l \phi \exp\left(\frac{\mu-\alpha_Q}{RT}\right)
\left(\begin{array}{cc}
1 & \frac{1}{2}\alpha_Q \\
\frac{1}{2}\alpha_Q  & \frac{1}{4}\alpha_Q^2
\end{array}\right)
\left(\begin{array}{l}
\nabla \left(\frac{\mu}{RT}\right) \\
\nabla\left(-\frac{1}{RT}\right)
\end{array}\right) \, .
\end{equation}

The heat flux $J_Q$ should be supplemented by the heat transport
together with particles, $\sum_i \mu_i J_i$, \cite{LandauLifshitzV6}. The total heat flux is
\begin{equation}\label{totalHeatFlux}
\mathbf{q}=J_U=J_Q+\sum_i \mu_i J_i \, .
\end{equation}
To describe the energy balance properly we have to include the work of various forces.
The proper framework for modeling of the energy transport gives continuum mechanics. In its simplest form,
fluid mechanics, we present these equation in the next Section.

\subsection{Momentum and Center of Mass Conservation}

\subsubsection{Mass Transfer and Heat Transfer}

In this subsection, we briefly discuss coupling of the
diffusion and thermal conductivity with fluid dynamics. The
heat and mass transfer should satisfy the general laws of
mechanics and, in particular, does not violate the Newton laws.
The diffusion and heat transfer equations do not present the
complete theory and should be included into the context of
continuum mechanics.

First of all, let us introduce the {\em mass average velocity}.
Let $m_i$ be the mass of mole (gram-molecule) for the component
$A_i$. For each diffusion flux $J_i$ the associated flux of
mass is  $m_i J_i$. We introduced the fluxes $J_i$ with respect
to a frame, which is connected to our cells. For continuum
motion, this frame should also move and we have to introduce
the velocity of the frame, $\mathbf{w}$. The flux of $A_i$
associated with $\mathbf{w}$ is $c_i \mathbf{w}$. The
corresponding flux of mass is $m_i c_i \mathbf{w}$. The total
flux of $A_i$ caused by diffusion and the frame motion is
$$\check{J}_i=J_i+c_i \mathbf{w}\, .$$
The mass density is
$$\rho={\sum_i m_i c_i}\, ;$$
the momentum density is
$$\sum_i m_i \check{J}_i\, ;$$
the average mass velocity is
$$\mathbf{v}=\frac{\sum_i m_i \check{J}_i}{\sum_i m_i c_i}=\mathbf{w}+\frac{\sum_i m_i {J}_i}{\sum_i m_i c_i} \, .$$

Both the frame velocity $\mathbf{w}$ and the average mass velocity $\mathbf{w}$ are unknown.
They are connected by the simple identity
$$\mathbf{v}=\mathbf{w}+ \frac{\sum_i m_i J_i}{\sum_i m_i c_i}\, ,$$
where the individual diffusion fluxes $J_i$ are given by the mechanism of diffusion.

The standard definition of the diffusion fluxes includes fluxes in the local center of mass frame where the average mass velocity is zero.
Therefore, let us introduce the ``proper fluxes":

$$\mathcal{J}_i=\check{J}_i- \mathbf{v} c_i=\check{J}_i-c_i \frac{\sum_i m_i \check{J}_i}{\sum_i m_i c_i}=J_i-c_i \frac{\sum_i m_i {J}_i}{\sum_i m_i c_i}\, .$$
These fluxes do not depend on the frame velocity. They are not independent and are connected by the momentum conservation: the total momentum is zero,
$$\sum m_i \mathcal{J}_i =0\, .$$

The heat flux in the local center of mass system is
$$\mathcal{J}_U=\left(J_Q -\frac{\sum_i m_i {J}_i}{\sum_i m_i c_i} u\right)+ \sum_i \mu_i \mathcal{J}_i\, ,$$
where $J_Q$ is given by the transport processes mechanism.

For the energy flux, the standard approach \cite{LandauLifshitzV6} gives
$$\mathbf{v}\left(\frac{\rho v^2}{2} + u + P\right) +\mathcal{J}_U \; +\mbox{viscosity terms}\, .$$

The conservation laws give:
\begin{equation}\label{FlDynEq}
\begin{split}
&\partial_t \rho+\mathrm{div}(\rho \mathbf{v})=0\, , \\
&\partial_t (\rho \mathbf{v})+\mathrm{div}(\rho \mathbf{v}\otimes \mathbf{v})+\nabla P = \mathrm{div} \sigma \, , \\
&\partial_t \left(\frac{\rho v^2}{2}+ u \right)+ \mathrm{div}\left[\mathbf{v} \left(\frac{\rho v^2}{2}+ u +P\right)+\mathcal{J}_U\right]
= \sigma : \nabla \mathbf{v} \, , \\
&\partial_t c_i+\mathrm{div}(\mathbf{v} c_i+ \mathcal{J}_i)=0 \, .
\end{split}
\end{equation}
Here, $\sigma$ is the viscous stress tensor and $\sigma :
\nabla \mathbf{v}=\sum_{ij}\sigma_{ij}(\partial v_i/\partial
x_j)$. The pressure $P$ should be defined in accordance with
thermodynamic properties of the mixture, for example,
$$P=-\frac{\partial F}{\partial V} =\sum_i c_i \frac{\partial f(c,T)}{\partial c_i }-f(c,T)\, ,$$
where $f(c,T)$ is the density of the free energy.

The viscous stress tensor should be derived in (\ref{FlDynEq}) from the additional closure assumption.
For the Newtonian liquid,
$$\sigma_{ij}=\mu \left(\frac{\partial v_i}{\partial x_j}+\frac{\partial v_j}{\partial x_i}
\right) +\delta_{ij} \left(\zeta - \frac{2}{3} \mu \right) \mathrm{div} \mathbf{v}\, ,$$
where $\mu$ (here and only here) is the shear viscosity and $\zeta$ is the bulk viscosity.

The individual equations in (\ref{FlDynEq}) are not
independent. For example, the sum of the equations for
conservation of $N_i$ with coefficients $m_i$ gives us the
first equation, the conservation of mass.

The elastic energy and the various viscoelastic terms may be added to this picture. This is necessary
to do and it is in our future plans.

\subsubsection{Mechanisms of Transport and the General Forms of Macroscopic Equation}

We developed a formalism of the mechanism of diffusion and heat
conduction represented by the system of stoichiometric
equations with the simple kinetic law $\exp(\alpha,\mu/RT)$ and
the balance condition (complex balance for the general Markov
microscopic kinetics and detailed balance for systems with
microreversibility). This formalism produces equations which
are particular cases of the general nonequilibrium
thermodynamic equations \cite{De Groot1962,Ottinger2005}. It is
a very simple task to demonstrate that our transport equations
are particular cases of the GENERIC formalism
\cite{GENERIC,Ottinger2005}. Due to this two--generator
formalism, evolution of any smooth function $A$ of the state
variables $x$ is given by
$$\frac{\D A}{\D t}= \{A,E\}+[A,S]\, $$
where $E$ and $S$ are the total energy and entropy, and $\{\cdot ,\cdot \}$ and $[\cdot ,\cdot ]$ are Poisson and
dissipative brackets, respectively.

The formulas for fluxes produced in this Section  have the form
of dissipative brackets:
$$[A,S] =\frac{\delta A}{\delta x} M \frac{\delta S}{\delta x}\, ,$$
where $M$ is a symmetric positively semidefinite operator, ``the friction matrix".

The general form of the ``dissipative brackets with
constraints" in application to multicomponent diffusion was
produced very recently \cite{Ottinger2009}. The flux of the
$i$th component $J_i$ in that formalism was presented by
formula (54) \cite{Ottinger2009}:
$$J_i=-\sum_{j}\Lambda_{ij}^c\left[\nabla\left(\frac{\mu_j}{T}\right)+\Lambda'_j\nabla\left(-\frac{1}{T}\right)\right]\, .$$
Our formulas belong to this type and give particular expressions for coefficients $\Lambda$.

In the paper \cite{Ottinger2009} a precise comparison of this
formula with the classical expressions from \cite{De Groot1962}
was presented and the equivalence of these general forms was
proven. Now, we can just refer to these results.

In addition to the general form, our approach gives the
possibility to build the model from the elementary processes.
This construction also satisfies the ``constrains"
(conservation laws) of diffusion because these conservation
laws are implemented in the algebra of the stoichiometric
coefficients (\ref{DiffCond}).

\section{Conclusion}

Chemical kinetics gave rise to the very seminal approach of the
modeling of processes. This is, the stoichiometric algebra
supplemented by the simple kinetic law. The results of this
invention are now applied in may areas of science, from
particle physics to sociology. In our work we extend the area
of applications to nonlinear multicomponent diffusion.

We demonstrated, how the mechanism based approach to
multicomponent diffusion can be included within the general
thermodynamic framework and proved the corresponding
dissipation inequalities. To satisfy the thermodynamic
restrictions, the kinetic law of an elementary process cannot
have an arbitrary form. For the general kinetic law $\phi
\exp(\alpha, \mu/RT)$ (the generalized Mass Action Law),
additional conditions on the set of kinetic factors $\phi$ are
needed. There are two main sets of these conditions. The
historically first of them, the condition of detailed balance,
follows from the special property of the underlying microscopic
dynamics, microreversibility. It was used by Boltzmann for his
equation and then by many researchers like Wegscheider
\cite{Wegscheider1901}, Einstein \cite{Einstein1916}, and
Onsager \cite{Onsager1}.  The second and more general property
we now call ``semi-detailed balance" or ``complex balance" was
proposed by Boltzmann in his discussion with Lorentz. The
theoretic principles underneath these conditions were
discovered by Stueckelberg in 1952 \cite{Stueckelberg1952} in
application to the Boltzmann equation. It received the name
``complex balance" ten years later in the works of Horn and
Jackson \cite{HornJackson1972} and Feinberg
\cite{Feinberg1972}. Recently, it was demonstrated how this
condition can be deduced from the quasiequilibrium and quasi
steady state approximations from Markov kinetics
\cite{GorbanShahzad2010}.

Explicit formulas for nonlinear multicomponent diffusion
combined from elementary processes can help in the practice of
modeling (and already helped \cite{GorBykYab1980}). The
cell--jump formalism gives an intuitively clear representation
of the elementary transport processes and, at the same time,
produces kinetic finite elements, a tool for numerical
simulation.

There remain many questions for the future work:
\begin{itemize}
\item{It is necessary to extend the experience of modeling of real systems;}
\item{The analysis of diffusion in solids should be
    properly coupled with the mechanics of solids. The
    detailed quantitative theory of the Kirkendall effect
    may be a proper challenge here;}
\item{The mechanism representation should be extended to
    viscosity and viscoelasticity;}
\item{The kinetic finite elements approach should be
    modified and extended to include continuum mechanics,
    perhaps, by proper joining with the lattice Boltzmann
    models approach;}
\item{Possible stoichiometric mechanisms of anomalous diffusion should be studied.}
\end{itemize}

\end{document}